\pdfoutput=1
\RequirePackage{fixltx2e}
\documentclass[12pt,letterpaper]{article}
\usepackage{putex}
\usepackage{graphicx}
\graphicspath{{Figures/}}
\usepackage{latexsym,amsmath,amsfonts,amssymb,amsthm}
\usepackage{bbm}
\usepackage{cite}
\usepackage{indentfirst}
\usepackage{booktabs,array}
\usepackage{placeins}
\usepackage{mathtools}
\usepackage{cancel}
\usepackage[mode=image|tex]{standalone}
\usepackage[dvipsnames]{xcolor}
\usepackage{tikz}
\usepackage{adjustbox}
\usepackage{genyoungtabtikz}
\usepackage{enumitem}
\usepackage{setspace}
\usepackage{mathdots}
\usetikzlibrary{decorations.markings}
\usepackage{graphicx}
\usepackage[margin=10pt,font=small,labelfont=bf]{caption}
\usepackage{subcaption}
\usepackage{xcolor}
\usepackage{float}
\theoremstyle{definition}
\newtheorem{conj}{Conjecture}
\usepackage{microtype}
\usepackage{hyperref}
\hypersetup{unicode}
\hypersetup{linktoc = all}
\hypersetup{pdfborderstyle={/S/U/W 0.5}}
\hypersetup{linkbordercolor = gray}
\hypersetup{bookmarksnumbered}
\newcommand{\eg}{\textsl{e.g.\@}}
\newcommand{\ie}{\textsl{i.e.\@}}

\numberwithin{equation}{section}

\newcommand{\cN}{\mathcal{N}}
\newcommand{\cQ}{\mathcal{Q}}

\DeclareMathOperator{\Tr}{Tr}
\DeclarePairedDelimiter{\abs}{\lvert}{\rvert}
\DeclarePairedDelimiter{\vev}{\langle}{\rangle}
\DeclareMathOperator*{\SumInt}{\ooalign{$\displaystyle\sum$\cr\hidewidth$\displaystyle\int$\hidewidth\cr}}

\newcommand{\nf}{n_{\text{f}}}
\newcommand{\zdz}{\vartheta}
\newcommand{\dimToda}{\Delta}
\newcommand{\anti}[1]{\texorpdfstring{{\widetilde{#1}}}{#1\003\140}}
\newcommand{\sch}{{\text{(s)}}} 
\newcommand{\uch}{{\text{(u)}}} 
\newcommand{\ssch}{{\text{(s,s)}}} 
\newcommand{\usch}{{\text{(u,s)}}} 

\newcommand{\beq}{\begin{equation}}
\newcommand{\eeq}{\end{equation}}

\begin{document}

\hypersetup{pdftitle=Intersecting Surface Defects and Two-Dimensional CFT}
\hypersetup{pdfauthor={Jaume Gomis, Bruno Le Floch, Yiwen Pan, Wolfger Peelaers}}


\title{\begin{LARGE}
Intersecting Surface Defects and Two\,-Dimensional CFT
\end{LARGE}\newline}

\date{Published 2 August 2017}
\authors{Jaume Gomis$^1$, Bruno Le Floch$^2$, Yiwen Pan$^3$ and Wolfger Peelaers$^4$
\medskip\medskip\medskip\medskip
 }

\institution{
PI}{${}^1$
Perimeter Institute for Theoretical Physics,\cr
$\;\,$ Waterloo, Ontario, N2L 2Y5, Canada}
\institution{PCTS}{${}^2$
Princeton Center for Theoretical Science, Princeton University,\cr
$\;\,$ Princeton, NJ 08544, USA}
\institution{UU}{${}^3$
Department of Physics and Astronomy, Uppsala University, \cr
$\;\,$ Box 516, SE-75120 Uppsala, Sweden}
\institution{NHETC}{${}^4$
New High Energy Theory Center, Rutgers University,  \cr
$\;\,$ Piscataway, NJ 08854, USA}

\abstract{\begin{onehalfspace}{
 We initiate the study of intersecting surface operators/defects in four-dimensional quantum field theories (QFTs). We characterize these defects by coupled 4d/2d/0d   theories constructed  by coupling the degrees of freedom localized at a point and on intersecting surfaces in spacetime to each other and to the four-dimensional QFT\@. We construct supersymmetric intersecting surface defects preserving just two supercharges in    $\cN=2$ gauge theories. These defects are amenable to exact analysis by localization of the partition function of the underlying 4d/2d/0d QFT\@. We identify the 4d/2d/0d QFTs that describe intersecting surface operators in $\cN=2$ gauge theories realized by intersecting M2-branes ending on $N$ M5-branes wrapping a Riemann surface. We conjecture and provide evidence for an explicit equivalence
between the squashed four-sphere partition function of these intersecting defects and correlation functions in Liouville/Toda CFT with the insertion of
arbitrary degenerate vertex operators, which are labeled by representations of $SU(N)$.}\end{onehalfspace}}

\preprint{UUITP-25/16}
\setcounter{page}{0}
\maketitle


{
\setcounter{tocdepth}{2}
\setlength\parskip{-0.7mm}
\tableofcontents
}

\section{Introduction}

The rich dynamics that a  Quantum Field Theory (QFT) can display may be probed with \textit{defects} of various dimensions. Classic examples  are the Wilson and 't~Hooft lines, which probe the state of the system through the response of an electrically and magnetically charged heavy particle respectively. In recent years, the construction of  novel defects of various (co)dimensions has significantly enlarged the probes available to quantum field theorists. Chief amongst  these are   codimension two  defects, which can discriminate  phases that are otherwise indistinguishable by the classic Wilson--'t~Hooft criterion~\cite{Gukov:2013zka}. Codimension two defects   define surface defects  in four dimensions (see   \cite{Gukov:2006jk,Braverman:2004vv,Gomis:2007fi,Gukov:2007ck,Witten:2007td,Buchbinder:2007ar,Beasley:2008dc,Gukov:2008sn,Drukker:2008wr,Gaiotto:2009fs} for early work) and vortex lines in three dimensions  \cite{Drukker:2008jm,Kapustin:2012iw,Drukker:2012sr,Assel:2015oxa}. For a recent review on     surface defects  see   \cite{Gukov:2014gja}.

 Defects in a QFT can be defined by coupling   the \textit{bulk} QFT to   additional  degrees of freedom that are localized on the support of the defect.  Canonically, the coupling is implemented by  gauging   global symmetries acting on the  defect degrees of freedom with   bulk gauge symmetries and/or  by identifying bulk and defect global symmetries through couplings between defect and bulk matter fields. A defect global symmetry   associated to the defect conserved current $J_\mu$ is gauged with a bulk gauge field $A_M$ through the following   coupling integrated over the defect:
 \beq
 \int_D \mathrm{d}x\, A_\mu(x,x_\perp=0) J^\mu(x) +\hbox{seagull terms}\,.
 \label{minimal}
 \eeq
 This construction realizes a defect operator as a  lower-dimensional QFT on the support~$D$ of the defect interacting with the bulk QFT and provides a uniform description of Wilson lines, vortex lines and surface   defects, among others.\footnote{It is not
 known how to realize a 't~Hooft line  by integrating out  localized degrees of freedom on the line defect.} The realization of defect operators as defect degrees of freedom coupled to the bulk QFT has played a key role in unraveling the action of various dualities on defect operators, see, \eg{}, \cite{Gomis:2006sb,Assel:2015oxa}.

The set of defects in a QFT can be enlarged by considering \textit{intersecting defects}. These are constructed intuitively by letting a collection of defects of various codimensions intersect in spacetime. This picture has a natural QFT realization. First,
 each defect comes equipped with its own localized degrees of freedom which couple to the bulk QFT as described above, just  as if the defect  were inserted in isolation. In the presence of multiple defects, this construction  can be further enriched by adding new intersection degrees of freedom along the intersection domain of the defects and letting them couple to the corresponding defect degrees of freedom (as well as the bulk). This is again accomplished by
 gauging the  flavor symmetries acting on the intersection degrees of freedom  with   gauge symmetries residing on the various defects (and/or bulk) and/or by identifying them with defect (and/or bulk) global symmetries.   Intersecting defects    exhibit quite a rich dynamics as they bring together under a single roof the intricate dynamics of QFTs in various dimensions.

\begin{figure}
  \centering
  \includegraphics{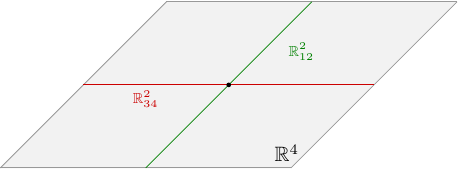}
  \caption{\label{fig:intersection-picture}   {Intersecting codimension two defects supported on planes $\mathbb{R}^2_{12}$ and $\mathbb{R}^2_{34}$}. There are localized degrees of freedom living on the planes $\mathbb{R}^2_{12}$ and $\mathbb{R}^2_{34}$ and at the origin; the latter couple to the former degrees of freedom, which in turn   couple  to the four-dimensional gauge theory living in the bulk $\mathbb{R}^4$.}
  \label{intersectfig}
\end{figure}

In this paper we initiate the study of intersecting surface defects in   four-dimensional gauge theories. More precisely, we consider the case of orthogonal planar surface defects intersecting at a point (see \autoref{fig:intersection-picture} for a pictorial representation). 
We focus our investigations  on  intersecting surface defects in four-dimensional~$\cN=2$   supersymmetric field theories that   preserve the zero-dimensional dimensional reduction of two-dimensional $\cN=(0,2)$ supersymmetry.  These intersecting surface operators on $\mathbb{R}^4$ are constructed by coupling an $\cN=(0,2)$ zero-dimensional theory\footnote{Namely zero-dimensional dimensional reductions of two-dimensional $\cN=(0,2)$ theories.}   at $x^1=x^2=x^3=x^4=0$ to a two-dimensional $\cN=(2,2)$ theory at $x^3=x^4=0$ and to a two-dimensional  $\cN=(2,2)$ theory at $x^1=x^2=0$. These two-dimensional theories are in turn coupled to the bulk four-dimensional~$\cN=2$ theory.\footnote{The zero-dimensional fields can also couple directly to the four-dimensional QFT\@.} This construction is very general, and defines a very large class of intersecting surface defects.

Pleasingly, the expectation values of these intersecting surface defects in the $\Omega$-background \cite{Nekrasov:2002qd} and on the squashed four-sphere~\cite{Pestun:2007rz,Hama:2012bg} are amenable to exact computation by supersymmetric localization, yielding novel non-perturbative results in four-dimensional QFTs.
Consider an intersecting defect on  the squashed four-sphere $S^4_b$ with the surface defects wrapping orthogonal two-spheres $S^2_L$ and $S^2_R$ that intersect at two points, the north pole and south pole of $S^4_b$. We show that the expectation value of the intersecting defect takes the  form 
\begin{equation}
\SumInt \ Z_{S^4_b}\ Z_{S^2_L} \ Z_{S^2_R} \   Z_{\text{0d}}^{\text{intersection}}\ \left|Z_{\text{instanton}}\right|^2 \,,
   \label{finalanswnewintro}
\end{equation}
where $Z_{S^4_b}$ is the  one-loop determinant of the bulk four-dimensional $\mathcal N=2$ theory together with the classical contribution, and $Z_{S^2_{L}}$ and $Z_{S^2_{R}}$ denote the  one-loop determinants and classical contributions of the two-dimensional $\mathcal N=(2,2)$ theories living on the respective surface defects, which are coupled to the four-dimensional theory.   $Z_{\text{0d}}^{\text{intersection}}$ is the one-loop determinant of the intersection degrees of freedom pinned at the poles and coupling to the two-dimensional (and four-dimensional) theories. Finally, $|Z_{\text{instanton}}|^2$ are two copies of the instanton partition function, one for the north pole and one for the south pole of ${S^4_b}$, encoding the contribution of instantons in the presence of the intersecting surface defects. The two-dimensional and zero-dimensional fields introduce new elements to the instanton computation, by specifying the allowed singular behavior of the four-dimensional gauge fields and by contributing extra zero-modes to the integral over the appropriate instanton moduli space. In this paper we perform the detailed computation of the expectation value of intersecting defects in four-dimensional theories without gauge fields (see \autoref{section:intersecting-partition-function-first-principle}).

\begin{figure}\centering
  \includegraphics{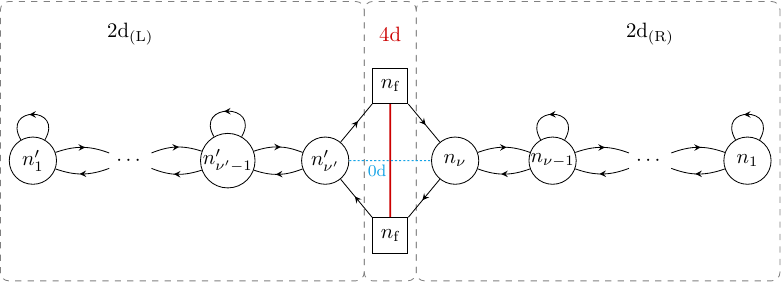}
\caption{\label{fig:proposalintersectingWVT}Joint 4d/2d/0d quiver diagram (later denoted $\mathcal{T}_{\text{Fermi}}$) describing the M2-brane intersection labeled by representations $(\mathcal{R}^\prime,\mathcal{R})$ ending on $n_\text{f}$ M5-branes wrapping a trinion with two full and one simple puncture. The four-dimensional degrees of freedom are denoted in $\cN=2$ quiver notation, the two-dimensional ones in $\cN=(2,2)$ quiver notation, and the zero-dimensional ones in the dimensional reduction of two-dimensional $\cN=(0,2)$ quiver notation, with dashed lines representing Fermi multiplets and solid arrows chiral multiplets. The ranks of the gauge groups are determined by the representations $(\mathcal{R}^\prime,\mathcal{R})$ as in \autoref{fig:quiver-young-diagram} and the complexified FI parameters of the innermost gauge group factors are opposite while the others vanish.  In both halves of the quiver the adjoint chiral multiplets are coupled through cubic superpotentials to their neighboring bifundamental chiral multiplets.  The two-dimensional chiral multiplets charged under $U(n_\nu)$ or $U(n'_{\nu'})$ are coupled through cubic and quintic superpotentials to the four-dimensional degrees of freedom, and appear in $E$~or $J$ terms for the 0d Fermi multiplet.  More generally, the 4d $SU(n_\text{f})\times SU(n_\text{f})\times U(1)$ symmetry can be partly or fully gauged to insert this 4d/2d/0d quiver in a larger 4d quiver gauge theory; we then call this 4d/2d/0d quiver ``local'' to insist on the presence of other 4d degrees of freedom.}
\end{figure}

We proceed to identify a family of  intersecting surface defects in four-dimensional~$\cN=2$  theories which admit an elegant  interpretation in two-dimensional non-rational conformal field theory (CFT) and     realize the low-energy dynamics of two intersecting sets of M2-branes ending on $n_\text{f}$ M5-branes wrapping a punctured Riemann surface. The configuration of intersecting M2-branes is labeled by a pair of irreducible representations $(\mathcal R^\prime,\mathcal R)$ of $SU(n_\text{f})$.
On the M5-branes resides a four-dimensional $\cN=2$   theory dictated by the choice of Riemann surface~\cite{Gaiotto:2009we} and the   M2-branes insert a    surface operator~\cite{Alday:2009fs,Gomis:2014eya}, whose field theory description we provide. Our construction realizes intersecting M2-brane surface operators   in    four-dimensional $\cN=2$   theories on M5-branes that admit  a choice of duality frame with an $SU(n_\text{f})\times SU(n_\text{f})\times U(1)$  symmetry,\footnote{This symmetry is   associated  to a trinion with two full and one simple puncture  in a  pants decomposition of the Riemann surface~\cite{Gaiotto:2009we}}
 which allows for the gauging of the corresponding global symmetries of the defect fields. This includes, among many other theories, $\cN=2$ $SU(n_\text{f})$ SQCD with $2n_\text{f}$ fundamental hypermultiplets and the $\cN=2^*$ theory, that is $\cN=2$ $SU(n_\text{f})$ super-Yang--Mills with a massive adjoint hypermultiplet.

We state, for clarity,  our results and conjectures for the simplest four-dimensional $\cN=2$ theory in this class: the theory of $n_\text{f}^2$ hypermultiplets, living on $n_\text{f}$ M5-branes wrapping a trinion with two full and one simple puncture.

\begin{conj}\label{conj:fermi}
The M2-brane intersection labeled by representations $(\mathcal{R}^\prime,\mathcal{R})$ of $SU(n_\text{f})$ ending on the $n_\text{f}$ M5-branes is described by the joint 4d/2d/0d quiver diagram in \autoref{fig:proposalintersectingWVT}.\footnote{There does not exist a unique quiver gauge description of intersecting surface defects as various dualities can take it to a different, but equivalent, one. Some of the duality frames may involve additional 0d fields. Indeed, we will encounter explicit examples of this.}
\end{conj}

The $SU(n_\text{f})\times SU(n_\text{f})\times U(1)$ global symmetries acting on the innermost chiral multiplets  of the right and left ${\cN=(2,2)}$ quiver gauge theories are identified with each other and with those acting on the bulk hypermultiplets via   defect, two-dimensional ${\cN=(2,2)}$ superpotentials, one localized in the $(x^1,x^2)$-plane and the other in the $(x^3,x^4)$-plane.  Quintic superpotentials identify the remaining $U(1)$ global symmetry of each two-dimensional theory to rotations transverse to the corresponding plane.  The $\cN=(0,2)$ Fermi multiplet localized at $x^1=x^2=x^3=x^4=0$ is gauged with the innermost gauge group  factor of the left and right $\cN=(2,2)$ quiver gauge theory.  The Fermi multiplet has an $E$-term or $J$-term superpotential\footnote{A Fermi multiplet is equivalent to its conjugate up to exchanging $E$-type and $J$-type superpotentials, thus we depict it in quivers as an unoriented (dashed) edge.} quadratic in the 0d $\cN=(0,2)$ restrictions of the 2d chiral multiplets.

\begin{figure}\centering
  \includegraphics{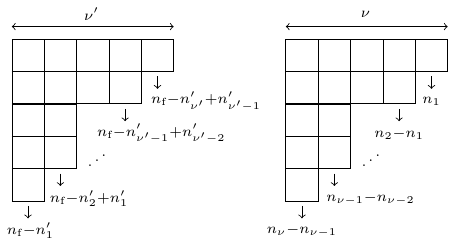}
  \caption{\label{fig:quiver-young-diagram}Gauge group ranks corresponding to Young diagrams of $(\mathcal{R}',\mathcal{R})$.}
\end{figure}

The representation data $(\mathcal{R}',\mathcal{R})$ labeling the intersecting M2-branes is encoded in the ranks of the gauge groups of the two-dimensional ${\cN=(2,2)}$ gauge theories on the left and right of the diagram by realizing $(\mathcal{R}^\prime,\mathcal{R})$ by a pair of Young diagrams, as in \autoref{fig:quiver-young-diagram}.
The number of boxes in each column of the Young diagram determine the rank of the gauge group of the corresponding ${\cN=(2,2)}$ gauge theory.\footnote{The dictionaries on the left and right only differ by conjugating the representation, which turns each column with $k$~boxes into a column with $(n_\text{f}-k)$ boxes.  Seiberg-like dualities of each 2d $\cN=(2,2)$ theory relate the quiver given here to $\nu!\nu'!$ quivers with permuted $(n_\kappa-n_{\kappa-1})$ and permuted $(n'_\kappa-n'_{\kappa-1})$.}

The complexified Fayet--Iliopoulos (FI) parameters
\begin{equation}
 \tau=i\xi+\frac{\vartheta}{2\pi}
 \label{FI}
\end{equation}
for the innermost gauge groups $U(n_\nu)$ and $U(n'_{\nu'})$ are opposite while the FI parameters for all other gauge groups vanish.\footnote{More precisely, $e^{2\pi i\tau_\iota}=(-1)^{n_{\iota-1}+n_{\iota+1}}$ for $1\leq\iota<\nu$ and $e^{2\pi i\tau'_\iota}=(-1)^{n'_{\iota-1}+n'_{\iota+1}}$ for $1\leq\iota<\nu'$.} The surviving complexified FI parameter encodes the position on the Riemann surface where the intersecting M2-branes end. For the precise brane configuration see \autoref{sec: M2-brane surface defects}.

The same quiver with $\nu+\nu'$ arbitrary FI parameters corresponds to the insertion of $\nu$~sets of M2-branes labeled by antisymmetric representations\footnote{We denote symmetric/antisymmetric powers of the fundamental representation $\protect\yngs(.4,1)$ by $\operatorname{sym}^n \protect\yngs(.4,1)$ and $\wedge^n \protect\yngs(.4,1)$.} $(1,\mathop{\wedge^{n_{\kappa}-n_{\kappa-1}}}\yngs(.4,1))$ and $\nu'$~sets labeled by $(\mathop{\wedge^{n_{\text{f}}-n'_{\kappa}+n'_{\kappa-1}}}\yngs(.4,1),1)$.  Their respective positions on the Riemann surface are encoded in the FI parameters.\footnote{By taking some of the FI parameters to vanish, one can bring subsets of the $\nu+\nu'$ branes together at different points on the Riemann surface and hence realize an arbitrary family of M2-brane intersections labeled by arbitrary representations.}

\begin{conj}\label{conj:Wconfblocks}
The instanton partition function in the $\Omega$-background  $\mathbb{R}^4_{\epsilon_1,\epsilon_2}$ of the family of intersecting defects captured by the  4d/2d/0d quiver diagram in \autoref{fig:proposalintersectingWVT} equals the $W_{n_\text{f}}$ conformal block on the four-punctured sphere with two full punctures, one simple puncture and an arbitrary degenerate puncture.  The choice of internal momentum labeling the conformal block maps to a choice of boundary condition for the vector multiplet scalars of the innermost gauge group factors in the intersecting defect theory.
\end{conj}
A degenerate puncture of the $W_{n_\text{f}}$ algebra is labeled by two dominant weights $(\Omega',\Omega)$ of $SU(n_\text{f})$ through the momentum vector
 \beq
 \alpha=-b\,\Omega -\frac{1}{b} \Omega'\,,
 \eeq
  where $b$ parametrizes the Virasoro central charge.\footnote{In detail, $c=(n_\text{f}-1)\bigl[1+n_\text{f}(n_\text{f}+1)(b+b^{-1})^2\bigr]$.}
 The data of the degenerate puncture is realized in the quiver diagram through  the  irreducible representations $(\mathcal{R}^\prime,\mathcal{R})$, which have highest weights $(\Omega',\Omega)$. The $\mathbb R^4_{\epsilon_1,\epsilon_2}$ deformation parameters are given in terms of the Virasoro central charge by $\epsilon_1=b$ and $\epsilon_2=1/b$ with $b>0$.\footnote{Our results apply more generally for $\Re(\epsilon_1/\epsilon_2)\geq 0$, see \autoref{foot:other-b-values} for details.} The masses of the four-dimensional and two-dimensional matter fields are encoded in the momenta of the two full punctures and the simple puncture (see \autoref{section:dictionary}).

\begin{conj}\label{conj:toda}
The expectation value  on the squashed four-sphere $S^4_b$
  \beq
  \frac{x_0^2}{r^2}+\frac{x_1^2+x_2^2}{\ell^2}+\frac{x_3^2+x_4^2}{\tilde \ell^2}=1
  \eeq
  of the intersecting surface theory in \autoref{fig:proposalintersectingWVT}, with the right $\cN=(2,2)$ quiver on the squashed two-sphere at $x_3=x_4=0$, the left $\cN=(2,2)$ quiver on the squashed two-sphere at $x_1=x_2=0$, and with the bifundamental $\cN=(0,2)$ Fermi multiplet localized at the North and South poles of $S^4_b$ at $x_0=r$ and $x_0=-r$  respectively, is given by the
  $A_{n_\text{f}-1}$ Toda CFT correlator on the four-punctured sphere with   two full punctures, one simple puncture and an arbitrary degenerate puncture labeled by $(\Omega',\Omega)$.  The Toda CFT central charge parameter is given by $b^2=\ell/\tilde \ell$.
\end{conj}

\begin{figure}\centering
  \includegraphics{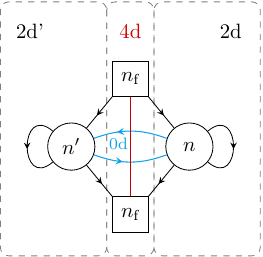}
\caption{\label{fig:proposalintersectingWVTsymmetrics} Joint 4d/2d/0d quiver diagram describing the M2-brane intersection labeled by the $n$- and $n^\prime$-fold symmetric representations  ending on $n_\text{f}$ M5-branes wrapping a trinion with two full and one simple puncture. The complexified FI parameters of the gauge group factors are equal. Cubic and quartic superpotentials coupling the four-dimensional degrees of freedom to the two-dimensional ones are turned on.  The 0d chiral multiplets on the intersection appear in $E$~and $J$-type superpotentials for 0d $\cN=(0,2)$ Fermi multiplet components of the 2d $\cN=(2,2)$ (anti)fundamental chiral multiplets.  As in \autoref{fig:proposalintersectingWVT}, the 4d $SU(n_\text{f})\times SU(n_\text{f})\times U(1)$ symmetry can be global or gauged.}
\end{figure}

\begin{conj}\label{conj:symmetric}
The M2-brane intersection labeled by representations $(\operatorname{sym}^{n^\prime}\yngs(.4,1),\operatorname{sym}^{n}\yngs(.4,1))$ of $SU(n_\text{f})$ ending on the $n_\text{f}$ M5-branes allows for an alternative description in terms of the joint 4d/2d/0d quiver diagram in \autoref{fig:proposalintersectingWVTsymmetrics}.\footnote{\autoref{fig:quivermanysymm} in \autoref{sec: M2-brane surface defects} gives the quiver for any number of M2-brane intersections labeled by symmetric representations.} Similarly to \autoref{conj:Wconfblocks}, the instanton partition function of the 4d/2d/0d gauge theory coincides with a $W_{\nf}$ conformal block on the four-punctured sphere with two full puncture, one simple puncture and a degenerate puncture labeled by the symmetric representations $(\operatorname{sym}^{n^\prime}\yngs(.4,1),\operatorname{sym}^{n}\yngs(.4,1))$.  Similarly to \autoref{conj:toda}, the $S^4_b$~expectation value coincides with the $A_{\nf-1}$~Toda CFT correlator with these four punctures.
\end{conj}

These results enrich the fascinating connections uncovered by AGT~\cite{Alday:2009aq}   between four-dimensional
theories  (see also \cite{Wyllard:2009hg}) and   between two-dimensional theories~\cite{Gomis:2014eya}    and two-dimensional  Toda CFT\@. Our   mapping of the intersecting defects in \autoref{fig:proposalintersectingWVT} with the most general Toda degenerate field insertion, which is labeled by the pair of representations $(\mathcal{R}',\mathcal{R})$, completes~\cite{Gomis:2014eya}, where one of the representations was taken to be trivial (see also \cite{Alday:2009fs,Dimofte:2010tz,Taki:2010bj,Bonelli:2011fq,Bonelli:2011wx,Doroud:2012xw}). Realizing the most general degenerate insertion crucially requires considering intersecting defects, with degrees of freedom localized along intersecting  surfaces and points on spacetime.

\begin{figure}
  \centering
  \includegraphics{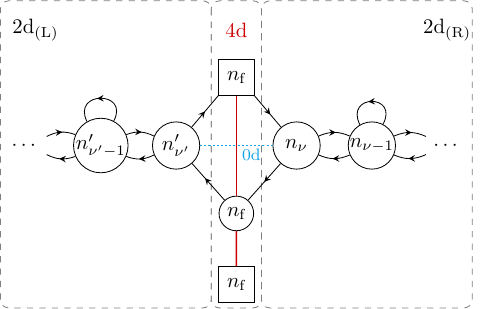}
  \caption{\label{fig:proposalintersectingSQCD} Joint 4d/2d/0d quiver realizing an M2-brane surface operator in $\cN=2$ SQCD\@.  This has the same matter content as the quiver in \autoref{fig:proposalintersectingWVT}.}
\end{figure}

Extending our story to other four-dimensional $\cN=2$ theories  with the properties described above is   straightforward. In the field theory, we
gauge the $SU(n_\text{f})\times SU(n_\text{f})\times U(1)$ global symmetry  of the 2d/0d degrees of freedom with an $SU(n_\text{f})\times SU(n_\text{f})\times U(1)$ symmetry of the four-dimensional theory. In the correspondence with Toda CFT, we insert an extra degenerate puncture labeled by $(\Omega',\Omega)$ on the punctured Riemann surface realizing the four-dimensional $\cN=2$ theory under consideration.
 As an example, the 4d/2d/0d quiver diagram for an M2-brane intersecting surface operator  in   four-dimensional SQCD is given by \autoref{fig:proposalintersectingSQCD}.\footnote{Equivalently, the defect $SU(n_\text{f})\times SU(n_\text{f})\times U(1)$ symmetry could be gauged using the bottom two nodes of the SQCD quiver. The two descriptions
are dual to each other and related by hopping duality~\cite{Gadde:2013ftv,Gomis:2014eya}.}
The partition function of this theory is conjecturally computed by the Toda CFT five-point function on the sphere, with two full punctures, two simple punctures and a degenerate puncture that encodes the choice of intersecting surface operator.

The paper is organized as follows. In \autoref{section: intersecting surface defects} we provide a general framework for the construction of quarter-supersymmetric intersecting defects in $\cN=2$ QFTs. In \autoref{section:intersecting-partition-function-first-principle} we compute exactly the expectation value of intersecting surface defects on the squashed four-sphere. \autoref{sec: M2-brane surface defects} discusses the M-theory realization of the intersecting surface defects of interest to this paper. Here we also show how the proposed 4d/2d/0d quiver gauge theories of \autoref{fig:proposalintersectingWVT} and \autoref{fig:proposalintersectingWVTsymmetrics} naturally arise in theories admitting a type IIA description. \autoref{section:dictionary} states the conjectured relation with Liouville/Toda degenerate correlators precisely. It describes the concrete and non-trivial verifications of our conjectures done in Appendix~\ref{appendix:Liouville-check} and Appendix~\ref{appendix:fermi-quiver-check}. We conclude with some interesting open questions and future directions.

\section{\label{section: intersecting surface defects}Coupling Intersecting Defects}

A planar, half-supersymmetric   surface defect in a four-dimensional $\cN=2$ theory can preserve either two-dimensional $\cN=(2,2)$ or $\cN=(0,4)$ supersymmetry. Indeed, the supercharges\footnote{We denote the four-dimensional $\cN=2$ Poincar\'e supercharges as $Q^A_\alpha, \overline Q^A_{\dot \alpha}$, with $A$ an $SU(2)_{\mathcal{R}}$ index and $\alpha, \dot\alpha$ Lorentz spinor indices.} of the bulk supersymmetry algebra
\beq
\{Q_\alpha^A,\overline Q_{\dot \alpha}^B\}= \epsilon^{AB} P_{\alpha\dot\alpha}
\eeq
preserved by a half-supersymmetric defect spanning the $(x^1,x^2)$-plane generate either a two-dimensional $\cN=(2,2)$ supersymmetry algebra, say, $(Q_+^1,Q_-^2,\overline Q_{\dot +}^2,\overline Q_{\dot -}^1)$, or an $\cN=(0,4)$ algebra, \eg{}, $(Q_+^A,\overline Q_{\dot +}^A)$.

Surface defects preserving these symmetries can be constructed by coupling a two-dimensional   $\cN=(2,2)$ or $\cN=(0,4)$ QFT supported on the defect to the four-dimensional theory.  This is done by gauging global symmetries of the defect QFT with bulk gauge or global symmetries and by additional potential terms.\footnote{\label{foot:global-remains}When a global $U(1)$ symmetry is gauged using a $U(1)$ gauge field of another theory (in our case, the bulk theory), there remains a global $U(1)$ symmetry (eliminated in our case by superpotentials).  A toy model of this property is as follows.  Start with two theories: $N$~free chiral multiplets, and a $U(1)$ vector multiplet coupled to $N$~charge $-1$ chiral multiplets.  Gauging the $U(1)$ flavor symmetry of the free chiral multiplets using the $U(1)$ vector multiplet yields SQED, which has $SU(N)\times SU(N)\times U(1)$ global symmetry.  The $U(1)$ factor stems from the original $U(1)$ flavor symmetry of the first theory up to gauge redundancy.} The minimal coupling~\eqref{minimal} and potential terms must be supersymmetrized.
 A   strategy to write down the action of these surface defects which makes manifest the supersymmetry of the defect theory is to rewrite the four-dimensional $\cN=2$ theory as a two-dimensional  $\cN=(2,2)$ or $\cN=(0,4)$ theory.\footnote{For a sample of references of this approach see, \eg{}, \cite{ArkaniHamed:2001tb,Constable:2002xt,Gaiotto:2009fs,Mintun:2014aka,Lamy-Poirier:2014sea}.} Indeed, by decomposing the four-dimensional multiplets in terms of the two-dimensional  $\cN=(2,2)$ or $\cN=(0,4)$ ones,
the bulk Lagrangian can be reproduced from the action constructed out of the lower-dimensional multiplets.\footnote{The four-dimensional Lorentz invariance of the bulk theory is reproduced after terms in the Lagrangian of different lower-dimensional multiplets are combined.}
The coordinates transverse to the defect appear  from the lower-dimensional viewpoint as continuous labels of the multiplets. The advantage of this approach is that it is now straightforward and manifestly two-dimensional $\cN=(2,2)$ or $\cN=(0,4)$ supersymmetric to couple   the bulk theory to a two-dimensional  $\cN=(2,2)$ or $\cN=(0,4)$ theory by gauging the flavor symmetries of the defect theory with bulk symmetries.
 The matter multiplets of the four-dimensional $\cN=2$ theory (\ie{}, hypermultiplets) can also be coupled via a localized $\cN=(2,2)$ or $\cN=(0,4)$ superpotential to the matter multiplets on the defect,  thus identifying the defect flavor symmetries with either bulk   gauge  or global symmetries.
In this way, the surface defect coupled to the bulk is represented   as a two-dimensional $\cN=(2,2)$ or $\cN=(0,4)$ QFT\@. Schematically,  the action describing the surface defect takes the form
\beq
S= S_{\text{4d}}+S_{\text{2d}}+S_{\text{2d/4d}}\,.
\eeq
This leads to a large family of surface operators in four-dimensional $\cN=2$ theories.

\begin{figure}
  \centering
  \includegraphics{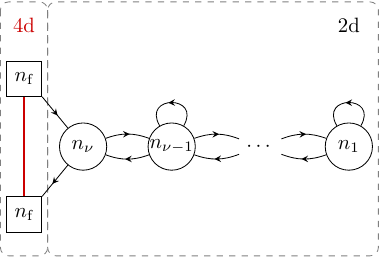}
  \caption{\label{fig:one-side-quiver} Local 4d/2d quiver diagram describing a class of $\cN=(2,2)$ preserving surface defects. The  4d $SU(n_\text{f})\times SU(n_\text{f})\times U(1)$ symmetry can be global or gauged.}
\end{figure}

The class of $\cN=(2,2)$ preserving surface defects that will be most relevant for us is encoded by the ``local'' 4d/2d quiver diagram of \autoref{fig:one-side-quiver}.\footnote{We call this quiver diagram ``local'' to emphasize that it only shows the four-dimensional fields to which the two-dimensional theory couples, and that these four-dimensional fields may be part of a larger quiver gauge theory.}  These surface defects were studied in detail in~\cite{Gomis:2014eya} and   given a two-dimensional CFT interpretation. Related $\cN=(2,2)$ surface defects were analyzed in~\cite{Gadde:2013ftv}.
The $n_\text{f}$ fundamental and antifundamental chiral multiplets on the inner end of the two-dimensional quiver couple  to the $n_\text{f}^2$ hypermultiplets via a localized cubic superpotential preserving two-dimensional $\cN=(2,2)$ supersymmetry. The superpotential identifies the $SU(n_\text{f})\times SU(n_\text{f})\times U(1)$ flavor symmetry acting on the chiral multiplets with a subgroup of the symmetry acting on the hypermultiplets.  The hypermultiplet scalars $(Q,\tilde{Q})$, which transform in conjugate representations of $SU(n_\text{f})\times SU(n_\text{f})\times U(1)$, are bottom components of 2d $\cN=(2,2)$ chiral multiplets which we denote $(Q^{\text{2d}},\tilde{Q}^{\text{2d}})$.\footnote{This decomposition looks analogous to the decomposition into a pair of 4d $\cN=1$ chiral multiplets, but differs in which fermions appear in each multiplet.  The four-dimensional $\cN=2$ vector multiplet decomposes into a two-dimensional $\cN=(2,2)$ vector multiplet and an adjoint chiral multiplet.} If we denote by $q$ and~$\tilde{q}$ the fundamental and anti-fundamental two-dimensional chiral multiplets, the relevant defect superpotential is
\begin{equation}
 S^{\text{cubic}}_{\text{2d/4d}}=\int  \mathrm{d}^4x\,  \delta (x^3) \delta (x^4)  \int \mathrm{d}^2\theta\, q\tilde{q}Q^{\text{2d}}\,.
 \label{cubicone}
\end{equation}
This manifestly two-dimensional $\cN=(2,2)$ supersymmetric superpotential couples a  gauge invariant meson operator of the  two-dimensional theory  to the hypermultiplets. Since masses in four-dimensional $\cN=2$ and two-dimensional $\cN=(2,2)$ theories are vevs for background vector multiplets for the flavor symmetries, the superpotential fixes the masses  of the  hypermultiplets in terms of the sum of the masses of the two-dimensional fundamental and anti-fundamental chiral multiplets (see \autoref{section:intersecting-partition-function-first-principle}).  In addition to~\eqref{cubicone}, a quintic superpotential couples the (next-to) innermost bifundamental chiral multiplets $q^{\text{bif}}$ and~$\tilde{q}^{\text{bif}}$ to $q$ and $\tilde{q}$ and to the chiral multiplet whose bottom component is a transverse derivative of~$Q$:
\begin{equation}
 S^{\text{quintic}}_{\text{2d/4d}}=\int  \mathrm{d}^4x\,  \delta (x^3) \delta (x^4)  \int \mathrm{d}^2\theta\, qq^{\text{bif}}\tilde{q}^{\text{bif}}\tilde{q}\bigl((\partial_3-i\partial_4)Q^{\text{2d}}\bigr)\,.
 \label{quinticone}
\end{equation}
It identifies the remaining two-dimensional flavor symmetry $U(1)$ (under which adjoint and bifundamental chiral multiplets have charges $2$ and $-1$ respectively) to rotations transverse to the defect.

In this paper we study intersecting surface defects  in four-dimensional $\cN=2$ theories constructed from $\cN=(2,2)$ planar surface defects spanning the $(x^1,x^2)$-plane and the $(x^3,x^4)$-plane. The  defects intersect at the origin of $\mathbb{R}^4$.  These intersecting surface defects can preserve two supercharges\footnote{Indeed, an $\cN=(2,2)$ surface defect supported on the $(x^3,x^4)$-plane can be chosen to preserve $(Q_+^1,Q_-^2,\overline Q_{\dot -}^2,\overline Q_{\dot +}^1)$. Note that the choice of $\mathcal N=(2,2)$ subalgebras preserved by the individual defects must be correlated to ensure the intersecting system is quarter-BPS.} of  the four-dimensional $\cN=2$ theory: $(Q_+^1,Q_-^2)$.  The  field theory description of these intersecting defects is invariant under the zero-dimensional dimensional reduction of two-dimensional $\cN=(0,2)$ supersymmetry.  When the intersecting defect is superconformal it preserves the following subalgebra of the four-dimensional $\cN=2$ superconformal algebra
\begin{equation}\label{fullSubalgebraIntersecting}
\mathfrak{su}(1|1)_1\oplus \mathfrak{su}(1|1)_2\oplus \mathfrak{u}(1)_3  \subset  \mathfrak{su}(2,2|2) \;.
\end{equation}

The field theory  construction of these intersecting surface defects allows for the insertion of a two-dimensional $\cN=(0,2)$ QFT dimensionally reduced to zero dimensions   at the intersection point. This defect  $\cN=(0,2)$ QFT can now be coupled to the two-dimensional  $\cN=(2,2)$ QFTs living in the $(x^1,x^2)$ and $(x^3,x^4)$-planes. The global symmetries of the zero-dimensional intersection QFT  can be gauged with those of the  two-dimensional  $\cN=(2,2)$ QFTs or four-dimensional $\cN=2$ QFT\@. This gauging can be explicitly carried out by first writing down the two-dimensional  $\cN=(2,2)$ QFTs living in the $(x^1,x^2)$ and $(x^3,x^4)$-planes as zero-dimensional $\cN=(0,2)$ theories in the spirit explained above. This  requires decomposing a two-dimensional $\cN=(2,2)$ vector multiplet into a zero-dimensional $\cN=(0,2)$ vector multiplet and chiral multiplet and   a  two-dimensional $\cN=(2,2)$
chiral multiplet into a zero-dimensional $\cN=(0,2)$   chiral multiplet and Fermi multiplet.
In this way, the two-dimensional  $\cN=(2,2)$ QFTs can now be rewritten as zero-dimensional $\cN=(0,2)$ theories and gauging the flavor symmetries of the zero-dimensional $\cN=(0,2)$ theory at the intersection  with those of the $\cN=(2,2)$ theories  in the $(x^1,x^2)$ and $(x^3,x^4)$-planes becomes standard.  In general, it is possible to add  zero-dimensional $\cN=(0,2)$ superpotentials coupling the   various matter multiplets in zero, two and four-dimensions while preserving all the symmetries. Each $\cN=(0,2)$ Fermi multiplet admits so-called $E$-type and $J$-type superpotentials (see \cite{Witten:1993yc} for more  background material on $\cN=(0,2)$ theories). This construction furnishes the Lagrangian description of our quarter-supersymmetric surface defects. Schematically it looks like
\begin{equation}\label{schemAction}
  S = S_{\text{4d}} + S_{\text{2d}}^{(L)} + S_{\text{2d}}^{(R)} + S_{\text{0d}} + S_{\text{2d/4d}}^{(L)} + S_{\text{2d/4d}}^{(R)} + S_{\text{0d/2d}}^{(L)} + S_{\text{0d/2d}}^{(R)} + S_{\text{0d/2d/4d}} \,.
\end{equation}
The schematic action \eqref{schemAction} captures a large class of intersecting surface operators.  We now describe two cases of importance for brane systems later in the paper.  In both cases the 0d theories  involve $\cN=(0,2)$ Fermi or chiral multiplets (no vector multiplets).

The first class of intersecting surface defects we will focus on in this paper is
 neatly summarized by the local 4d/2d/0d quiver diagram of \autoref{fig:proposalintersectingWVT}.
The left and right two-dimensional $\cN=(2,2)$ theories couple via cubic and quintic superpotentials to the four-dimensional hypermultiplets.  If we denote by $(q_{(L)},\tilde{q}_{(L)},q_{(L)}^{\text{bif}},\tilde{q}_{(L)}^{\text{bif}})$ and $(q_{(R)}, \tilde{q}_{(R)},q_{(R)}^{\text{bif}},\tilde{q}_{(R)}^{\text{bif}})$ the inner fundamental, anti-fundamental, and bifundamental chiral multiplets of the left and right $\cN=(2,2)$ quivers with respect to their corresponding gauge group, and by $Q^{\text{2d}}_{(L)}$ and $\tilde{Q}^{\text{2d}}_{(R)}$ the two-dimensional chiral multiplets whose bottom components are the hypermultiplet scalars $Q$ and~$\tilde{Q}$, then the superpotential couplings are
\begin{align}
  S^{(R)}_{\text{2d/4d}}&=\int \mathrm{d}^4x\,  \delta (x^3) \delta (x^4) \int \mathrm{d}^2\theta_{(R)}\, \Bigl( q_{(R)}\tilde{q}_{(R)}Q^{\text{2d}}_{(R)} + q_{(R)}q^{\text{bif}}_{(R)}\tilde{q}^{\text{bif}}_{(R)}\tilde{q}_{(R)}\bigl((\partial_3-i\partial_4)Q^{\text{2d}}_{(R)}\bigr)\Bigr)
\label{defectW1}\\
  S^{(L)}_{\text{2d/4d}}&=\int \mathrm{d}^4x\,  \delta (x^1) \delta (x^2) \int \mathrm{d}^2\theta_{(L)}\, \Bigl( q_{(L)}\tilde{q}_{(L)} \tilde{Q}^{\text{2d}}_{(L)} + q_{(L)}q^{\text{bif}}_{(L)}\tilde{q}^{\text{bif}}_{(L)}\tilde{q}_{(L)}\bigl((\partial_1-i\partial_2)\tilde{Q}^{\text{2d}}_{(L)}\bigr)\Bigr) \,.
\label{defectW2}
\end{align}
The cubic superpotentials identify the $SU(n_\text{f})\times SU(n_\text{f})\times U(1)$ flavor symmetries acting on the inner fundamental  and anti-fundamental chiral multiplets of the left and right
$\cN=(2,2)$ quiver to each other and to a subgroup of the symmetry acting on the hypermultiplets. The quintic superpotentials identify the remaining $U(1)$ flavor symmetries acting on bifundamental and adjoint chiral multiplets of each two-dimensional theory to rotations transverse to that plane. In \autoref{section:intersecting-partition-function-first-principle} we shall explore the consequences of this identification for the masses and $\mathcal{R}$-charges of the various fields.

The zero-dimensional $\cN=(0,2)$ Fermi multiplet $\Lambda$ has an $S[U(n'_{\nu'})\times U(n_\nu)]$ flavor symmetry, which is gauged with the innermost gauge group factors of the left and right $\cN=(2,2)$ theories.\footnote{Note that $\Lambda$ is neutral under the diagonal $U(1)$.} The couplings of $\Lambda$ with the two-dimensional fields can be obtained by embedding a zero-dimensional $S[U(n'_{\nu'})\times U(n_\nu)]$~$\cN=(0,2)$ vector multiplet in the corresponding two-dimensional $\cN=(2,2)$ vector multiplets.  As explained in \autoref{foot:global-remains}, gauging does not eliminate the $U(1)$~flavor symmetry acting only on~$\Lambda$, and a background vector multiplet for this symmetry could be added.  This is prevented by a zero-dimensional $\cN=(0,2)$ $E$-type or $J$-type superpotential, for instance $E[\Lambda]=\tilde{q}_{(L)}q_{(R)}$ restricted to zero dimensions.  Since the $S^4_b$ partition function we compute is only sensitive to superpotentials through the global symmetries that they identify, our methods do not fix them.

The second class of intersecting surface defects we will study in this paper is given by the local 4d/2d/0d quiver diagram of \autoref{fig:proposalintersectingWVTsymmetrics}.
In this case the superpotential couplings are
\begin{align}
&S^{(R)}_{\text{2d/4d}}=\int \mathrm{d}^4x\,  \delta (x^3) \delta (x^4) \int \mathrm{d}^2\theta_{(R)}\, \Bigl( q_{(R)}\tilde{q}_{(R)}Q^{\text{2d}}_{(R)} + q_{(R)}\varphi_{(R)}\tilde{q}_{(R)}\bigl((\partial_3-i\partial_4)Q^{\text{2d}}_{(R)}\bigr)\Bigr) \label{symmcouplinga}\\
&S^{(L)}_{\text{2d/4d}}=\int \mathrm{d}^4x\,  \delta (x^1) \delta (x^2)  \int \mathrm{d}^2\theta_{(L)}\,  \Bigl( q_{(L)}\tilde{q}_{(L)}Q^{\text{2d}}_{(L)} + q_{(L)}\varphi_{(L)}\tilde{q}_{(L)}\bigl((\partial_1-i\partial_2)Q^{\text{2d}}_{(L)}\bigr)\Bigr)\,,  \label{symmcouplingb}
\end{align}
where $\varphi_{(L)}$ and~$\varphi_{(R)}$ denote the adjoint chiral multiplets.  This again identifies the flavor symmetries of the left and right two-dimensional quiver with the one of the four-dimensional hypermultiplets and with transverse rotations.

The zero-dimensional $\cN=(0,2)$ chiral multiplets $\chi$ and~$\tilde{\chi}$ each have an $S[U(n'_{\nu'})\times U(n_\nu)]$ flavor symmetry.  Both of these $S[U(n'_{\nu'})\times U(n_\nu)]$ global symmetries are gauged with the innermost gauge group factors of the left and right $\cN=(2,2)$ theories.  As before, gauging does not eliminate global $U(1)$ symmetries acting only on $\chi$ and~$\tilde{\chi}$ and there should exist $E$ or $J$-type superpotentials identifying those symmetries to bulk symmetries.  The analysis is complicated by $J$-type superpotentials due to two-dimensional superpotentials and $E$-type superpotentials capturing derivatives in transverse dimensions: the added zero-dimensional superpotentials must fulfill the overall constraint $\Tr(E\cdot J)=0$ for supersymmetry.  Since our computations are not sensitive to the precise superpotential, we will not pursue it here.

\section{\label{section:intersecting-partition-function-first-principle}Localization on \texorpdfstring{$S^4_b$}{Squashed Sphere} of Intersecting Defects}

In this section we perform the exact computation of the expectation value of quarter-supersymmetric intersecting surface defects on the squashed four-sphere $S^4_b$
 \beq
  \frac{x_0^2}{r^2}+\frac{x_1^2+x_2^2}{\ell^2}+\frac{x_3^2+x_4^2}{\tilde \ell^2}=1\,,
  \label{squashsph}
  \eeq
where $b^2=\ell/\tilde \ell$ is a dimensionless squashing parameter. A four-dimensional theory on the round four-sphere $S^4$ has  an  $OSp(2|4)$ supersymmetry algebra~\cite{Pestun:2007rz}. Upon squashing the sphere to $S^4_b$, the symmetry of the theory is reduced to $SU(1|1)$. Any four-dimensional $\cN=2$ theory can be placed on $S^4_b$ while preserving this symmetry~\cite{Hama:2012bg}.

\begin{figure}
  \centering
  \includegraphics{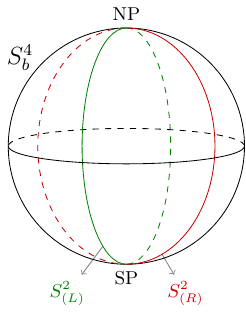}
  \caption{\label{fig:intersectingS2} Intersecting surface defects supported on two intersecting two-spheres $S^2_{(L)}$ and $S^2_{(R)}$. There are localized degrees of freedom living on the two-spheres $S^2_{(L)}$ and $S^2_{(R)}$ and at their intersection points, \ie{}, the north pole (NP) and south pole (SP); the latter couple to the former degrees of freedom, which in turn couple to the four-dimensional gauge theory living in the bulk~$S^4_b$.}
\end{figure}

A two-dimensional $\cN=(2,2)$ theory  on the round  $S^2$ preserves $OSp(2|2)$\cite{Benini:2012ui,Doroud:2012xw,Gomis:2012wy,Doroud:2013pka}. When the sphere is squashed to $S^2_b$, the symmetry of the theory is  $SU(1|1)$ \cite{Gomis:2012wy}. A two-dimensional
$\cN=(2,2)$ theory  on the round  $S^2$ can be coupled to a four-dimensional $\cN=2$ theory on $S^4$ while preserving $OSp(2|2)$ \cite{Lamy-Poirier:2014sea}.
Upon squashing the four-sphere to $S^4_b$, the combined 4d/2d system preserves $SU(1|1)$, provided the two-dimensional theory is placed either on the $S^2_b$ at $x_3=x_4=0$ or at $x_1=x_2=0$, which we call $S^2_{(R)}$ and $S^2_{(L)}$ respectively. In fact, we can place  a two-dimensional $\cN=(2,2)$ theory at $x_3=x_4=0$ \textit{and}   another one  at  $x_1=x_2=0$ while preserving $SU(1|1)$. This allows us to couple the four-dimensional $\cN=2$ theory on $S^4_b$ to a two-dimensional $\cN=(2,2)$ theory on $S^2_{(R)}$ and to a two-dimensional $\cN=(2,2)$ theory on $S^2_{(L)}$. This setup can be further enriched by adding localized degrees of freedom at the intersection of the two-dimensional theories, that is the North and South poles of $S^4_b$ at $x_0=r$ and $x_0=-r$ with $x_1=x_2=x_3=x_4=0$ respectively, see \autoref{fig:intersectingS2} for a cartoon. The localized degrees of freedom, pinned at the poles, are the dimensional reduction of a two-dimensional $\cN=(0,2)$ theory down to zero dimensions.  Consistently coupling the $\cN=(0,2)$ multiplets to the four-dimensional and two-dimensional degrees of freedom on $S^4_b$   requires turning on a background field for a    flavor symmetry of the zero-dimensional theory that includes the $U(1)\times U(1)$ rotations  of $S^4_b$. This background field is necessary for the zero-dimensional $\cN=(0,2)$ theories at the poles  of $S^4_b$ to be invariant under the $SU(1|1)$ symmetry of the combined system (see below). In this way, the quarter-supersymmetric intersecting defects we have introduced in the previous sections can be placed on $S^4_b$ while preserving $SU(1|1)$.

Our primary goal is to compute the   $S^4_b$ partition function of the intersecting defects in \autoref{fig:proposalintersectingWVT}.   We  accomplish this by supersymmetric localization with respect to the supercharge $\cQ$ in $SU(1|1)$. It is precisely this supercharge that was used to compute  the $S^4_b$ partition function of a four-dimensional $\cN=2$ theory~\cite{Hama:2012bg} and the $S^2_b$ partition function of a two-dimensional $\cN=(2,2)$ theory~\cite{Gomis:2012wy}. We localize the path integral by choosing the ``Coulomb branch localization'' $\cQ$-exact deformation terms of the four-dimensional and two-dimensional theories in \cite{Hama:2012bg,Gomis:2012wy}. In the absence of four-dimensional gauge fields, the saddle points of the four-dimensional and two-dimensional fields are the same as if the theories were considered in isolation.  Finally, the  North and South pole $\cN=(0,2)$ Fermi multiplet action coupled to the saddle points of the two-dimensional and four-dimensional fields can be easily integrated out using the computation  of the  index of one-dimensional $\cN=(0,2)$ supersymmetric quantum mechanics~\cite{Hori:2014tda}.

Putting all these  facts  together we arrive at the following integral representation\footnote{In order not to clutter formulas, we leave implicit  the dependence  of the various ingredients on the masses of the matter multiplets and the dependence of two-dimensional contributions on complexified FI parameters.} of the partition function of the intersecting defects in \autoref{fig:proposalintersectingWVT},
\begin{multline}
  Z=Z_{S^4_b}^{\text{free HM}}\sum_{B^{(L)}} \sum_{B^{(R)}} \int  \frac{\mathrm{d}\sigma^{(L)} }{(2\pi)^{\operatorname{rank}G^{(L)}}}  \frac{\mathrm{d}\sigma^{(R)}}{(2\pi)^{\operatorname{rank}G^{(R)}}}
  Z_{S^2_{(L)}}(\sigma^{(L)},B^{(L)}) \ Z_{S^2_{(R)}}(\sigma^{(R)},B^{(R)}) \,
  \\ \times Z_{\text{0d}}^{\text{intersection}}(\sigma^{(L)}, B^{(L)},\sigma^{(R)},B^{(R)})\,.
   \label{finalansw}
\end{multline}
Here $Z_{S^4_b}^{\text{free HM}}$ is the $S^4_b$ partition function \cite{Hama:2012bg}\footnote{The function $\Upsilon_b(x)$ is related to the Barnes double-Gamma function.} of the $n_\text{f}^2$ hypermultiplets with dimensionless masses ${M_{js}}$, measured in units of $1/\sqrt{\ell\tilde \ell}$:
 \beq
 Z_{S^4_b}^{\text{free HM}}=\prod_{j,s=1}^{n_\text{f}}\frac{1}{\Upsilon_b\left(\frac{b}{2}+\frac{1}{2b}-i M_{js}\right)}\,.
 \eeq
Furthermore, $G^{(L), (R)}$ denote the total gauge groups of the left/right two-dimensional theories while $Z_{S^2_b}(\sigma^{(L/R)},B^{(L/R)})$ is the integrand of the $S^2_b$ partition function of the two-dimensional $\cN=(2,2)$ theory on the left/right of the quiver diagram. The integrand is given by~\cite{Benini:2012ui,Doroud:2012xw,Gomis:2012wy}\footnote{$\Gamma(x)$ is Euler's Gamma function.}
\begin{equation}
Z_{S^2_b}(\sigma,B)= \frac{1}{\mathcal{W}} z^{i\sigma+\frac{B}{2}} \, \bar{z}^{i\sigma-\frac{B}{2}} \prod_{\alpha>0} \left[(-1)^{\alpha B} \left[(\alpha\sigma)^2+\frac{(\alpha B)^2}{4} \right]\right] \prod_{w\in R}\frac{\Gamma\bigl(-w\bigl(im+i\sigma+\frac{B}{2}\bigr)\bigr)}{\Gamma\bigl(1+w\bigl(im+i\sigma-\frac{B}{2}\bigr)\bigr)} \,
\label{twodpart}
\end{equation}
with $z=e^{-2\pi\xi_\text{FI}+i\vartheta}$, where $\xi_\text{FI}$ is the FI parameter   and $\vartheta$ its corresponding topological angle. $B$~and $\sigma$ take values in the Cartan subalgebra of the gauge group and $\alpha$ and $w$ are the roots of the gauge group and weights of the representation of the chiral multiplets respectively, while $\mathcal{W}$ is the order of the gauge Weyl group.  We will use conventions adapted to quiver gauge theories, \ie{}, fundamental chiral multiplets transform anti-fundamentally under their flavor symmetry and \textsl{vice versa}. The parameter $m$ in~\eqref{twodpart} is complex: the real part measures the mass and the imaginary part the $\mathcal{R}$-charge of the two-dimensional chiral multiplet through~\cite{Benini:2012ui,Doroud:2012xw,Gomis:2012wy}
\begin{equation}
\begin{aligned}
  m^{(R)}&=\ell\, \mathrm{m}^{(R)}-\frac{i}{2}\mathcal{R}_{\text{2d}}^{(R)}[q_{(R)}]\;, \qquad &
  m^{(L)}&={\tilde \ell}\, \mathrm{m}^{(L)}-\frac{i}{2}\mathcal{R}_{\text{2d}}^{(L)}[q_{(L)}]\;, \\
  \anti m^{(R)}&=\ell\, \anti{\mathrm{m}}^{(R)}+\frac{i}{2}\mathcal{R}_{\text{2d}}^{(R)}[\tilde{q}_{(R)}]\;, \qquad &
  \anti m^{(L)}&={\tilde \ell}\, \anti{\mathrm{m}}^{(L)}+\frac{i}{2}\mathcal{R}_{\text{2d}}^{(L)}[\tilde{q}_{(L)}]\;,
\end{aligned}
  \label{twodimmass}
\end{equation}
where $\mathrm{m}$ are masses of fundamental chiral multiplets while $\anti{\mathrm{m}}$ denote masses of antifundamental chiral multiplets. The dimensionless ``masses'' $(m^{(R)},\anti m^{(R)})$ and $(m^{(L)},\anti m^{(L)})$ are measured in units of $1/\ell$  and $1/\tilde \ell$ respectively for the right and left $\cN=(2,2)$ theories. This  is because the corresponding squashed two-spheres $S^2_b$ on which the two-dimensional theories live, which are  embedded in $S^4_b$,  have equatorial radii $\ell$ and $\tilde \ell$ respectively.

 Since the two-dimensional $\cN=(2,2)$ theories   are coupled to a four-dimensional $\cN=2$ theory in $S^4_b$, the canonical two-dimensional $\mathcal{R}$-charges are induced by the four-dimensional $SU(1|1)$ supersymmetry algebra. This is a consequence of the $SU(1|1)$-invariant coupling of the left and right two-dimensional $\cN=(2,2)$ theories on the two $S^2_b$'s with the four-dimensional $\cN=2$ theory on $S^4_b$. While $SU(1|1)$ acts on four-dimensional $\cN=2$ multiplets as \cite{Hama:2012bg}
\beq
\cQ_{\text{4d}}^2=\frac{1}{\ell}\mathcal M_{12}+\frac{1}{\tilde \ell}\mathcal M_{34}-\frac{1}{2}\left(\frac{1}{\ell}+ \frac{1}{\tilde \ell} \right)J_3^\mathcal{R}\,,
\label{fourdsupercharge}
\eeq
 $SU(1|1)$ acts on two-dimensional $\cN=(2,2)$ multiplets on an $S^2_b$ with equatorial radius $\ell$ as \cite{Gomis:2012wy}
\beq
\cQ_{\text{2d}}^2=\frac{1}{\ell}\mathcal M_{12}- \frac{1}{2 \ell}\mathcal R_{\text{2d}}\,.
\eeq
Here $\mathcal M_{ij}$ denotes the $U(1)$ generator that acts on the $(x_i,x_j)$ coordinates defining the squashed sphere, $J_3^\mathcal{R}$ is the Cartan generator of the $SU(2)$ $\mathcal{R}$-symmetry of the four-di\-men\-sion\-al $\cN=2$ theory in flat space\footnote{The charge  is normalized such that $\cQ_{\text{4d}}$ has charge one under $J_3^\mathcal{R}$, the same as that of $(Q,\tilde{Q})$, the two four-dimensional chiral multiplets that represent a hypermultiplet.} and $\mathcal R_{\text{2d}}$ is the vector $\mathcal{R}$-symmetry of a two-dimensional   $\cN=(2,2)$ theory. Since the right $\cN=(2,2)$ theory is on the $S^2_b$ at $x_3=x_4=0$ and the left $\cN=(2,2)$ theory is on the $S^2_b$ at $x_1=x_2=0$,   common $SU(1|1)$-invariance implies  that the $\mathcal{R}$-charge generators for the right and left $\cN=(2,2)$ theories are
\beq
\mathcal{R}^{(R)}_{\text{2d}}=\left(1+b^2\right)J_3^\mathcal{R}-2b^2 \mathcal M_{34} , \qquad\qquad
\mathcal{R}^{(L)}_{\text{2d}}=\left(1+b^{-2}\right)J_3^\mathcal{R}-2b^{-2} \mathcal M_{12}\,.
\label{relarcharge}
\eeq
The formula~\eqref{relarcharge} determines the $\mathcal{R}$-charges under $\mathcal{R}^{(R)}_{\text{2d}}$ and $\mathcal{R}^{(L)}_{\text{2d}}$ of the four-dimensional hypermultiplet scalars $(Q,\tilde{Q})$ restricted to each~$S^2_b$. Recall that chiral multiplets of the right and left $\cN=(2,2)$ theories couple to the corresponding $\cN=(2,2)$ ``bulk'' chiral multiplets with bottom components $Q$ and~$\tilde{Q}$.

The cubic defect superpotentials in~\eqref{defectW1} and~\eqref{defectW2} coupling bulk hypermultiplets with  innermost chiral multiplets   identify     their  respective $SU(n_\text{f})\times SU(n_\text{f})\times U(1)$ global symmetries. This   implies that the masses of the hypermultiplets and  the innermost chiral multiplets obey a relation, which follows from the  common  $SU(n_\text{f})\times SU(n_\text{f})\times U(1)$  symmetry acting on these fields. Another constraint follows from the $SU(1|1)$ symmetry of $S^2_b$. A two-dimensional $\cN=(2,2)$  superpotential on $S^2_b$ is supersymmetric if and only if the   $\mathcal{R}$-charge of the superpotential is two~\cite{Gomis:2012wy}.  This gives two relations, one arising from \eqref{defectW1} requiring that $\mathcal{R}^{(R)}_{\text{2d}}[Q^{\text{2d}}_{(R)}q_{(R)}\tilde{q}_{(R)}]=2$ and the other from
\eqref{defectW2} requiring that $\mathcal{R}^{(L)}_{\text{2d}}[\tilde{Q}^{\text{2d}}_{(L)}q_{(L)}\tilde{q}_{(L)}]=2$.
The hypermultiplet scalars $Q,\tilde{Q}$ have $\mathcal{R}^{(R)}_{\text{2d}}[Q]=1+b^2$ and $\mathcal{R}^{(L)}_{\text{2d}}[\tilde{Q}]=1+b^{-2}$, since $J^3_{\mathcal{R}}[Q]=J^3_{\mathcal{R}}[\tilde{Q}]=1$ and they are Lorentz scalars.
In total, the $SU(n_\text{f})\times SU(n_\text{f})\times U(1)$ global symmetry constraints and $\mathcal{R}$-symmetry superpotential constraints neatly combine into the following relation between the four-dimensional masses $M_{js}$ and the two-dimensional complexified masses \eqref{twodimmass} $m_j$ and $\anti m_s$ for the fundamental and anti-fundamental chiral multiplets
\beq
  \left[\frac{M_{js}}{\sqrt{\ell\tilde \ell}}+\frac{i}{2 \ell}+ \frac{i}{2\tilde \ell}  \right]+\frac{-m^{(R)}_j+\anti m^{(R)}_s}{\ell}=\frac{i}{\ell}\,,
  \label{massrelaone}
\eeq
and
\beq
  \left[ -{\frac{{{M_{js}}}}{{\sqrt {\ell \tilde \ell } }} + \frac{i}{{2\ell }} + \frac{i}{{2\tilde \ell }}} \right] + \frac{{ - m_s^{(L)} + \anti m_j^{(L)}}}{{\tilde \ell }} = \frac{i}{{\tilde \ell }}\,.
  \label{massrelatwo}
\eeq
The real part of these equations encodes the $SU(n_\text{f})\times SU(n_\text{f})\times U(1)$ global symmetry constraints on the masses and the imaginary part the $\mathcal{R}$-charge constraints.
The first relation \eqref{massrelaone} fixes the four-dimensional masses~$M_{js}$, which appear in $Z_{S^4_b}^{\text{free HM}}$ in \eqref{finalansw}, in terms of the two-dimensional masses $ m^{(R)}_j$ and $\anti m^{(R)}_s$.
Adding  \eqref{massrelaone}  and \eqref{massrelatwo} we find the following system of equations
\beq
  \frac{{ - m_j^{(R)}  + \anti m_s^{(R)}}}{\ell } + \frac{{ - m_s^{(L)} + \anti m_j^{(L)}}}{{\tilde \ell }} = 0\,,
\eeq
whose solution is
\beq\label{mass-relation-fermi-from-superpotential}
  b^{-1} \anti m_s^{(R)} = b m_s^{(L)} + c\;, \qquad b^{-1}  m_j^{(R)} = b \anti m_j^{(L)} + c\;,
\eeq
for some constant~$c$ which we set to zero by shifting the vector multiplet scalars in the left theory by $c/b$.  This relation is consistent with the $\mathcal{R}$-charges above. We can use this relation to express in terms of $(m_j^{(R)},{\anti m}_s^{(R)})$ the masses of the innermost (fundamental and antifundamental) chirals of the right and left $\cN=(2,2)$ theories that appear in $Z_{S^2_b}(\sigma^{(R)},B^{(R)})$ and $Z_{S^2_b}(\sigma^{(L)},B^{(L)})$ in \eqref{finalansw}.

The quintic superpotentials in \eqref{defectW1} and \eqref{defectW2} yield relations similar to \eqref{massrelaone} and~\eqref{massrelatwo} which force the (next-to) innermost bifundamental chiral multiplets to have zero twisted mass and $\mathcal{R}$-charges $\mathcal{R}^{(R)}_{\text{2d}}[q^{(R)}_{\text{bif}}] = -b^2$ and $\mathcal{R}^{(L)}_{\text{2d}}[q^{(L)}_{\text{bif}}] = - b^{-2}$.  The cubic superpotentials of each two-dimensional theory then proceed to set all twisted masses to zero and $\mathcal{R}$-charges to $-b^2$ and $2+2b^2$ for bifundamental and adjoint chiral multiplets of the theory on the right and $-b^{-2}$ and $2+2b^{-2}$ for the one on the left.

Once the path integrals for the four-dimensional and two-dimensional theories have been localized to zero-mode integrals, we must still integrate out the  fields of the zero-dimensional $\cN=(0,2)$ theories at the poles of $S^4_b$, captured by two matrix integrals, one for the theory at the North pole and one for the theory at the South pole. This requires first understanding how to couple the  zero-dimensional $\cN=(0,2)$ theories to the other fields on $S^4_b$ in an $SU(1|1)$-invariant way.  A ``flat space'' zero-dimensional $\cN=(0,2)$   theory, obtained by trivial dimensional reduction from two-dimensions, has nilpotent supercharges. The supersymmetry algebra can be  deformed by turning on a supersymmetric zero-dimensional $\cN=(0,2)$ vector multiplet background for a flavor symmetry $G_F$ of the theory. The deformed algebra acts on the fields as
\beq
\cQ^2_{\text{0d}}=i\frac{u_F}{\sqrt{\ell\tilde \ell}}\ Q_F\,,
\eeq
where $u_F$ is a constant background value for the dimensionless complex combination of scalars in the zero-dimensional $\cN=(0,2)$ vector multiplet invariant under supersymmetry,\footnote{The scalar is made dimensionless with a factor $\sqrt{\ell\tilde\ell}$.} and $Q_F$ is the charge under $G_F$. Therefore, in order to consistently couple a zero-dimensional $\cN=(0,2)$ theory at a pole with the rest of the fields of the intersecting defect theory on $S^4_b$ in an $SU(1|1)$-invariant way, comparison with the four-dimensional supersymmetry algebra \eqref{fourdsupercharge} requires that we turn on a constant background
\beq
u_F=-i
\eeq
for the zero-dimensional flavor symmetry
\beq
Q_F=b^{-1}\mathcal M_{12}+b\mathcal M_{34}-\frac{1}{2}\left(b+b^{-1}\right)J_3^\mathcal{R}\,.
\eeq

Now that we know how to couple the zero-dimensional $\cN=(0,2)$ theories at the poles to $S^4_b$ we can easily compute their path integrals. The  result is obtained by keeping the zero-mode along the circle  of the index computation of  $\cN=(0,2)$ supersymmetric quantum mechanics in \cite{Hori:2014tda}. The formula for the path integral over a zero-dimensional $\cN=(0,2)$ Fermi multiplet coupled to a background vector multiplet through a  representation $\mathbf{r}$ and to  a background vector multiplet for a flavor symmetry $G_F$ with charge $Q_F$ is
\beq
Z_{\text{0d}}^{\text{Fermi}}=\prod_{w\in\mathbf{r}}\left({w(i u)+ i Q_F u_F}\right)\,.
\eeq
Here $u$ are the (dimensionless) scalars in the dynamical vector multiplet and  $u_F$  the background value for the $G_F$ global symmetry.

We can now determine the contribution of the zero-dimensional $\cN=(0,2)$ Fermi multiplets at the North and South poles of $S^4_b$ depicted in \autoref{fig:proposalintersectingWVT} to the intersecting defect partition function \eqref{finalansw}. It is given by
\begin{equation}
\label{intersect}
Z_{\text{intersection}}(\sigma^{(L)}, B^{(L)},\sigma^{(R)},B^{(R)}) = \prod_{a  = 1}^{n_\nu } \prod_{b  = 1}^{n^\prime_{\nu^\prime} } \Delta_{ab}^+ \ \Delta_{ab}^-  \;,
\end{equation}
with $\Delta_{ab}^\pm =b^{-1}\Bigl(i\sigma^{(R)}_a \pm \frac{B^{(R)}_a}{2}\Bigr)- b \Bigl(i\sigma^{(L)}_b \pm \frac{B^{(L)}_b}{2}\Bigr)$.
The factors with $\Delta_{ab}^+$ originate from the $\cN=(0,2)$ Fermi at the North pole while the factors  $\Delta_{ab}^-$ come from the South pole.\footnote{The gauge equivariant parameters on the North and South poles of $S^2_b$ are the complex conjugate  of each other~\cite{Benini:2012ui,Doroud:2012xw,Gomis:2012wy}, which explains the sign difference between North and South pole contributions. In \eqref{intersect} we substitute the equivariant parameters at the poles with their values at the saddle points (see \cite{Benini:2012ui,Doroud:2012xw,Gomis:2012wy} for more details).}
The $S[U(n_\nu)\times U(n'_{\nu'})]$ symmetry is gauged with the innermost gauge group factor of the left and right $\cN=(2,2)$ theories. This explains the appearance of $\sigma^{(R)}$ and $\sigma^{(L)}$ in \eqref{intersect}. We have also used the fact that the $\cN=(0,2)$ Fermi multiplets are uncharged under the flavor symmetry~$G_F$: this can be enforced for instance by the $E$-type superpotential $E[\Lambda]=\tilde{q}_{(L)}q_{(R)}$ for the Fermi multiplet~$\Lambda$ put forward above already (see below \eqref{defectW2}).  Indeed, the cubic defect superpotentials in \eqref{defectW1} and~\eqref{defectW2} constrain the $\mathcal{R}$-charges of $q_{(R)}$, $\tilde{q}_{(R)}$, $q_{(L)}$ and~$\tilde{q}_{(L)}$ hence their charge under $\cQ^2/\sqrt{\ell\tilde{\ell}}$, and the  $E$-type superpotential fixes the charge of~$\Lambda$.  The $\cQ^2/\sqrt{\ell\tilde{\ell}}$ charges are given in \autoref{tab:qsquarecharges} up to mixing with two-dimensional $U(1)$ gauge symmetries namely shifting the integration contour of~$\sigma^{(L/R)}$ in the imaginary direction.
The $E$-type superpotential also identifies the $U(1)$ flavor symmetry of~$\Lambda$ with a combination of two-dimensional gauge symmetries.

\begin{table}\centering
\begin{tabular}{*5{>{$\displaystyle}c<{$}}}
  \toprule
  \vphantom{\Big|} & Q, \tilde{Q} & q_{(R)}, \tilde{q}_{(R)} & q_{(L)}, \tilde{q}_{(L)} & \Lambda \\\hline
  \vphantom{\Bigg|} \frac{\cQ^2}{\sqrt{\ell\tilde{\ell}}} & \frac{-(b+b^{-1})}{2} & \frac{b-b^{-1}}{4} & \frac{-(b-b^{-1})}{4} & 0
  \\\bottomrule
\end{tabular}
\caption{\label{tab:qsquarecharges}Charges of various fields under $\cQ^2/\sqrt{\ell\tilde{\ell}}$.}
\end{table}

Similarly, we can determine the integral representation of the partition function of the intersecting defects in \autoref{fig:proposalintersectingWVTsymmetrics},
\begin{multline}
  Z=Z_{S^4_b}^{\text{free HM}}\sum_{B^{(L)}} \sum_{B^{(R)}} \int_{\text{JK}}  \frac{\mathrm{d}\sigma^{(L)} }{(2\pi)^{\operatorname{rank}G^{(L)}}} \frac{\mathrm{d}\sigma^{(R)}}{(2\pi)^{\operatorname{rank}G^{(R)}}}
  \ Z_{S^2_{(L)}}(\sigma^{(L)},B^{(L)}) \ Z_{S^2_{(R)}}(\sigma^{(R)},B^{(R)}) \,
  \\ \times \widetilde Z_{\text{0d}}^{\text{intersection}}(\sigma^{(L)}, B^{(L)},\sigma^{(R)},B^{(R)})\,,
   \label{finalanswsymm}
\end{multline}
where again $G^{(L), (R)}$ denote the total gauge groups of the two 2d theories. The symbol $\int_{\text{JK}}$ stands for taking a Jeffrey--Kirwan-like residue prescription (see definition below). Similarly as above, the superpotential couplings \eqref{symmcouplinga}--\eqref{symmcouplingb} impose relations among the complexified mass parameters. In this case they read
\beq
  \left[\frac{M_{js}}{\sqrt{\ell\tilde \ell}}+\frac{i}{2 \ell}+ \frac{i}{2\tilde \ell}  \right]+\frac{-m^{(R)}_j+\anti m^{(R)}_s}{\ell}=\frac{i}{\ell}\,,
  \label{massrelaonesymm}
\eeq
and
\beq
  \left[ {\frac{{{M_{js}}}}{{\sqrt {\ell \tilde \ell } }} + \frac{i}{{2\ell }} + \frac{i}{{2\tilde \ell }}} \right] + \frac{{ - m_j^{(L)} + \anti m_s^{(L)}}}{{\tilde \ell }} = \frac{i}{{\tilde \ell }}\,.
  \label{massrelatwosymm}
\eeq
As before, the real part of these equations encode the flavor symmetry constraints on the masses and the imaginary part the $\mathcal{R}$-charge constraints. The four-dimensional masses $M_{js}$ can be determined in terms of the dimensional masses $ m^{(R)}_j$ and $\anti m^{(R)}_s$ in precisely the same way as above. Moreover, subtracting  \eqref{massrelaone}  and \eqref{massrelatwo} one obtains
\beq
  \frac{{ - m_j^{(R)}  + \anti m_s^{(R)}}}{\ell } - \frac{{ - m_j^{(L)} + \anti m_s^{(L)}}}{{\tilde \ell }} = \frac{i}{\ell } - \frac{i}{{\tilde \ell }}\,,
\eeq
with solution
\beq
  b^{-1} \bigl(m_j^{(R)}+i/2\bigr) = b\bigl(m_j^{(L)}+i/2\bigr) + \tilde{c},
  \qquad
  b^{-1} \bigl(\anti{m}_s^{(R)}-i/2\bigr) = b\bigl(\anti{m}_s^{(L)}-i/2\bigr) + \tilde{c}\,,
  \label{mass-relation-chiral-from-superpotential}
\eeq
for some constant $\tilde{c}$, which can be absorbed by shifting the vector multiplet scalars, allowing one to express the masses of the left quiver in terms of those of the right quiver.  The quartic superpotential sets the real twisted masses of the adjoint chiral multiplets to zero and their $\mathcal{R}$-charges to be $-2b^2$ and $-2b^{-2}$ respectively.

Using that the formula for the path integral over a zero-dimensional $\cN=(0,2)$ chiral multiplet coupled to a background vector multiplet through a  representation $\mathbf{r}$ and to  a background vector multiplet for $G_F$ with charge $Q_F$ is
\beq
Z_{\text{0d}}^{\text{chiral}}=\prod_{w\in\mathbf{r}}\frac{1}{w(i u)+ i Q_F u_F}\,.
\eeq
we can easily determine the contribution of the zero-dimensional $\cN=(0,2)$ chiral multiplets
at the North and South poles of $S^4_b$ depicted in \autoref{fig:proposalintersectingWVTsymmetrics} to the intersecting defect partition function \eqref{finalanswsymm}. It is given by
\begin{multline}
\label{intersectsymm}
\widetilde Z_{\text{intersection}}(\sigma^{(L)}, B^{(L)},\sigma^{(R)},B^{(R)}) \\
= \prod_{a  = 1}^{n } \prod_{b  = 1}^{n^\prime} \Biggl[ \left(\Delta_{ab}^+ + \frac{b+b^{-1}}{2}\right)\left(\Delta_{ab}^+ - \frac{b+b^{-1}}{2}\right) \left(\Delta_{ab}^- + \frac{b+b^{-1}}{2}\right) \left(\Delta_{ab}^- - \frac{b+b^{-1}}{2}\right) \Biggr]^{-1} \;,
\end{multline}
with $\Delta_{ab}^\pm =b^{-1}\Bigl(i\sigma^{(R)}_a \pm \frac{B^{(R)}_a}{2}\Bigr)- b \Bigl(i\sigma^{(L)}_b \pm \frac{B^{(L)}_b}{2}\Bigr)$ as before.
The factors with $\Delta_{ab}^\pm$ originate from the $\cN=(0,2)$ chirals at the North and South pole respectively. The terms in \eqref{intersect} proportional to $\frac{b+b^{-1}}{2}$ indicate that the $\cN=(0,2)$ chiral multiplets carry charge $\frac{b+b^{-1}}{2}$ under the global symmetry~$G_F$.  This should be explained by a zero-dimensional superpotential but we have not worked it out.

Let us conclude this section with a brief discussion of the Jeffrey--Kirwan-like residue prescription \cite{1993alg.geom..7001J} used in \eqref{finalanswsymm}. We note that in the absence of the zero-dimensional chiral multiplets, our prescription coincides with the standard one in \cite{Benini:2012ui,Doroud:2012xw,Gomis:2012wy}  to close the contour according to the sign of the FI parameter.
Let $N=n+n^\prime$ denote the total rank of the gauge groups in the quiver depicted in \autoref{fig:proposalintersectingWVT}, and let $\mathfrak S$ be the notation for the combined $N$ integration variables $(\sigma^{(R)}, \sigma^{(L)})$. The pole equations of the integrand \eqref{twodpart} corresponding to the right and left quiver are of the form
\begin{equation}
w^{(R)}(i\sigma^{(R)}) + \ldots = 0\;, \qquad w^{(L)}(i\sigma^{(L)}) + \ldots = 0\;,
\end{equation}
where $w^{(R/L)}$ is any weight of the representations of the chiral multiplets in the respective quiver. Denoting by $\mathfrak w$ the collection of combined weights, which take the form $(w^{(R)},0)$ or $(0,w^{(L)})$, it can be written as $\mathfrak w(i\mathfrak S) + \ldots =0$. The pole equations of all four factors in the intersection factor \eqref{intersect} can be written similarly as\footnote{Note that the definition of $\mathfrak u_{ab}$ does not respect the naive charge assignments of the 0d bifundamental chiral multiplets.}
\begin{equation}
\mathfrak u_{ab} (i\mathfrak S) + \ldots = b^{-1}i\sigma^{(R)}_a - b i\sigma^{(L)}_b + \ldots =0
\end{equation}
for all $a=1,\ldots, n$ and $b=1,\ldots,n^\prime$. We collectively denote the charges $\mathfrak w$ and $\mathfrak u_{ab}$ thus defined by $\mathfrak W$. A collection of $N$ linearly independent pole equations,\footnote{If more than $N$ of the hyperplanes defined by pole equations intersect one must locally decompose $F(\mathfrak S)$ as a sum of terms that each have only $N$~singular factors at $\mathfrak S^\star$, and apply the JK residue to each term.} associated to charge vectors $\mathfrak W^I$ for $I=1,\ldots, N$, define a pole solution $\mathfrak S^\star$, whose residue we define to be
\begin{equation}
\text{JK-Res}_{\eta}\  F(\mathfrak S^\star) = \begin{cases}
\underset{\mathfrak S\rightarrow \mathfrak S^\star}{\text{Res}}F(\mathfrak S) & \text{if}\ \eta \in C(\mathfrak W^{I=1,\ldots,N})  \\
0 & \text{otherwise}
\end{cases}
\end{equation}
where $\eta = (\xi^{(R)},\xi^{(L)})$ is the combined FI parameter understood as an $N$-dimensional vector, and  $C(\mathfrak W^{I=1,\ldots,N})$ is the positive cone spanned by the vectors $\mathfrak W^I$. Finally, $\underset{\mathfrak S\rightarrow \mathfrak S^\star}{\text{Res}}$ denotes the usual residue at the pole $\mathfrak S = \mathfrak S^\star$, with a sign determined by the contour.

In this section we have obtained the formula that computes  the exact partition function of the intersecting defects in \autoref{fig:proposalintersectingWVT} and \autoref{fig:proposalintersectingWVTsymmetrics}.

\section{\label{sec: M2-brane surface defects}M2-Brane Surface Defects}

Despite our very incomplete understanding of M-theory, it is known that M2-branes can end on a collection of $n_\text{f}$ M5-branes along a surface. When the M5-branes wrap a punctured Riemann surface, the UV-curve, the M2-branes define  a half-supersymmetric surface defect  in a four-dimensional $\cN=2$ theory.
Under favorable circumstances, this surface defect admits a Lagrangian description in the manner described in the previous section.

\begin{table}\centering
  \begin{tabular}{lccccccccccc}
    \toprule
          & $1$ & $2$ & $3$ & $4$ & $5$ & $6$ & $7$ & $8$ & $9$ & $10$ & $11$\\
    \hline
    M5    & --- & --- & --- & --- &     &     & --- &     &     &      & --- \\
    M5$'$ & --- & --- & --- & --- & --- & --- &     &     &     &      &     \\
    M2    & --- & --- &     &     &     &     &     &     &     & ---  &     \\
    \bottomrule
  \end{tabular}
  \caption{\label{tab:m2defect}Intersection of M2 and M5 branes defining a half-supersymmetric surface operator.}
\end{table}
The brane configuration that realizes this half-supersymmetric surface defect is given in \autoref{tab:m2defect}.
The M2-brane endings on $n_\text{f}$ M5-branes are labeled by a representation $\mathcal{R}$ of $SU(n_\text{f})$. The M5$^\prime$-branes are codimension two defects for the M5-branes that encode the flavor symmetries of the four-dimensional $\cN=2$ theory  and that are realized by the punctures on the Riemann surface~\cite{Gaiotto:2009we}.\footnote{The Riemann surface lies along $(x^7,x^{11})$.}

\begin{figure}\centering
  \includegraphics{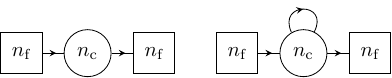}
  \caption{\label{fig:minuscule-quivers}Quiver description of the 2d $\cN=(2,2)$ surface defects corresponding to the rank $n_\text{c}$ antisymmetric and symmetric representations, respectively.}
\end{figure}
\begin{figure}\centering
  \includegraphics{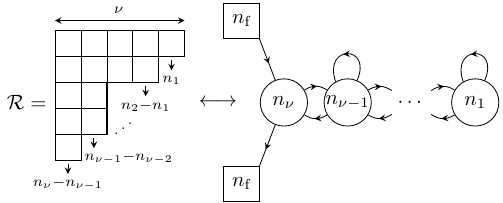}
  \caption{\label{fig:single2d}The 2d $\cN=(2,2)$ quiver gauge theory corresponding to the Young diagram of a given representation~$\mathcal{R}$.}
\end{figure}

As argued in~\cite{Gomis:2014eya}, when $\mathcal{R}$ is the rank $n_\text{c}$ antisymmetric representation, the two-di\-men\-sion\-al
$\cN=(2,2)$ theory  description of the surface defect is given by the first quiver diagram in \autoref{fig:minuscule-quivers}.
If $\mathcal{R}$ is the rank $n_\text{c}$ symmetric representation, the corresponding 2d $\cN=(2,2)$ theory is the second quiver diagram in \autoref{fig:minuscule-quivers}.
For a representation $\mathcal{R}$ described by a generic Young diagram the two-dimensional $\cN=(2,2)$ theory has the quiver diagram representation given in \autoref{fig:single2d}\footnote{Note that the symmetric representation admits two descriptions. The two descriptions share, at the very least, the value of the two-sphere partition function. This is akin to the giant and dual giant description of Wilson loops~\cite{Drukker:2005kx,Yamaguchi:2006tq,Gomis:2006sb,Gomis:2006im}.}.
The complexified  FI  parameters for all gauge group factors except the one that couples to the $n_\text{f}$ fundamentals and anti-fundamentals must be set to zero.

These two-dimensional $\cN=(2,2)$ theories can be coupled to a four-dimensional $\cN=2$ theory by gauging the $SU(n_\text{f})\times SU(n_\text{f})\times U(1)$ flavor symmetries acting on the $n_\text{f}$ fundamental and anti-fundamental chiral multiplets with gauge and/or global symmetries of the four-dimensional theory.  The simplest four-dimensional $\cN=2$ theory in which to consider these surface operators  is the theory of $n_\text{f}^2$ hypermultiplets. This corresponds to compactifying $n_\text{f}$ M5-branes on a trinion with two full and one simple puncture, which makes manifest an $SU(n_\text{f})\times SU(n_\text{f})\times U(1)$ flavor symmetry acting on the hypermultiplets, which gets identified via the  cubic superpotential  \eqref{cubicone} with the corresponding  defect flavor symmetry. For other four-dimensional theories, such as for conformal SQCD with $SU(n_\text{f})$ gauge group and $2n_\text{f}$ hypermultiplets or the $\cN=2^*$ theory, one or both of the defect  $SU(n_\text{f})$ symmetry factors is gauged with a dynamical bulk gauge field.

\begin{table}\centering
  \begin{tabular}{lccccccccccc}
    \toprule
          & $1$ & $2$ & $3$ & $4$ & $5$ & $6$ & $7$ & $8$ & $9$ & $10$ & $11$ \\
    \hline
    M5    & --- & --- & --- & --- &     &     & --- &     &     &      & ---  \\
    M5$'$ & --- & --- & --- & --- & --- & --- &     &     &     &      &      \\
    M2    & --- & --- &     &     &     &     &     &     &     & ---  &      \\
    M2$'$ &     &     & --- & --- &     &     &     &     &     & ---  &
    \\\bottomrule
  \end{tabular}
  \caption{\label{tab:m2m2defect}Intersection of M2 and M5 branes defining quarter-supersymmetric intersecting surface defects on the M5-branes.}
\end{table}
A richer class of surface defects on M5-branes can be constructed by letting two sets of M2-branes end on the M5-branes as in \autoref{tab:m2m2defect}.
This configuration preserves one-quarter of the supersymmetry and defines intersecting surface defects on the M5-branes. When the M5-branes wrap a punctured Riemann surface, the brane configuration engineers an intersecting surface defect in the corresponding
four-dimensional $\cN=2$  theory of precisely the kind described in the previous section. The configuration of intersecting M2-branes is now labeled by a pair $(\mathcal{R}^\prime,\mathcal{R})$ of representations of $SU(n_\text{f})$.

\begin{table}\centering
  \begin{tabular}{l | cccccccccc}
    \toprule
              & $1$ & $2$ & $3$ & $4$ & $5$ & $6$ & $7$ & $8$ & $9$ & $10$ \\
    \hline
    NS5       & --- & --- & --- & --- & --- & --- &     &     &     &     \\
    NS5$'$    & --- & --- &     &     & --- & --- &     & --- & --- &     \\
    NS5$''$   &     &     & --- & --- & --- & --- &     & --- & --- &     \\
    \ \ D4    & --- & --- & --- & --- &     &     & --- &     &     &     \\
    \ \ D2    & --- & --- &     &     &     &     &     &     &     & --- \\
    \ \ D2$'$ &     &     & --- & --- &     &     &     &     &     & --- \\
    \bottomrule
  \end{tabular}
  \caption{\label{tab:IIAbranes}IIA brane realization of intersecting surface defects arising from M-theory brane intersections.  See \autoref{fig:IIAbranes} for details on which branes intersect.}
\end{table}
We propose that the field theory description of these intersecting surface defects is precisely the one detailed in the previous section, and encoded in the quiver diagram  in  \autoref{fig:proposalintersectingWVT}. For a class of four-dimensional $\cN=2$ theories, the intersecting defects admit a type IIA brane realization given in \autoref{tab:IIAbranes}.
In these cases, we can deduce the low-energy effective field theory description of the intersecting defect.

\begin{figure}\centering
  \includegraphics{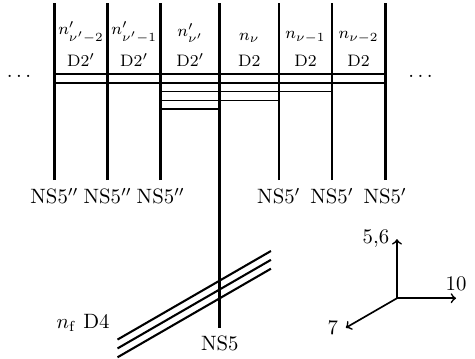}
  \caption{\label{fig:IIAbranes}IIA brane realization of intersecting surface defects arising from M-theory brane intersections.  See \autoref{tab:IIAbranes} for brane directions.}
\end{figure}
As an example, when the four-dimensional $\cN=2$ theory is that of $n_\text{f}^2$ hypermultiplets, the intersecting defect realized by the M-theory brane array in \autoref{tab:IIAbranes} has the type IIA description given in \autoref{fig:IIAbranes}.
The NS5$^\prime$-branes and NS5$^{\prime\prime}$-branes on which the D2 and D2$^\prime$-branes end respectively are away from the main stack and give rise to the two-dimensional gauge theories in the quiver in \autoref{fig:proposalintersectingWVT}. The  two-dimensional $\cN=(2,2)$ theories at $x^3=x^4=0$ and labeled by a representation $\mathcal{R}$ and at $x^1=x^2=0$ and labeled by a representation $\mathcal{R}'$ live  on the D2-branes and D2$^\prime$-branes respectively. The zero-dimensional bifundamental $\cN=(0,2)$ Fermi multiplet arise from quantizing the open strings stretching between the D2 and the D2$^\prime$-branes. The gaugings and superpotential couplings encoded in the quiver in \autoref{fig:proposalintersectingWVT} can be inferred from the brane construction.\footnote{The brane setup is not perturbative in string theory, as it involves NS5-branes. The rules for reading off the light degrees of freedom and couplings generalize the more supersymmetric constructions in~\cite{Hanany:1996ie}.
For instance, consider a D2 and a D2$^\prime$-brane ending on different sides of an NS5-brane, or spanning between two parallel NS5-branes.  The three types of branes preserve 0d $\cN=(0,4)$ supersymmetry hence massless modes of strings stretching from D2 to D2$^\prime$ can be either in a hypermultiplet or a Fermi multiplet: in 0d $\cN=(0,2)$ language, a Fermi multiplet or a pair of chiral multiplets. To find out which multiplet appears in either situation, we can T-dualize the standard ADHM construction describing instantons in an $\mathcal N=2$ quiver gauge theory and involving the subsequence of branes D4-NS5-D4/D0-NS5-D4 along two spatial directions to bring it into the form D2-NS5-D2/D2$^\prime$-NS5-D2. Borrowing from the ADHM dictionary, we then conclude that strings stretching between D2 and a D2$^\prime$-branes ending on different sides of an NS5-brane constitute a Fermi multiplet, while those stretching between D2 and a D2$^\prime$-branes suspended between two parallel NS5-branes make up a hypermultiplet.} The intersection degrees of freedom are thus coupled to the two $\cN=(2,2)$ theories.

The FI parameter $\xi_\text{FI}$ corresponding to the $\ell$-th gauge group factor of the right two-dimensional $\cN=(2,2)$  gauge theories is encoded in the separation between the $\ell$-th and $(\ell+1)$-th NS5$^\prime$-brane along the $x^7$ coordinate. We take the NS5$^\prime$-branes to coincide in their location along $x^7$. Thus, all the FI parameters for gauge group factors with $\ell\geq 2$ vanish.\footnote{The setup where these FI parameters do not vanish corresponds to multiple insertions of degenerate fields in Toda CFT~\cite{Gomis:2014eya}.} Similarly, the separation in the $x^7$ direction of the NS5$^{\prime\prime}$-branes encode the FI parameters of the left quiver, all of which vanish for $\ell\geq 2$ when we take the branes to have the same $x^7$ coordinate.\footnote{The complexified FI parameters are taken to vanish for all these nodes.} The complexified FI parameter~\eqref{FI} for the innermost gauge group factor for the left and right quiver are non-zero and encode the position of the respective defect on the UV-curve. The case that has the simplest Toda CFT interpretation is when they are opposite, \ie{}, when\footnote{When they are opposite, we conjecture that the partition  function of the intersecting defect is computed by the insertion of a single degenerate field in Toda CFT, with momentum $\alpha=-b\Omega-\Omega'/b$. The partition function when  $\xi^{(L)}\neq -\xi^{(R)}$ is expected to correspond to the insertion of two degenerate fields, one with momentum $\alpha= -b\Omega$ and the other with momentum $\alpha=-\Omega'/b$.}
\beq
\xi^{(L)}=-\xi^{(R)}\,.
\eeq
We thus end up with precisely the QFT encoded in the 4d/2d/0d quiver diagram in \autoref{fig:proposalintersectingWVT}. The brane construction  can be easily generalized to other $\cN=2$ theories.\footnote{Moving NS5$'$-branes along $x^{10}$ does not affect the IR description.  In particular, the two-dimensional $\cN=(2,2)$ Seiberg-like duality for a gauge factor $U(n_\kappa)$ with $1\leq\kappa<\nu$ is realized by exchanging the $x^{10}$ positions of neighboring NS5$'$-branes.  On the other hand, moving an NS5$'$-brane past the middle NS5-brane realizes a Seiberg-like duality on the gauge factor $U(n_\nu)$; it would be interesting to clarify how the 0d Fermi multiplet transforms under such a duality, and correspondingly what matter content to expect from a brane configuration with NS5$'$ and NS5$''$-branes on the same side of the brane configuration depicted above.}

\begin{figure}\centering
  \includegraphics{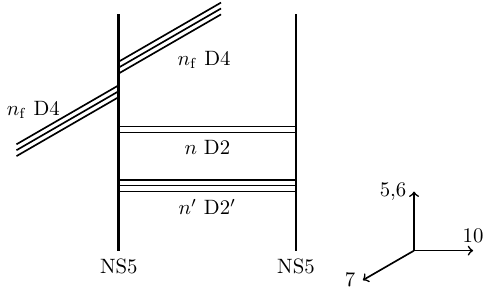}
  \caption{\label{fig:IIAbranessymm}IIA brane diagram for the case of symmetric representations, namely the quiver in \autoref{fig:proposalintersectingWVTsymmetrics}.}
\end{figure}
\begin{figure}\centering
  \includegraphics{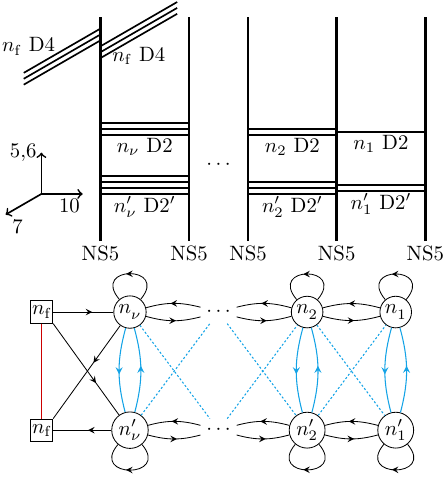}
\caption{\label{fig:quivermanysymm} IIA brane diagram and joint 4d/2d/0d quiver diagram description for multiple M2-brane intersections labeled by symmetric representations ending on $n_\text{f}$ M5-branes wrapping a trinion with two full and one simple puncture.  Each $(\operatorname{sym}^{n^\prime}\protect\yngs(.4,1),\operatorname{sym}^{n}\protect\yngs(.4,1))$ M2-brane intersection is encoded as an NS5-brane (parallel to that on which D4-branes end) on which $n'$~D2$'$ and $n$~D2-branes end.  Gauge group ranks in the quiver description are given by the numbers of D2$'$~and D2-branes stretching in each interval.  These ranks decrease: $n_\nu\geq\cdots\geq n_1$ and $n'_\nu\geq\cdots\geq n'_1$ (otherwise supersymmetry is broken) and their differences give the orders $(n',n)$ of symmetric representations labeling M2-brane intersections.  The FI parameters of gauge group factors are pairwise equal and equal to distances between consecutive NS5-branes.  In each 2d theory, cubic superpotentials couple adjoint and bifundamental chiral multiplets.  Cubic and quartic superpotentials couple the 4d and 2d fields.  The intersection features pairs of chiral multiplets corresponding to strings stretching between D2 and D2$'$ in the same interval and Fermi multiplets corresponding to strings stretching between D2 and D2$'$ in neighboring intervals.  Apart from the D4-branes the brane setup preserves 0d $\cN=(0,4)$ supersymmetry hence the 0d and 2d fields that are neutral under the $SU(\nf)\times SU(\nf)\times U(1)$ flavor symmetry are coupled through quadratic $E$-term and $J$-term superpotentials (see~\cite{Tong:2014yna}).}
\end{figure}

The brane picture describing the 4d/2d/0d quiver diagram in \autoref{fig:proposalintersectingWVTsymmetrics} is given in \autoref{fig:IIAbranessymm}.
The right and left two-dimensional theories live on the D2 and D2$^\prime$ respectively. The open strings stretching between the D2 and D2$^\prime$-branes provide the 0d $\cN=(0,2)$ chiral multiplets. There is a unique FI parameter measuring the distance between the NS5-branes in the $x^7$ direction.  The brane system readily generalizes to D2 and D2$'$-branes stretching between any number of parallel NS5-branes, as depicted in \autoref{fig:quivermanysymm}.

\section{\label{section:dictionary}Liouville/Toda Degenerate Correlators}

It is now time to test in detail our conjectures on the quiver description of intersecting M2-brane surface operators.  We give here the precise dictionary between the partition functions computed in \autoref{section:intersecting-partition-function-first-principle} and Liouville/Toda CFT degenerate correlators.  We begin in \autoref{section:dictionary-fundamental} with the simplest non-trivial case: a Liouville correlator ($n_\text{f}=2$) with two generic, one semi-degenerate, and a degenerate operator labeled by $(\yngs(.4,1),\yngs(.4,1))$.  We move on to Toda CFT in \autoref{section:dictionary-conjecture-fermi} devoted to the quiver in \autoref{fig:proposalintersectingWVT}, and in        \autoref{section:dictionary-conjecture-symmetric} to \autoref{fig:proposalintersectingWVTsymmetrics}.  In each case we describe the evidence worked out in the appendices.

\subsection{\label{section:dictionary-fundamental}Liouville Fundamental Degenerate}

\begin{figure}\centering
\includegraphics{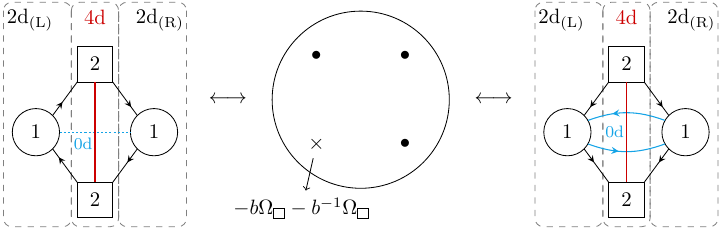}
\caption{\label{fig:Liouville} The left quiver (denoted $\mathcal{T}_{\text{Fermi}}$ in the text due to its 0d matter content) shows the worldvolume theory of two intersecting surface defects, both labeled by the fundamental representation of $A_1$, coupled to the four-dimensional theory of four free hypermultiplets and to a Fermi multiplet on their intersection. The coupling introduces cubic superpotentials. In the middle, the corresponding UV-curve is depicted: it features three punctures $\bullet$ and an additional marked point corresponding to the defect and labeled by its defining representations. In the AGT correspondence, the latter corresponds to the insertion of a degenerate vertex operator with the indicated momentum in Liouville theory. The right quiver ($\mathcal{T}_{\text{chirals}}$) depicts a dual realization of the same intersecting defect, in which the intersection features a pair of bifundamental chiral multiplets. Note that the free $U(1)$ adjoint chiral multiplets have been omitted in the latter quiver.}
\end{figure}

We focus here on the setting of $A_1$ theories ($\nf=2$) for the case of a degenerate operator with Liouville momentum\footnote{We use standard Liouville CFT notation.  Vertex operators $\hat{V}_\alpha$ are labeled by their complex momentum~$\alpha$ and their conformal dimension is equal to twice their (anti)holomorphic conformal weight $\dimToda(\alpha)=\alpha(Q-\alpha)$ where $Q=b+\frac{1}{b}$.  In the appendix we describe a normalization $\hat{V}_\alpha=N^{(\alpha)}V_\alpha$ for which $\hat{V}_{Q-\alpha}=\hat{V}_{\alpha}$.  We use a different normalization to define $\hat{V}$ for degenerate vertex operators ($\alpha=-m\frac{b}{2}-n\frac{1}{2b}$ with $m,n\geq 0$) because $N^{(\alpha)}$ is singular.  Liouville and Toda CFT notations are related by measuring momenta in units of the positive root of~$A_1$.} $\alpha = -b \Omega_{\yngs(.4,1)} - b^{-1} \Omega_{\yngs(.4,1)} = -\frac{b}{2} - \frac{1}{2b} = -Q/2$.  The two conjectured quiver descriptions of the intersecting M2-brane surface operators are depicted in \autoref{fig:Liouville}, as well as the UV-curve.  We prove in Appendix~\ref{appendix:Liouville-check} that the two descriptions have equal $S^4_b$~expectation values and check up to fifth order in vortex expansions that they match a degenerate Liouville correlator.
Namely,
\begin{equation}\label{Liouville-matching}
  \vev{\hat{V}_{-\frac{Q}{2}}(x,\bar{x})  \hat{V}_{\alpha_1}(0)  \hat{V}_{\alpha_2}(1)  \hat{V}_{\alpha_3}(\infty)}
  = \frac{1}{A_1(x,\bar{x})} Z_{S^4_b}\left[\mathcal{T}_{\text{Fermi}}\right]
  = \frac{1}{A_2(x,\bar{x})} Z_{S^4_b}\left[\mathcal{T}_{\text{chirals}}\right]
\end{equation}
where the prefactors $A_1(x,\bar{x})$ and $A_2(x,\bar{x})$ given in Appendix~\ref{appendix:Liouville-check} can be associated to ambiguities in the definition of the gauge theory partition function, as explained in~\cite{Gomis:2014eya}.  The position~$x$ gives the FI and theta parameters through $e^{-2\pi\xi+i\vartheta}=1/x$ for the left $U(1)$ of the first quiver and~$x$ for all other gauge groups.

We will denote the complexified twisted masses of (anti)fundamental chiral multiplets of the right and left theories by $(m^{(R)}_j, \anti{m}^{(R)}_s)$ and $(m^{(L)}_s, \anti{m}^{(L)}_j)$ for the first quiver and $(m_j, \anti{m}_s)$ and $(m'_j, \anti{m}'_s)$ for the second.  The 4d/2d superpotentials relate twisted masses of the two 2d theories in each quiver as \eqref{mass-relation-fermi-from-superpotential} and~\eqref{mass-relation-chiral-from-superpotential} respectively.  In fact, twisted masses of the two quivers are related by Seiberg-like duality as we will later see:
\begin{gather}
  \label{Liouville-mrel1}
  b^{-1}i m^{(R)}_j = b i\anti{m}^{(L)}_j = b^{-1} (i m_j-1/2) = b(i m'_j-1/2) \qquad \text{for $j=1,2$,} \\
  \label{Liouville-mrel2}
  b^{-1}i\anti{m}^{(R)}_s = b i m^{(L)}_s = b^{-1} (i\anti{m}_s+1/2) = b(i\anti{m}'_s+1/2) \qquad \text{for $s=1,2$.}
\end{gather}
Liouville CFT momenta can then be written in terms of twisted masses of any of the four 2d theories\footnote{We subtracted $Q/2$ from each momentum to make the symmetries $\alpha_i\to Q-\alpha_i$ more manifest.  The odd choice of sign for $\alpha_3-Q/2$ is chosen to match our $\nf>2$ result.},
\begin{subequations}
  \label{Liouville-momenta}
  \begin{align}
    & \begin{aligned}
      \alpha_1 -\frac{Q}{2}
      & = \frac{1}{2b}(i m^{(R)}_1-i m^{(R)}_2)
      = \frac{b}{2}(i\anti{m}^{(L)}_1-i\anti{m}^{(L)}_2) \\
      & = \frac{1}{2b}(i m_1-i m_2)
      = \frac{b}{2}(i m'_1-i m'_2) \,,
    \end{aligned}
    \displaybreak[1]\\
    & \begin{aligned}
      \alpha_2 - \frac{Q}{2}
      & = \frac{1}{2b}(i\anti{m}^{(R)}_1 + i\anti{m}^{(R)}_2 - i m^{(R)}_1 - i m^{(R)}_2)
      = \frac{b}{2}( i m^{(L)}_1 + i m^{(L)}_2 - i\anti{m}^{(L)}_1 - i\anti{m}^{(L)}_2) \\
      & = \frac{1}{b} + \frac{1}{2b}(i\anti{m}_1 + i\anti{m}_2 - i m_1 - i m_2)
      =  b + \frac{b}{2}(i\anti{m}'_1 + i\anti{m}'_2 - i m'_1 - i m'_2) \,,
    \end{aligned}
    \displaybreak[1]\\
    & \begin{aligned}
      \alpha_3 - \frac{Q}{2}
      & = \frac{1}{2b}(i\anti{m}^{(R)}_2 - i\anti{m}^{(R)}_1)
      = \frac{b}{2}( i m^{(L)}_2 - i m^{(L)}_1) \\
      & = \frac{1}{2b}(i\anti{m}_2 - i\anti{m}_1)
      = \frac{b}{2}( i\anti{m}'_2 - i\anti{m}'_1) \,,
    \end{aligned}
  \end{align}
\end{subequations}

Let us describe salient aspects of the relation, leaving details for Appendix~\ref{appendix:Liouville-check}.  The operator product expansion (OPE) of a generic operator $\hat{V}_{\alpha}$ with the degenerate operator $\hat{V}_{-\frac{Q}{2}}$ is given by
\begin{equation}
  \hat{V}_{-\frac{Q}{2}}(x,\bar{x}) \hat{V}_{\alpha_1}(0) \sim \sum_{s_1=\pm, s_2=\pm}  (x\bar{x})^{\dimToda(\alpha_{s_1s_2})-\dimToda(\alpha_1)-\dimToda(-Q/2)} \ \hat{C}_{\alpha_1,-\frac{Q}{2}}^{\alpha_{s_1s_2}} \  \bigl(\hat{V}_{\alpha_{s_1s_2}}(0) + \cdots\bigr)\;,
\end{equation}
where $\alpha_{s_1s_2}=\alpha_1+s_1b/2+s_2/(2b)$, the structure constants $\hat{C}_{\alpha,-Q/2}^{\alpha\pm b/2\pm1/(2b)}$ are known and the $\cdots$ denotes Virasoro descendant fields multiplied by powers of $x$ or~$\bar{x}$.  In the limit $x\to 0$ the Liouville correlator~\eqref{Liouville-matching} thus admits an s-channel decomposition as a sum of four terms with leading powers of $x\bar{x}$ equal to
\begin{equation}\label{Liouville-sch-expo}
  \begin{aligned}
    & \dimToda\bigl(\alpha_1+s_1b/2+s_2/(2b)\bigr) - \dimToda(\alpha_1) - \dimToda(-Q/2)
    \\
    & \qquad = Q^2/2-(\alpha_1-Q/2)(s_1 b+s_2 b^{-1})+(1-s_1s_2)/2 \,.
  \end{aligned}
\end{equation}
Correspondingly, each of the two gauge theory partition functions can be written as a sum of contributions from four Higgs branches in this limit.

In the first quiver, $x\to 0$ is the limit of large positive FI parameter for the right $U(1)$ and negative FI parameter for the left $U(1)$ and Higgs branches are located at $\sigma^{(R)}=m^{(R)}_j$ and $\sigma^{(L)}=\anti{m}^{(L)}_k$ for $j,k=1,2$.  The leading power of $(x\bar{x})$ of the $(j,k)$ Higgs branch contribution is $i\sigma^{(R)}-i\sigma^{(L)}=i m^{(R)}_j-i\anti{m}^{(L)}_k$ with a sign due to the FI parameter of the left theory being opposite to that of the right theory.  In fact, for $j=k$ the 0d Fermi multiplet contribution makes zero-vortex terms in the series vanish, so that the leading power of $(x\bar{x})$ is $i m^{(R)}_j-i\anti{m}^{(L)}_j+1$ instead.  The partition function thus takes the form
\begin{equation}
  Z = \sum_{j=1}^{2} \sum_{k=1}^{2} (x\bar{x})^{i m^{(R)}_j-i\anti{m}^{(L)}_k+\delta_{jk}} (\text{series in $x$}) (\text{series in $\bar{x}$}) \,.
\end{equation}
The identification~\eqref{Liouville-momenta} of momenta with twisted masses ensures that the four gauge theory exponents match the Liouville ones up to the prefactor $A_1(x,\bar{x})$.  In particular the shift by~$\delta_{jk}$ due to the 0d Fermi multiplet reflects the term $(1-s_1s_2)/2$ in~\eqref{Liouville-sch-expo}.

In the second quiver, $x\to 0$ is the limit of large positive FI parameters and the Higgs branches are located at $\sigma=m_j$ and $\sigma'=m'_k$.  The 0d chiral multiplet contribution~\eqref{intersectsymm} has poles that induce additional terms, in effect decreasing the leading power of $(x\bar{x})$ by~$1$ for terms with $j=k$.  The partition function takes the form
\begin{equation}\label{Liouville-gaugetheory-sch}
  Z = \sum_{j=1}^{2} \sum_{k=1}^{2} (x\bar{x})^{i m_j+i m'_k-\delta_{jk}} (\text{series in $x$}) (\text{series in $\bar{x}$}) \,.
\end{equation}
Again, gauge theory and Liouville exponents match.  The shift of the exponent by~$\delta_{jk}$ has opposite signs in the first and second quivers, which may seem inconsistent.  However, Liouville CFT internal momenta $\alpha_1\pm b/2\pm 1/(2b)$ are identified with different terms $(j,k)$ for the two quivers: $k=1$ and $k=2$ are interchanged.  The two quivers are in fact related by a Seiberg-like duality of the left 2d theory and we leverage this observation in the appendix to prove that their partition functions are equal.

The Liouville correlator of interest to us has been worked out in \cite{Fateev:2009me} by solving the fourth order differential equation associated with the degenerate puncture.  The leading coefficients in~\eqref{Liouville-gaugetheory-sch} reproduce expected Liouville three-point functions and we checked up to fifth order that vortex partition functions of the intersecting surface defects coincide with conformal blocks.  We performed the same checks (exponents, leading coefficients, vortex partition functions) in the limit $x\to\infty$ using the OPE of $\hat{V}_{-Q/2}$ with~$\hat{V}_{\alpha_3}$.

Pleasingly, the dictionary has all the expected symmetries.
\begin{itemize}
\item Exchanging the flavors $(m^{(R)}_1,\anti{m}^{(L)}_1,m_1,m'_1)\leftrightarrow(m^{(R)}_2,\anti{m}^{(L)}_2,m_2,m'_2)$ corresponds to mapping $\alpha_1\to Q-\alpha_1$, which leaves the normalized vertex operator $\hat{V}_{\alpha_1}$ invariant.  Similarly $(\anti{m}^{(R)}_1,m^{(L)}_1,\anti{m}_1,\anti{m}'_1)\leftrightarrow(\anti{m}^{(R)}_2,m^{(L)}_2,\anti{m}_2,\anti{m}'_2)$ is $\alpha_3\to Q-\alpha_3$.
\item The conformal map $x\to 1/x$ which exchanges $\alpha_1\leftrightarrow\alpha_3$ corresponds to charge conjugation for all gauge groups, which exchanges fundamental and antifundamental chiral multiplets, changing their signs as well as those of FI and theta parameters.  The conformal factor $(x\bar{x})^{2\dimToda(-Q/2)}$ coincides with a change in $A_1(x,\bar{x})$ and $A_2(x,\bar{x})$.
\item For each quiver, the $b\to 1/b$ symmetry of the Liouville correlator exchanges the two two-dimensional theories (up to charge conjugation for the case of the first quiver).
\end{itemize}

\begin{figure}\centering
  \includegraphics{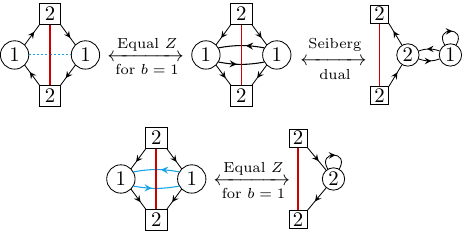}
  \caption{\label{fig:b1case}First row: the partition function of the first intersecting surface defect coincides for $b=1$ with that of a single surface defect, Seiberg-dual to the one expected from~\cite{Gomis:2014eya}.  Second row: likewise, the partition function of the second intersecting surface defect reduces for $b=1$ to that of a single surface defect.}
\end{figure}

For $b=1$ the $\hat{V}_{-Q/2}$ degenerate operator coincides with $\hat{V}_{-b}$ already studied in~\cite{Gomis:2014eya} and the partition functions reduce to that of a single two-dimensional theory coupled to the four-dimensional free hypermultiplets.
More precisely, up to a shift of theta angles by~$\pi$, the 0d Fermi multiplet contribution in the first quiver can be written for $b=1$ as the one-loop determinant of a pair of bifundamental 2d chiral multiplets of $\mathcal{R}$-charge~$2$:
\begin{equation}
  \begin{aligned}
    &
    (-1)^{B^{(R)}+B^{(L)}}
    \prod_{\pm} \pm (i \sigma^{(R)} \pm B^{(R)}/2 - i\sigma^{(L)} \mp B^{(L)}/2)
    \\
    & \quad =
    \frac{\Gamma(1+i\sigma^{(R)}+B^{(R)}/2 - i\sigma^{(L)}-B^{(L)}/2)}{\Gamma(-i\sigma^{(R)}+B^{(R)}/2 + i\sigma^{(L)}-B^{(L)}/2)}
    \frac{\Gamma(1-i\sigma^{(R)}-B^{(R)}/2 + i\sigma^{(L)}+B^{(L)}/2)}{\Gamma(i\sigma^{(R)}-B^{(R)}/2 - i\sigma^{(L)}+B^{(L)}/2)} \,.
  \end{aligned}
\end{equation}
As depicted in \autoref{fig:b1case}, the partition function is thus equal to that of a single 2d $U(1)\times U(1)$ gauge theory (coupled to free hypermultiplets), which is itself equivalent under a Seiberg-like duality to the quiver corresponding to $\hat{V}_{-b}$ in~\cite{Gomis:2014eya}.  Importantly the bifundamental 2d chiral multiplets in the last quiver have $\mathcal{R}$-charge $-1=-b^2$.  A $U(2)$ gauge theory description of $\hat{V}_{-b}$ in~\cite{Gomis:2014eya} matches the second quiver for $b=1$ (ignoring free chiral multiplets), as depicted in \autoref{fig:b1case}.  There, the adjoint chiral multiplet has $\mathcal{R}$-charge $-2=-2b^2$.  Indeed, its one-loop determinant combines with the $U(2)$ vector multiplet one-loop determinant to give the 0d chiral multiplet contribution of the intersecting defects.

As we will see in the next section in a more general setting, the identification of the first quiver with a Liouville correlator still holds if the FI and theta parameters of the two gauge groups are taken to be arbitrary rather than opposite.  The partition function then matches a five-point function with two degenerate operators $\hat{V}_{-b/2}(x,\bar{x})$ and $\hat{V}_{-1/(2b)}(x',\bar{x}')$ and the three generic~$\hat{V}_{\alpha_i}$.  The FI and theta parameters are given by $\exp(-2\pi\xi^{(R)}+i\vartheta^{(R)})=x$ and $\exp(-2\pi\xi^{(L)}+i\vartheta^{(L)})=1/x'$ and other parameters are unchanged.  The quiver with 0d chiral multiplets does not have the same property: making FI parameters distinct does not reproduce the Liouville five-point function.  This is not surprising, both in view of the $b=1$ case where the surface defect reduces to one with a single gauge group, and in view of the IIA realization where D2 and D2$'$-branes stretch between the same pair of NS5-branes.

We now move on to arbitrary intersecting defects for any number of flavors~$\nf$.

\subsection{\label{section:dictionary-conjecture-fermi}Quiver with 0d Fermi Multiplet}

This section presents the quiver description of intersecting defects corresponding to an arbitrary set of Toda CFT degenerate operators.  We focus on degenerate operators labeled by antisymmetric representations, because all degenerate operators can be obtained as the dominant term in the OPE of such degenerate operators (see \autopageref{page:ope-dominant-term}).  Besides comparing leading terms in several channels as in the last section, we prove in Appendix~\ref{appendix:fermi-quiver-check} that some braiding matrices coincide.

The main statement is
\begin{equation}
  \label{general-matching-sphere}
  Z\left[\mathcal{T}_{\text{Fermi}}\right]
  = A_1(x,x';\bar{x},\bar{x}') \biggl\langle \hat{V}_{\alpha_{\infty}}(\infty) \hat{V}_{\lambda\omega_1}(1) \hat{V}_{\alpha_{0}}(0)
  \prod_{\iota=1}^{\nu} \hat{V}_{-b\omega_{K_{\iota}}}(x_{\iota},\bar{x}_{\iota}) \prod_{\iota=1}^{\nu'} \hat{V}_{-b^{-1}\omega_{\nf-K'_{\iota}}}(x'_{\iota},\bar{x}'_{\iota})
  \biggr\rangle \,.
\end{equation}
The left-hand side is the expectation value of the intersecting surface defect of \autoref{fig:proposalintersectingWVT} in the theory of $\nf^2$~free 4d hypermultiplets: a $U(n_1)\times\cdots\times U(n_\nu)$ gauge theory on $S^2_b$, a $U(n'_1)\times\cdots\times U(n'_{\nu'})$ gauge theory on $S^2_{1/b}$, and on their intersection a single 0d $\cN=(0,2)$ Fermi multiplet in the bifundamental representation of $U(n_\nu)\times U(n'_{\nu'})$ with $\mathcal{R}$-charge zero.  Couplings are explained in previous sections.

The right-hand side\footnote{The prefactor $A_1$ given explicitly in~\eqref{conj1-axx} is an ambiguity of the sphere partition function~\cite{Gomis:2014eya}.} is a Toda CFT correlator of two generic vertex operators at $\infty$ and~$0$, one semi-degenerate at~$1$, and $\nu+\nu'$ degenerates at $x_\iota$ and~$x'_\iota$.  Vertex operators $\hat{V}_{\alpha}$ are labeled by their momentum~$\alpha$, a linear combination of the weights $h_1,\ldots,h_{\nf}$ of the fundamental representation of $SU(\nf)$ (the~$h_i$ sum to~$0$).  We normalize the generic vertex operators $\hat{V}_{\alpha_\infty}$ and~$\hat{V}_{\alpha_0}$ such that they are invariant under $\alpha\to 2Q-\alpha$ and under the Weyl group, which permutes components of $\alpha-Q$, where $Q=(b+b^{-1})\rho$ with $\rho$ the half-sum of positive roots.\footnote{Liouville and Toda CFT conventions unfortunately differ by factors of~$2$, for instance the conformal weight of~$\hat{V}_{\alpha}$ is $\dimToda(\alpha)=\vev{Q,\alpha}-\frac{1}{2}\vev{\alpha,\alpha}$.  The Killing form is such that $\vev{h_i,h_j}=\delta_{ij}-1/\nf$.}  One degenerate operator has momentum $-b\omega_K$ proportional to the highest weight $\omega_K=h_1+\cdots+h_K$ of the $K$-th antisymmetric representation, and the other has momentum $-b^{-1}\omega_{\nf-K'}$, corresponding to the conjugate of the $K'$-th antisymmetric representation.

In short, the dictionary is that mass parameters of the $SU(\nf)\times SU(\nf)\times U(1)$ flavor symmetry are encoded in $\alpha_0$, $\alpha_\infty$ and $\lambda$ respectively, complexified FI parameters give positions of degenerate punctures, and gauge group ranks determine the antisymmetric representations.

We find $K_{\kappa}=n_{\kappa}-n_{\kappa-1}$ and $K'_{\kappa}=n'_{\kappa}-n'_{\kappa-1}$ (and $K_1=n_1$ and $K'_1=n'_1$),
\begin{equation}
  x_{\kappa} = \prod_{\iota=\kappa}^{\nu} \hat{z}_{\iota} \quad \text{and} \quad x'_{\kappa} = \prod_{\iota=\kappa}^{\nu'} \frac{1}{\hat{z}'_{\iota}} \,.
\end{equation}
Here, $\hat{z}_{\nu}=(-1)^{n_{\text{f}}+n_{\nu-1}-n_{\nu}+1}z_\nu$ and $\hat{z}_{\iota} = (-1)^{n_{\iota+1}+n_{\iota-1}} z_{\iota}$ for $1\leq \iota\leq \nu-1$ in terms of $z_{\kappa} = e^{-2\pi\xi_{\kappa}+i\vartheta_{\kappa}}$ and similarly for $\hat{z}'$ and~$z'$.

In quiver conventions, we recall that twisted masses $\mathrm{m}$ and $\mathcal{R}$-charges of (anti)fun\-da\-men\-tal chiral multiplets combine as $i m^{(R)}_j = i \mathrm{m}^{(R)}_j/\ell + \mathcal{R}^{(R)}_{\text{2d}}[q^{(R)}_j]/2$ and $i\anti{m}^{(R)}_s = i\anti{\mathrm{m}}^{(R)}_s/\ell - \mathcal{R}^{(R)}_{\text{2d}}[\anti{q}^{(R)}_s]/2$ for the right theory and $i m^{(L)}_s=i\mathrm{m}^{(L)}_s/\tilde{\ell}+\mathcal{R}^{(L)}_{\text{2d}}[q^{(L)}_s]/2$ and $i\anti{m}^{(L)}_j=i\anti{\mathrm{m}}^{(L)}_j/\tilde{\ell}-\mathcal{R}^{(L)}_{\text{2d}}[\anti{q}^{(L)}_j]/2$ for the left one.  As explained in \autoref{section:intersecting-partition-function-first-principle}, two-dimensional masses are related by~\eqref{mass-relation-fermi-from-superpotential},
\begin{equation}
  b^{-1} m^{(R)}_j = b\anti{m}^{(L)}_j
  \quad \text{and} \quad
  b^{-1}\anti{m}^{(R)}_s = bm^{(L)}_s \,,
\end{equation}
and four-dimensional hypermultiplets have masses $M_{js} = \frac{b-b^{-1}}{2i}+b^{-1}m^{(R)}_j-b^{-1}\anti{m}^{(R)}_s$.  In the theory on the right, bifundamental chiral multiplets have $i m^{(R)}_{\text{bif}}=b^2/2$ namely $\mathcal{R}$-charge $-b^2$ and adjoint chiral multiplets have $i m^{(R)}_{\text{adj}}=-1-b^2$ namely $\mathcal{R}$-charge $2+2b^2$.  Similarly, $i m^{(L)}_{\text{bif}}=b^{-2}/2$ and $i m^{(L)}_{\text{adj}}=-1-b^{-2}$.

The $S[U(\nf)\times U(\nf)]$ mass parameters correspond to Toda CFT momenta as\footnote{We subtract~$Q$ from generic momenta and $\nf(b+b^{-1})/2$ from the semi-degenerate momentum to make symmetry under Toda CFT conjugation (discussed shortly) more manifest.  Due to the 4d/2d superpotential the right-hand sides have real parts $0$, $0$, and $n_\nu b-n'_{\nu'}b^{-1}$ which vanish in the absence of surface defect.}
\begin{align}\label{general-matching-alpha}
  \alpha_0 - Q & = \sum_{j=1}^{\nf} b^{-1} i m^{(R)}_j h_j
  = \sum_{j=1}^{\nf} bi\anti{m}^{(L)}_j h_j
  \\
  \alpha_\infty - Q & = - \sum_{s=1}^{\nf} b^{-1} i\anti{m}^{(R)}_s h_s
  = - \sum_{s=1}^{\nf} bi m^{(L)}_s h_s
  \\
  \lambda - \frac{b+b^{-1}}{2}\nf & = \biggl(n_{\nu}-\frac{\nf}{2}\biggr) b + \biggl(\frac{\nf}{2}-n'_{\nu'}\biggr)b^{-1} + \sum_{s=1}^{\nf} b^{-1} i\anti{m}^{(R)}_s - \sum_{j=1}^{\nf} b^{-1} i m^{(R)}_j
  \,.
\end{align}

We can immediately perform simple consistency checks.
\begin{itemize}
\item Permutations of flavors $1\leq j\leq\nf$ permute components of $\alpha_0-Q$ namely perform a Weyl reflection of this momentum; this leaves the normalized vertex operator~$\hat{V}_{\alpha_0}$ invariant.  Similarly, permutations of flavors $1\leq s\leq\nf$ leave $\hat{V}_{\alpha_{\infty}}$ invariant.
\item The conformal map $x\to 1/x$ exchanging $0\leftrightarrow\infty$ corresponds on the gauge theory side to conjugating charges of every gauge group.  The change in~$A_1$ precisely cancels the conformal factor $\prod_{\kappa=1}^{\nu}\abs{x_{\kappa}}^{4\dimToda(-b\omega_{K_{\kappa}})} \prod_{\kappa=1}^{\nu'}\abs{x'_{\kappa}}^{4\dimToda(-b^{-1}\omega_{\nf-K'_{\kappa}})}$.
\item If we set $n_1=0$ then $A_1$, given explicitly in \eqref{conj1-axx} is independent of~$x_1$; similarly, $x'_1$~factors disappear when setting $n'_1=0$.  The matching also reproduces results of~\cite{Gomis:2014eya} for $\nu'=0$ (or $\nu=0$) namely for a single 2d theory.  In that case, conjugating all Toda CFT momenta ($\omega_{K}^C=\omega_{\nf-K}$) is known to correspond to a sequence of 2d $\cN=(2,2)$ Seiberg dualities.  Unfortunately, for intersecting defects it is not known how 0d~matter behaves under Seiberg dualities.
\item Combining $x\to 1/x$, $b\to b^{-1}$, Toda CFT conjugation and Weyl reflections give rise to a symmetry of the gauge theory setup: the two 2d theories are interchanged.  To see this it is useful to note that the conjugate of $\lambda\omega_1$ is $(\nf(b+b^{-1})-\lambda)\omega_1$ up to a Weyl reflection.
\item For $b=1$ there is no distinction between degenerate operators with momenta $-b\omega_K$ and $-b^{-1}\omega_K$.  As in the Liouville case from the previous section, the equality~\eqref{general-matching-sphere} reduces to (a Seiberg dual of) the matching for a single 2d theory with $\nu+\nu'$ gauge groups corresponding to $\nu+\nu'$ antisymmetric degenerate operators.
\end{itemize}

While we have found the dictionary and the prefactors by comparing expansions of the sphere partition function and the Toda correlator in several limits $\hat{z}_{\iota}\to 0,1,\infty$ and $\hat{z}'_{\kappa}\to 0,1,\infty$, we only write details explicitly for $\nu=\nu'=1$ (see Appendix~\ref{appendix:fermi-quiver-check}).

A major piece of evidence in this case is that braiding matrices relating conformal blocks in different Toda CFT channels (different operator product expansions) match the analogous matrices in gauge theory.  This is proven schematically as follows.  The 0d Fermi contribution is recast as a differential operator acting on the product of (a generalization of) partition functions of the left and right 2d theories.  Braiding (analytically continuing) $x$ around~$1$ commutes with this differential operator, thus the braiding matrix coincides with that of the right 2d theory in isolation, itself known to coincide with the Toda CFT braiding matrix.  More precisely, the presence of an additional degenerate vertex operator shifts momenta slightly, and this translates in gauge theory to a different normalization.

\phantomsection\label{page:ope-dominant-term}
To conclude this section, we determine the dominant term in the OPE of degenerate vertex operators.\footnote{\label{foot:other-b-values}We assume $b>0$, as this is the case realized by the geometric background~$S^4_b$.  The arguments easily generalize to $\Re b^2\geq 0$.  For other~$b$, such as the self-dual $\Omega$-background $b^2=-1$, different terms dominate; as a result, degenerate vertex operators labeled by arbitrary representations may appear as the dominant term in the OPE of symmetric degenerate vertex operators, captured by the quiver in \autoref{fig:quivermanysymm}.} The OPE of two degenerate vertex operators $\hat{V}_{-b^{-1}\Omega'_i-b\Omega_i}$ labeled by $(\mathcal{R}'_1,\mathcal{R}_1)$ and $(\mathcal{R}'_2,\mathcal{R}_2)$ is known to be
\begin{equation}
  \!\begin{aligned}
    & \hat{V}_{(\mathcal{R}'_1,\mathcal{R}_1)}(x_1,\bar{x}_1) \hat{V}_{(\mathcal{R}'_2,\mathcal{R}_2)}(x_2,\bar{x}_2) \\
    & = \!\! \sum_{\mathcal{R}',\mathcal{R}} \abs{x_1-x_2}^{2[\dimToda(-\Omega'/b-b\Omega)-\dimToda(-\Omega_1'/b-b\Omega_1)-\dimToda(-\Omega_2'/b-b\Omega_2)]} \hat{C}_{(\mathcal{R}'_1,\mathcal{R}_1),(\mathcal{R}'_2,\mathcal{R}_2)}^{(\mathcal{R}',\mathcal{R})} \bigl( \hat{V}_{(\mathcal{R}',\mathcal{R})} + \cdots \bigr)
  \end{aligned}
\end{equation}
where $\Omega'$ and~$\Omega$ are the highest weights of $\mathcal{R}'$ and $\mathcal{R}$ and the sum ranges over irreducible representations $\mathcal{R}'$ in $\mathcal{R}'_1\otimes\mathcal{R}'_2$ and $\mathcal{R}$ in $\mathcal{R}_1\otimes\mathcal{R}_2$.  The dominant term in this OPE is that with the most negative $\dimToda(-b^{-1}\Omega'-b\Omega)$, and we will see that it is given by the highest weights $\Omega=\Omega_1+\Omega_2$ and $\Omega'=\Omega'_1+\Omega'_2$ of the tensor products.  We sum over irreducible representations~$\mathcal{R}$ in $\mathcal{R}_1\otimes\mathcal{R}_2$, whose highest weights take the form
\begin{equation}
  \Omega = \Omega_1 + \Omega_2 - \sum_{i=1}^{\nf-1} k_i (h_i-h_{i+1}) \qquad \text{for integers $k_i\geq 0$}
\end{equation}
where $h_i-h_{i+1}$ are the simple roots.  They must also be dominant:
\begin{equation}\label{omega-is-dominant}
  \vev{h_i-h_{i+1},\Omega}\geq 0 \qquad \text{for all $1\leq i<\nf$.}
\end{equation}
The highest weight~$\Omega'$ is parametrized similarly by integers $k'_i\geq 0$.  We prove that $\dimToda(-b^{-1}\Omega'-b\Omega)$ is minimal for $k_i=0$ and $k'_i=0$ by allowing real $k_i,k'_i\geq 0$ and showing that derivatives are positive in the region~\eqref{omega-is-dominant}:
\begin{align}
  \partial_{k_i} \dimToda(-b^{-1}\Omega'-b\Omega)
  & = \vev*{\partial_{k_i}(-b^{-1}\Omega'-b\Omega),Q+b^{-1}\Omega'+b\Omega} \\
  & = \vev*{h_i-h_{i+1},Q+b^{-1}\Omega'+b\Omega} \geq b+b^{-1}\,.
\end{align}
We conclude by noting that the space carved out by~\eqref{omega-is-dominant} is convex.

From the pairwise OPE of degenerate vertex operators we deduce that the dominant term in the OPE of any number of degenerate vertex operators has a momentum equal to the sum of all momenta.  Given that any weight is a sum of fundamental weights~$\omega_K$, any vertex operator is the dominant term in the OPE of some set of antisymmetric degenerate vertex operators.  Explicitly in the case where we fuse all $\nu+\nu'$ degenerate operators,
\begin{equation}
  \prod_{\iota=1}^{\nu} \hat{V}_{-b\omega_{K_\iota}}(x_\iota,\bar{x}_\iota) \prod_{\iota=1}^{\nu'} \hat{V}_{-b^{-1}\omega_{\nf-K'_\iota}}(x'_\iota,\bar{x}'_\iota)
  = a(\{x_\iota,x'_\iota\})a(\{\bar{x}_\iota,\bar{x}'_\iota\}) \hat{V}_{-b^{-1}\Omega'-b\Omega}(x) + \cdots
\end{equation}
as $x_\iota,x'_\iota\to x$ (we suppressed subleading terms), where the prefactor~$a$ consists of powers of position differences (the three-point functions turn out to be~$1$),
\begin{equation}\label{omega-as-sum}
  \Omega = \sum_{\iota=1}^{\nu} \omega_{K_\iota} \qquad \text{and} \qquad
  \Omega' = \sum_{\iota=1}^{\nu'} \omega_{\nf-K'_\iota} \,.
\end{equation}
The Young diagram associated to~$\Omega$ has $\nu$~columns with $K_1,\ldots,K_\nu$ boxes in some order.  The Young diagram associated to~$\Omega'$ has columns with $\nf-K'_\iota$ boxes, or equivalently the conjugate representation has a Young diagram with $K'_\iota$-box columns.

Translating to gauge theory, the fusion limit corresponds to $\hat{z}_{\nu}=x$, $\hat{z}'_{\nu'}=1/x$ and all other $\hat{z}_{\iota}=\hat{z}'_{\kappa}=1$.  Selecting the leading term in the OPE corresponds to ignoring vacua that go to infinity along the Coulomb branch when setting FI parameters to~$0$.

In the case depicted in the introduction, namely $K_1\leq\cdots\leq K_\nu$ and $K'_1\leq\cdots\leq K'_{\nu'}$, many factors in the prefactors $A_1$ and~$a$ cancel.  The limit $x_\kappa\to x$ and $x'_\kappa\to x'$ is then smooth, and in the limit $x'\to x$ the partition function behaves as
\begin{equation}
  Z(x,x') \sim \abs{x-x'}^{2\sum_{\iota=1}^{\nu'}\sum_{\kappa=1}^{\nu}\max(0,K'_\iota+K_\kappa-\nf)} A_1(x;\bar{x}) \vev*{\hat{V}_{\alpha_{\infty}}(\infty) \hat{V}_{\lambda\omega_1}(1) \hat{V}_{\alpha_{0}}(0) \hat{V}_{-b^{-1}\Omega'-b\Omega}(x,\bar{x})} \,.
\end{equation}
While the simplicity of the factor is convenient for calculations, and in particular allows one to write an explicit formula for the Toda CFT four-point function, one should remember that the prefactors depend on the renormalization scheme.

\subsection{\label{section:dictionary-conjecture-symmetric}Quiver with 0d Chiral Multiplets}

We give in this section a dual quiver description of the intersecting defect labeled by a pair of symmetric representations.  The main statement is
\begin{equation}
  \label{conj2-matching}
  Z\left[\mathcal{T}_{\text{chirals}}\right] = A_2(x;\bar{x}) \Bigl\langle\hat{V}_{\alpha_{\infty}}(\infty) \hat{V}_{\lambda\omega_1}(1) \hat{V}_{\alpha_{0}}(0) \hat{V}_{-(n'b^{-1}+nb)\omega_1}(x,\bar{x})\Bigr\rangle \,.
\end{equation}
The left-hand side is the expectation value, in the theory of $\nf^2$~free hypermultiplets on~$S^4_b$, of the intersecting surface defect of \autoref{fig:proposalintersectingWVTsymmetrics} described by a $U(n)$ theory on one two-sphere and a $U(n')$ theory on the other, coupled through a pair of bifundamental 0d chiral multiplets on their intersection.  Both the $U(n)$ and the $U(n')$ theories have one adjoint, $\nf$~fundamental and $\nf$~antifundamental chiral multiplets.  Twisted masses obey
\begin{equation}
    b^{-1}(i m_j-1/2) = b (i m'_j-1/2) \qquad \text{and} \qquad b^{-1} (i\anti{m}_s+1/2) = b (i\anti{m}'_s+1/2)
\end{equation}
due to cubic superpotential couplings with the free 4d hypermultiplets.  Adjoint chiral multiplets of the $U(n)$ and $U(n')$ theories have $\mathcal{R}$-charges $-2b^2$ and $-2/b^2$ respectively due to 0d/2d superpotential terms.  The two theories have equal FI and theta parameters.

The prefactor~$A_2$ given in~\eqref{conj2-A2} is as before an ambiguity of the $S^4_b$~partition function, and the Toda CFT correlator features two generic and one semi-degenerate operators.  The degenerate vertex operator is labeled by the $n'$-th and the $n$-th symmetric representations of $SU(\nf)$ and placed at $x=(-1)^{\nf}e^{-2\pi\xi+i\vartheta}$.  Momenta encode twisted masses as follows:\footnote{The 4d/2d superpotential implies that the right-hand sides have real parts $0$, $0$, and $n b+n'b^{-1}$.}
\begin{align}\label{symm-matching-alpha}
  \alpha_0 - Q & = \sum_{j=1}^{\nf} b^{-1} i m_j h_j
  = \sum_{j=1}^{\nf} bi m'_j h_j
  \\
  \alpha_\infty - Q & = - \sum_{s=1}^{\nf} b^{-1} i\anti{m}_s h_s
  = - \sum_{s=1}^{\nf} bi\anti{m}'_s h_s
  \\
  \lambda - \frac{b+b^{-1}}{2}\nf
  & = \biggl(n - \frac{\nf}{2}\biggr) b + \biggl(n'+\frac{\nf}{2}\biggr) \frac{1}{b} + \frac{1}{b}\sum_{s=1}^{\nf} i\anti{m}_s - \frac{1}{b}\sum_{j=1}^{\nf} i m_j \\
  \nonumber & = \biggl(n+\frac{\nf}{2}\biggr) b + \biggl(n'-\frac{\nf}{2}\biggr) \frac{1}{b} + \sum_{s=1}^{\nf} b i\anti{m}'_s - \sum_{j=1}^{\nf} b i m'_j \,.
\end{align}
Contrarily to the previous section, the two 2d theories must share the same FI and theta parameters for the partition function to coincide with a Toda CFT correlator.  In the IIA brane construction, this is understood by noting that all D2 and D2$'$-branes stretch between a single pair of NS5-branes, whose separation gives a single FI parameter.

We can immediately perform consistency checks similar to the previous conjecture.
\begin{itemize}
\item For $\nf=2$ and $n=n'=1$ this reduces to the Liouville matching we discussed earlier.
\item Permutations of flavors correspond to Weyl reflections of momenta.
\item The conformal map $x\to 1/x$ corresponds to conjugating gauge theory charges.
\item For $n=0$ or $n'=0$ the matching reduces to previously known results of~\cite{Gomis:2014eya}.
\item A combination of $b\to b^{-1}$ and Weyl reflections exchanges the two 2d theories.
\item For $b=1$ the partition function is equal to that of a single 2d theory with gauge group $U(n+n')$ and one adjoint, $\nf$~fundamental and $\nf$~antifundamental chiral multiplets.
\item For the cases where $n_\mathrm{f} = 2, 3, 4$, $n = n' = 1$ in the quivers with 0d chiral, and $n^R = 1, n^L = n_\mathrm{f} - 1$ in the quivers with 0d Fermi multiplets, we checked up to second order in $x$ that the partition functions of two types of quivers agree.
\end{itemize}
In the limits $x\to 0$ and $x\to\infty$ both the partition function and the Toda CFT correlator decompose into a sum of $\binom{\nf+n-1}{n}\binom{\nf+n'-1}{n'}$ terms.  For each of these terms the leading coefficient and leading exponent of $x\bar{x}$ can be compared.

A detailed discussion of constructing these intersecting surface operators from vortices will appear elsewhere~\cite{Pan:2016fbl}.

\section{Discussion}

In this paper we have initiated the study of intersecting surface operators in four-dimensional QFTs. When intersecting at a point, these can  be constructed  by coupling together 4d/2d/0d degrees of freedom by gauging the global symmetries of defect fields with symmetries acting on higher-dimensional fields.
 In the context of four-dimensional $\cN=2$ supersymmetry, we have shown how to couple the 4d/2d/0d degrees of freedom so as to preserve two supercharges. We have shown that these surface operators are amenable to supersymmetric localization on the $\Omega$-background and the squashed four-sphere.

We have also identified a class of intersecting surface operators that describe M2-brane surface operators ending on a collection of M5-branes wrapping a punctured Riemann surface. It is this class of intersecting surface operators whose squashed four-sphere partition function we conjecturally relate  to correlation functions in Toda CFT in the presence of a general degenerate vertex operator. We have provided rather non-trivial quantitative evidence of this connection by showing that the squashed four-sphere partition function of an intersecting defect in the theory of hypermultiplets matches in detail the correlation function in Toda/Liouville  CFT\@.

The explicit computation of the expectation value of our intersecting defects in a general four-dimensional $\cN=2$ gauge theory becomes more challenging, as the four-dimensional instanton equations are modified by the pair of  two-dimensional and the zero-dimensional degrees of freedom. The explicit 4d/2d/0d quiver diagram realizing the intersecting surface operator gives a   definition of the allowed gauge field singularities
along the two $ \mathbb R^2$'s and of how these singularities merge at the origin, where the zero-dimensional fields are inserted. The partition function of an intersecting defect obtained
by coupling zero-dimensional theories to two-dimensional $\cN=(2,2)$ theories and in turn to a four-dimensional $\cN=2$ gauge theory takes the following form, with $G^{(L), (R)}$ denoting the total gauge groups of the two 2d theories,\footnote{Once again the dependence on masses and FI parameters is left implicit to avoid cluttering formulas.}
\begin{multline}
  \int \mathrm{d}a \sum_{B^{(L)}} \sum_{B^{(R)}} \int_{\mathrm{JK}}   \frac{\mathrm{d}\sigma^{(L)} }{(2\pi)^{\operatorname{rank}G^{(L)}}}  \frac{\mathrm{d}\sigma^{(R)}}{(2\pi)^{\operatorname{rank}G^{(R)}}}
  Z_{S^2_b}(\sigma^{(L)},B^{(L)},a) \ Z_{S^2_b}(\sigma^{(R)},B^{(R)},a) \  Z_{S^4_b}(a)
  \\ \times Z_{\text{0d}}^{\text{intersection}}(\sigma^{(L)}, B^{(L)},\sigma^{(R)},B^{(R)})
  \left|Z_{\text{instanton}}(a,\sigma^{(L)},B^{(L)},\sigma^{(R)},B^{(R)})\right|^2 \,.
   \label{finalanswnew}
\end{multline}
There are new ingredients in addition to those appearing in the analysis in \autoref{section:intersecting-partition-function-first-principle}, where the formulas for
$Z_{S^2_b}(\sigma^{(L/R)},B^{(L/R)},a)$ and  $Z_{\text{0d}}^{\text{intersection}}(\sigma^{(L)}, B^{(L)},\sigma^{(R)},B^{(R)})$ can be found. For a general four-dimensional $\cN=2$ gauge theory we must also localize the four-dimensional gauge dynamics, which results in an integral over the vector multiplet scalar zero mode $a$ in \eqref{finalanswnew}, where $a$ takes values in the Cartan of the four-dimensional gauge group.  $Z_{S^4_b}(a)$ is the familiar classical and one-loop factor in the computation of the $S^4_b$ partition function~\cite{Pestun:2007rz,Hama:2012bg}.
In this more general case, the masses of the innermost chiral multiplets in the 4d/2d/0d quiver diagram can be fixed in terms of the four-dimensional Coulomb branch parameter $a$ by the  localized superpotential. $Z_{\text{instanton}}(a,\sigma^{(L)},B^{(L)},\sigma^{(L)},B^{(L)})$ is the instanton partition function of the 4d/2d/0d theory in the $\Omega$-background. It can be computed by   an ADHM-like matrix integral, which computes the equivariant volume of the instanton moduli space in the presence of the codimension two singularities induced by the two-dimensional fields and codimension four singularities induced by zero-dimensional fields. The ADHM  matrix model has new additional fields in the presence of defect fields (see \cite{Gaiotto:2014ina}). The extra fields in the ADHM matrix model arise from the two-dimensional fields that couple directly to the four-dimensional gauge group, that is the innermost chiral multiplets.\footnote{These contribute to the ADHM matrix model zero-dimensional $\cN=(0,2)$ chiral and Fermi multiplets.} It would be interesting to explicitly compute the partition function of our intersecting defects for gauge theories such as SQCD\@.
For the computation of instanton calculus in the $\Omega$-background for the theory living on stacks of intersecting D3-branes see~\cite{Nekrasov:2016qym}.

We proposed that the partition function of our intersecting defects in gauge theories computes the correlation function in Toda CFT in the presence of a degenerate vertex operator. In this dictionary, the expansion of the CFT correlator in conformal blocks  is  obtained after integrating  over the partition function of the two-dimensional and zero-dimensional fields. This is a rather non-trivial prediction that stems from our analysis.

\begin{figure}\centering
  \includegraphics{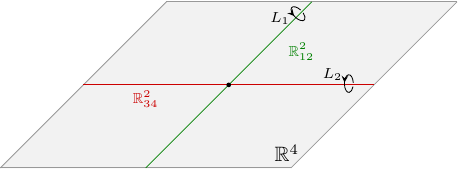}
  \caption{\label{fig:LeviIntersectingCartoon} Intersecting Levi-type defects supported on planes $\mathbb{R}^2_{12}$ and $\mathbb{R}^2_{34}$. The gauge group $G$ is broken to $L_1$ in the plane $\mathbb{R}^2_{12}$ and to $L_2$ in the plane $\mathbb{R}^2_{34}$. }
\end{figure}

\begin{figure}\centering
\includegraphics{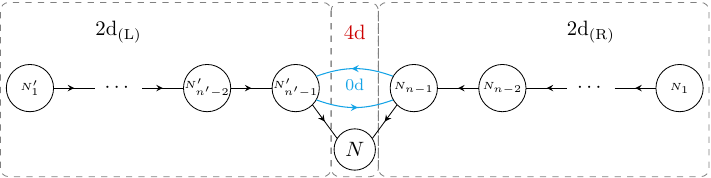}
\caption{\label{fig:LeviIntersecting} An example of a 4d/2d/0d quiver gauge description of intersecting Levi-type surface defects inserted in pure $\cN=2$ $SU(N)$ super-Yang--Mills. The Levi subgroup $L_1$ is determined by a non-decreasing partition of $N$, \ie{}, $N = K_1 + K_2 + \ldots K_n$ and $K_i \leq K_{i+1}$. The ranks of the gauge groups are then $N_j = \sum_{i=1}^j K_i$. The ranks $N^\prime_j$ are similarly determined in terms of the data encoded in $L_2$. Other choices for the 0d $\cN=(0,2)$ theory are possible.}
\end{figure}

Our  discussion of intersecting defects can be applied to surface operators of Levi-type, where the four-dimensional gauge group $G$ is broken at a surface to a Levi subgroup $L$ of $G$ \cite{Gukov:2006jk}. These are naturally associated to surface operators engineered by M5-branes instead of M2-branes~\cite{Alday:2010vg}. Our 4d/2d/0d field theory construction allows a more general possibility. We can consider a four-dimensional theory where the gauge group $G$ is broken to $L_1$ in the plane $x^3=x^4=0$ and to $L_2$ in the plane $x^1=x^2=0$, see \autoref{fig:LeviIntersectingCartoon}.
These two singularities are then glued at the origin, in a way determined by the zero-dimensional fields supported there. An interesting example to consider using our formalism is an intersecting surface defect in four-dimensional $\cN=2$ $SU(N)$ super-Yang--Mills characterized by a pair of Levi groups $(L_1,L_2)$. Using that one can associate to each Levi group a canonical two-dimensional $\cN=(2,2)$ theory (see \eg{}, \cite{Gukov:2006jk,Gadde:2013ftv,Gaiotto:2013sma}), we can consider as an example of such an intersecting defect the quiver diagram in \autoref{fig:LeviIntersecting}.

It is expected that for the choice of Levi groups $(L_1,G)$ obtained by coupling just one two-dimensional $\cN=(2,2)$ theory the partition function of the theory computes a correlation function in $W_\rho$ Toda CFT, where $\rho$ is the partition of $N$ associated to the Levi group $L_1$ \cite{Braverman:2004vv} (see also, \eg{}, \cite{Alday:2010vg,Kozcaz:2010yp,Wyllard:2010rp,Wyllard:2010vi,Tachikawa:2011dz,Kanno:2011fw,Belavin:2012qh,Tan:2013tq,Nawata:2014nca,Bullimore:2014upa,Frenkel:2015rda}). It would be interesting to find a two-dimensional CFT interpretation of the partition function of intersecting surface defects with Levi groups $(L_1,L_2)$ obtained by coupling, as we did in this paper, two two-dimensional $\cN=(2,2)$ and a zero-dimensional $\cN=(0,2)$ theory to each other and to the bulk.

The discussion of intersecting surface defects inserted in four-dimensional quantum field theories can be straightforwardly generalized to codimension two defects in five-dimensional theories. Trivially uplifting all dimensions by one unit, we expect our results to be relevant for the study of the five-dimensional AGT correspondence \cite{Awata:2009ur,Awata:2010yy,Taki:2014fva} as well as for the work in \cite{Nieri:2013yra,Nieri:2013vba}.

  The vacuum expectation value of intersecting surface defects labeled by symmetric representations on the four-sphere (or $S^4\times S^1$ or $S^5$) can be obtained alternatively via a Higgsing procedure \cite{Gaiotto:2012xa,Nieri:2013vba,Gaiotto:2014ina,Gaiotto:2015una} or, equivalently, from a Higgs branch localization computation \cite{Pan:2015hza,Chen:2015fta,Pan:2014bwa}.\footnote{Note that supersymmetric intersecting codimension two defects do not occur in the (Higgs branch localization) computations on ($S^1$-fibrations of) lower-dimensional manifolds, see \cite{Fujitsuka:2013fga,Benini:2013yva,Peelaers:2014ima}.} This computation heavily relies on massaging the instanton partition function and agrees with our proposal in this paper. It will be presented elsewhere~\cite{Pan:2016fbl}.

\section*{Acknowledgements}
The authors would like to thank Benjamin Assel, Davide Gaiotto, Hee-Cheol Kim, Heeyeon Kim, Chan Youn Park, Daniel Park, Jian Qiu, Jacob Winding and Maxim Zabzine for helpful conversations and useful suggestions. W.P. is grateful to Perimeter Institute, the Galileo Galilei Institute and the Simons Center for Geometry and Physics during the 2016 Simons Summer Workshop for their kind hospitality. This research was supported in part by Perimeter Institute for Theoretical Physics. Research at Perimeter Institute is supported by the Government of Canada through Industry Canada and by the Province of Ontario through the Ministry of Research and Innovation. Y.P. is supported in part by Vetenskapsr{\aa}det under grant {\#}2014- 5517, by the STINT grant and by the grant ``Geometry and Physics'' from the Knut and Alice Wallenberg foundation. The work of W.P. is supported in part by the DOE grant DOE-SC0010008.

\appendix

\section{\label{appendix:Liouville-check}Liouville Fundamental Degenerate}

In \autoref{section:dictionary-fundamental} we wrote \eqref{Liouville-matching} relating the partition functions of two 4d/2d/0d quiver gauge theories with 2d gauge groups $U(1)$ and $U(1)$ and a Liouville four-point function with three generic vertex operators and one degenerate vertex operator of momentum $\alpha = -b \Omega_{\yngs(.4,1)} - b^{-1} \Omega_{\yngs(.4,1)} = -\frac{b}{2} - \frac{1}{2b} = -Q/2$.  In this appendix we first discuss the Liouville correlator then match it to a partition function involving a 0d Fermi multiplet then to one involving a 0d chiral multiplet, and conclude with a proof that the two partition functions are equal up to some factors (in Appendix~\ref{appendix:dual-quivers}).

\subsection{\label{appendix:Liouville-correlator}The Liouville Correlator}
Let us start by writing down the Liouville correlation function we aim at reproducing from the gauge theory point of view:
\begin{equation}\label{Liouvillecorrelator}
\langle \hat V_{-\frac{Q}{2}}(z,\bar z)\  \hat V_{\alpha_1}(0)\  \hat V_{\alpha_2}(1)\  \hat V_{\alpha_3}(\infty) \rangle\;.
\end{equation}
It involves one degenerate vertex operator $\hat V_{-\frac{Q}{2}}$ with Liouville momentum $-\frac{1}{2}(b+b^{-1})= -\frac{Q}{2}$, and three generic vertex operators $\hat V_{\alpha_i},i=1,2,3$. Here the hats indicate that we normalized the operators as follows
\begin{subequations}
  \label{Liouville-normalization}
\begin{gather}
  \hat V_{-\frac{Q}{2}} = N_{\text{deg.}}^{(-\frac{Q}{2})} V_{-\frac{Q}{2}}\,, \qquad \hat V_{\alpha_i} = N^{(\alpha_i)} V_{\alpha_i}\,, \qquad
  N_{\text{deg.}}^{(-\frac{Q}{2})} = \bigl(\pi \mu \gamma(b^2) b^{2-2b^2} \bigr)^{-\frac{Q}{2b}} \, , \\
  N^{(\alpha_i)} = \bigl(\pi \mu \gamma(b^2) b^{2-2b^2} \bigr)^{\frac{\alpha_i-Q/3}{b}} \Bigm/ \bigl(\Upsilon_b^\prime(0)^{\frac{1}{3}}\ \Upsilon_b(2\alpha_i)\bigr)\,,
\end{gather}
\end{subequations}
where $\mu$ is the cosmological constant, $\gamma(x)=\frac{\Gamma(x)}{\Gamma(1-x)}$ and $\Upsilon^\prime(0)$ is the derivative of the Upsilon function evaluated at zero. Recall that the conformal weight of a Liouville vertex operator $V_{\alpha}$ is given by $\Delta(\alpha) = \alpha (Q-\alpha)$.\footnote{We follow the standard notation in the literature, where the equal weights $h(\alpha)=\bar h (\alpha)$ of the scalar operators $V_\alpha$ are denoted by $\Delta(\alpha)$. Its conformal dimension is twice that.}

The three-point function of three primary fields $V_{\beta_j}$ with generic momenta $\beta_j$ is given by the DOZZ formula \cite{Dorn:1994xn,Zamolodchikov:1995aa,Teschner:1995yf}
\begin{equation}\label{Liouville3ptfunction}
C(\beta_1,\beta_2,\beta_3) =  \left[ \pi \mu \gamma(b^2) b^{2-2b^2} \right]^{(Q-\beta)/b}
\frac{\Upsilon^\prime(0) \Upsilon(2\beta_1)\Upsilon(2\beta_2)\Upsilon(2\beta_3)}{\Upsilon(\beta-Q)\Upsilon(\beta-2\beta_1)\Upsilon(\beta-2\beta_2)\Upsilon(\beta-2\beta_3)}\,,
\end{equation}
where $\beta = \beta_1+\beta_2+\beta_3$. Including our normalization \eqref{Liouville-normalization}, it becomes
\begin{equation}\label{Liouville3ptfunctionnorm}
\hat C(\beta_1,\beta_2,\beta_3) = \frac{1}{\Upsilon(\beta-Q)\prod_{i=1}^{3}\Upsilon(\beta-2\beta_i)}\,.
\end{equation}
Furthermore, the operator product expansion of a generic operator $\hat V_{\alpha_1}$ with the degenerate operator $\hat V_{-\frac{Q}{2}}$ is given by
\begin{equation}\label{LiouvilleOPEdegen}
\hat V_{-\frac{Q}{2}}(z,\bar z)\ \hat V_{\alpha_1}(0)\ \sim\ \sum_{s_1=\pm, s_2=\pm}  (z\bar z)^{\Delta(\alpha_{s_1s_2})-\Delta(\alpha_1)-\Delta(-Q/2)} \  \hat C_{\alpha_1,-\frac{Q}{2}}^{\alpha_{s_1s_2}} \  (\hat V_{\alpha_{s_1s_2}}(0) + \ldots)\,,
\end{equation}
where $\alpha_{s_1s_2} = \alpha_1 +\frac{s_1 b+s_2 b^{-1}}{2}$, and the $\ldots$ denotes Virasoro descendant fields. The structure constants $\hat C_{\alpha_1,-\frac{Q}{2}}^{\alpha_{s_1s_2}}$ are computed by
\begin{align}
\hat C_{\alpha_1,-\frac{Q}{2}}^{\alpha_{s_1s_2}} = \frac{N^{(\alpha_1)} N_{\text{deg.}}^{(-Q/2)}}{N^{(\alpha_{s_1s_2})}} C\left(\alpha_1,-Q/2,Q-\alpha_{s_1s_2}\right)^\prime\,.
\end{align}
Here the prime on the last factor indicates that one should take the residue of the single pole one finds when inserting the arguments in \eqref{Liouville3ptfunction}.\footnote{Such prescription can be argued for by analytically continuing the operator product expansion of three generic operators to the situation where one of the operators is degenerate \cite{Ponsot:2000mt,Nieri:2013vba}.}

Given the OPE in \eqref{LiouvilleOPEdegen}, we deduce that the correlator \eqref{Liouvillecorrelator} has an s-channel conformal block decomposition involving four intermediate channels with intermediate momenta $\alpha_{s_1s_2} = \alpha_1 +(s_1 b+s_2 b^{-1})/2$,
\begin{equation}\label{Liouvillecorrelatorconfblock}
\langle \hat V_{-\frac{Q}{2}}(z,\bar z)\  \hat V_{\alpha_1}(0)\  \hat V_{\alpha_2}(1)\  \hat V_{\alpha_3}(\infty) \rangle = \sum_{s_1=\pm, s_2=\pm} \ \hat C_{\alpha_2,\alpha_3,\alpha_{s_1s_2}}\  \hat C_{\alpha_1,-\frac{Q}{2}}^{\alpha_{s_1s_2}}\  |G_{\alpha_{s_1s_2}}(z)|^2\,.
\end{equation}
The conformal blocks are normalized in the usual way $G_{\alpha_{s_1s_2}}(z) =z^{\Delta(\alpha_{s_1s_2})-\Delta(\alpha_1)-\Delta(-Q/2)} (1+c_1 z + \ldots)$.
They have been computed in closed form in \cite{Fateev:2009me} by solving the fourth order differential equation associated with the degenerate puncture. Before presenting them, we introduce various notations, following \cite{Fateev:2009me}. We denote
\begin{equation}
  p_1 = b(\alpha-2\alpha_1-Q/2)\,, \qquad p_2 = b(\alpha-2\alpha_2-Q/2)\,, \qquad p_3 = b(3Q/2-\alpha)\,, \qquad p^\prime_i = b^{-2}p_i\,,
\end{equation}
with $\alpha=\alpha_1+\alpha_2+\alpha_3$, use the notation $p_{ij}=p_i+p_j, p_{ijk}=p_i+p_j+p_k$, and finally define
\begin{align}
&\mathcal F_1(y_1,y_2,y_3,z) = {}_{2}F_1(1+y_3,2+y_1+y_2+y_3,2+y_1+y_3,z)\,,\\
&\mathcal F_2(y_1,y_2,y_3,z) = z^{-1-y_1-y_3} {}_2F_1(1+y_2,-y_1,-y_1-y_3,z)\,,
\end{align}
in terms of the hypergeometric function $ {}_2F_1$. Then the conformal blocks $G_{\alpha_{s_1s_2}}$ describing the exchange of momentum $\alpha_{s_1s_2}$ in the correlator \eqref{Liouvillecorrelatorconfblock} are given by\footnote{A minor typo in~\cite{Fateev:2009me} is that the exponent of $(1-z)$ reads $1+p_{13}+\alpha_1Q$, which is incompatible with the known behavior of the Liouville correlator as $z\to 1$.}
\begin{align}
& \qquad G_{\alpha_{s_1s_2}}(z) = z^{1+p_{13}+\alpha_1Q} (1-z)^{1+p_{23}+\alpha_2Q}\mathcal G_{\alpha_{s_1s_2}}(z)
\\
\mathcal G_{\alpha_{--}}(z) & = \mathcal F_{2}(p_1,p_2,p_3,z)\mathcal F_1(-p_1^\prime -1,-p_2^\prime, -p_3^\prime,z)\nonumber\\
& \qquad -\frac{(1-p^\prime_{123})p_1}{(1-p^\prime_{13})p_{13}}\mathcal F_{2}(p_1-1,p_2,p_3,z)\mathcal F_1(-p_1^\prime,-p_2^\prime, -p_3^\prime,z)
\\
\mathcal G_{\alpha_{++}}(z) & = \mathcal G_{\alpha_{--}}(z)\big|_{p_i \leftrightarrow - p^\prime_i}
\\
\mathcal G_{\alpha_{+-}}(z) & = \frac{(1+p_{123})(1-p^\prime_{13})}{p_2+p_2^\prime} \mathcal F_{1}(p_1,p_2,p_3,z)
\mathcal F_1(-p_1^\prime -1,-p_2^\prime, -p_3^\prime,z) + (p_i\leftrightarrow -p_i')
\\
\mathcal G_{\alpha_{-+}}(z) & =\frac{p_{13}(1-p_{13})(1+p^\prime_{13})}{p_3(p_{123}+p_{123}^\prime)}\Bigl(
\mathcal F_{2}(p_1-1,p_2,p_3,z)\mathcal F_2(-p_1^\prime ,-p_2^\prime, -p_3^\prime,z) - (p_i\leftrightarrow -p_i')\Bigr) \,.
\end{align}

\subsection{\label{appendix:gauge1}Gauge Theory Computation I}
The partition function of the 2d/0d part of the left quiver gauge theory in \autoref{fig:Liouville} is computed as
\begin{equation}\label{S2US2LiouvilleI}
\begin{aligned}
& Z_{S^2_{(R)}\cup S^2_{(L)}}^{(\yngs(.4,1),\yngs(.4,1))}
= \sum_{B^{(R)} \in \mathbb Z} \sum_{B^{(L)} \in \mathbb Z} \int \frac{\mathrm{d}\sigma^{(R)}}{2\pi} \frac{\mathrm{d}\sigma^{(L)}}{2\pi} \Biggl[
e^{-4\pi i \xi ( \sigma^{(R)} - \sigma^{(L)}) + i \vartheta (B^{(R)} - B^{(L)})} \Delta^+ \Delta^- \\
& \quad \times \prod_{j=1}^2 \frac{\Gamma\bigl(-i(\sigma^{(R)}-m_j^{(R)}) - B^{(R)}/2 \bigr)}{\Gamma\bigl(1+i(\sigma^{(R)}-m_j^{(R)}) - B^{(R)}/2 \bigr)} \prod_{s =1}^2 \frac{\Gamma\bigl(-i(-\sigma^{(R)}+\anti m_s^{(R)}) + B^{(R)}/2 \bigr)}{\Gamma\bigl(1+i(-\sigma^{(R)}+\anti m_s^{(R)}) + B^{(R)}/2 \bigr)}\\
& \quad \times \prod_{s=1}^2 \frac{\Gamma\bigl(-i(\sigma^{(L)}-m_s^{(L)}) - B^{(L)}/2 \bigr)}{\Gamma\bigl(1+i(\sigma^{(L)}-m_s^{(L)}) - B^{(L)}/2 \bigr)} \prod_{j =1}^2 \frac{\Gamma\bigl(-i(-\sigma^{(L)}+\anti m_j^{(L)}) + B^{(L)}/2 \bigr)}{\Gamma\bigl(1+i(-\sigma^{(L)}+\anti m_j^{(L)}) + B^{(L)}/2 \bigr)} \Biggr]
\end{aligned}
\end{equation}
with $\Delta^\pm =b^{-1}(i\sigma^{(R)} \pm B^{(R)}/2)- b(i\sigma^{(L)} \pm B^{(L)}/2)$. Recall the mass relations \eqref{mass-relation-fermi-from-superpotential} (with $ c = 0$)
\begin{equation}\label{paramidentificationS2US2I}
 b m_s^{(L)} = b^{-1}\anti m_s^{(R)}\,,\qquad \qquad b\anti m_j^{(L)} = b^{-1} m_j^{(R)} \,.
\end{equation}
In \eqref{S2US2LiouvilleI} we have   used $\xi_\text{FI}^{(R)}=-\xi_\text{FI}^{(L)}=\xi$ and similarly for $\vartheta$.  We also define $z=e^{-2\pi\xi+i\vartheta}$. For positive FI parameter, $\xi>0$, the naive poles are located at
\begin{equation}\label{naivepoles}
\begin{aligned}
i \sigma^{(R)} \pm B^{(R)}/2= i m_j^{(R)} + p_j^{\pm} \,, \ \text{with}\ p_j^{\pm}\geq 0 \ \text{for}\ j=1,2\,, \\
i \sigma^{(L)} \pm B^{(L)}/2= i \anti m_k^{(L)} - q_k^{\pm} \,, \ \text{with}\ q_k^{\pm}\geq 0 \ \text{for}\ k=1,2\,.
\end{aligned}
\end{equation}
Using the mass relations \eqref{paramidentificationS2US2I}, it is easy to see that the contribution of the Fermi multiplet provides zeros canceling the poles (or equivalently, setting their residue to zero) for $j=k$ and $p_j^+ = q_j^+ = 0$ or $p_j^- = q_j^- = 0$.

Introducing the quantities for $j,k=1,2$,
\begin{align}
Z_{\text{1-loop}}^{(R)}(j) &= \gamma\bigl(-i m_j^{(R)} + i m_{j^\prime\neq j}^{(R)}\bigr)\  \prod_{s=1,2} \gamma\bigl(i m_j^{(R)} - i \anti m_{s}^{(R)}\bigr)\,,\\
Z_{\text{1-loop}}^{(L)}(k) &= \gamma\bigl(i \anti m_k^{(L)} - i \anti m_{k^\prime \neq k}^{(L)}\bigr)\  \prod_{s=1,2} \gamma\bigl(-i \anti m_k^{(L)} + i  m_{s}^{(L)}\bigr)\,,
\end{align}
and for $\mathfrak{m}\in\mathbb{Z}_{\geq 0}$,
\begin{align}
Z_{\text{vortex}|(j)}^{\mathbb R^2, (R)}(\mathfrak m) &= \frac{\prod_{s=1}^2 \bigl( i m_j^{(R)} - i \anti m_{s}^{(R)}  \bigr)_\mathfrak{m}}{\mathfrak{m}! \bigl(1+im_j^{(R)}-im_{j^\prime\neq j}^{(R)}\bigr)_\mathfrak{m}}\,,\\
Z_{\text{vortex}|(k)}^{\mathbb R^2, (L)}(\mathfrak m) &= \frac{\prod_{s=1}^2 \bigl( -i \anti m_k^{(L)} + i  m_{s}^{(L)}  \bigr)_\mathfrak{m}}{\mathfrak{m}! \bigl(1-i\anti m_k^{(L)}+i\anti m_{k^\prime\neq k}^{(L)}\bigr)_\mathfrak{m}}\,,
\end{align}
one easily finds
\begin{align}
\label{hbexpressionS2US2LiouvilleI}
Z_{S^2_{(R)}\cup S^2_{(L)}}^{(\yngs(.4,1),\yngs(.4,1))} \!\!& =  \sum_{j,k=1}^2 e^{-4\pi i \xi \left(m_j^{(R)}-\anti m_k^{(L)}\right)}  Z_{\text{1-loop}}^{(R)}(j) Z_{\text{1-loop}}^{(L)}(k) \\
\nonumber
&\quad \times \!\! \sum_{p_j^{+},q_k^{+}\geq 0} \Bigl[\left( b^{-1}(i m_j^{(R)} + p_j^+) - b(i \anti m_k^{(L)} -q_k^+) \right) 
  z^{p_j^{+}} Z_{\text{vortex}|(j)}^{\mathbb R^2, (R)}(p_j^{+}) \ z^{q_k^{+}} Z_{\text{vortex}|(k)}^{\mathbb R^2, (L)}(q_k^{+})\Bigr] \\
\nonumber
&\quad \times \!\! \sum_{p_j^{-},q_k^-\geq 0} \Bigl[\left( b^{-1}(i m_j^{(R)} + p_j^-) - b(i \anti m_k^{(L)} -q_k^-) \right)
  \bar z^{p_j^{-}} Z_{\text{vortex}|(j)}^{\mathbb R^2, (R)}(p_j^{-}) \ \bar z^{q_k^{-}} Z_{\text{vortex}|(k)}^{\mathbb R^2, (L)}(q_k^{-})\Bigr]\,.
\end{align}
Next, we match this expression to the Liouville correlator after including
the contribution of the four free four-dimensional hypermultiplets,
\begin{equation}\label{PFoffreeHMI}
Z_{S^4_b}^{\text{free HM}} = \prod_{j,s=1}^2 \frac{1}{\Upsilon_b(\frac{b}{2}+\frac{1}{2b} -i M_{js})}\,,
\end{equation}
with masses fixed by \eqref{massrelaone}--\eqref{massrelatwo},
\begin{equation}
\begin{aligned}
&\Bigl[\frac{M_{js}}{\sqrt{\ell\tilde \ell}}+\frac{i}{2\ell}+ \frac{i}{2\tilde \ell}  \Bigr]+\frac{-m^{(R)}_j+\anti m^{(R)}_s}{\ell}=\frac{i}{\ell}\\
&\Bigl[-\frac{M_{js}}{\sqrt{\ell\tilde \ell}}+\frac{i}{2\ell}+ \frac{i}{2\tilde \ell}  \Bigr]+\frac{-m^{(L)}_s+\anti m^{(L)}_j}{\tilde \ell}=\frac{i}{\tilde \ell}\,.
\end{aligned}
\end{equation}

\subsection{\label{appendix:matching1}Matching Liouville to Gauge Theory I}
The detailed match between the Liouville correlator and the gauge theory computation of the previous section is given by\footnote{As explained in~\cite{Gomis:2014eya} the prefactor on the right-hand side can be associated to ambiguities in the definition of the gauge theory partition function.}
\begin{equation}\label{matchI}
Z_{S^4_b}^{\text{free HM}}\  Z_{S^2_{(R)}\cup S^2_{(L)}}^{(\yngs(.4,1),\yngs(.4,1))} = A \ |z|^{2\beta}\  |1-z|^{2\gamma}  \langle \hat V_{-\frac{Q}{2}}(z,\bar z)\  \hat V_{\alpha_1}(0)\  \hat V_{\alpha_2}(1)\  \hat V_{\alpha_3}(\infty) \rangle\,,
\end{equation}
where
\begin{align}\label{defAI}
&A = b^{-4Q(\alpha_2-Q/2)} \,, \\
&\gamma =  (b - b^{-1})\alpha_2-bQ \,, \\
&\beta = -\frac{Q^2}{2} + \frac{i}{2b}(b-b^{-1})(m_1^{(R)}+m_2^{(R)})\,, \label{defbetaI}
\end{align}
and with the parameters $\alpha_i$ identified with two-dimensional masses as
\begin{align}
\alpha_1 - \frac{Q}{2} & = \frac{i}{2b}(m^{(R)}_1-m^{(R)}_2) = \frac{ib}{2}(\anti m^{(L)}_1-\anti m^{(L)}_2)\,, \nonumber\displaybreak[1]\\
\alpha_2 - \frac{Q}{2} & = \frac{i}{2b}(\anti m^{(R)}_1 + \anti m^{(R)}_2 - m^{(R)}_1 - m^{(R)}_2) = \frac{ib}{2}( m^{(L)}_1 +  m^{(L)}_2 - \anti m^{(L)}_1 - \anti m^{(L)}_2)\,, \nonumber\displaybreak[1]\\
\alpha_3 - \frac{Q}{2} & = -\frac{i}{2b}(\anti m^{(R)}_1 - \anti m^{(R)}_2) = -\frac{ib}{2}( m^{(L)}_1 -  m^{(L)}_2)\,.
\label{LiouvilleparamidentificationsI}
\end{align}

More in detail, the sum over the four vacua $j,k=1,2$ of the gauge theory result \eqref{hbexpressionS2US2LiouvilleI} corresponds to the four internal channels of the Liouville correlator \eqref{Liouvillecorrelatorconfblock} as in \autoref{Tab:vacuacorrespondenceI}.
\begin{table}[t]\centering
\begin{tabular}{ccccc}
\toprule
$j,k$ & $1,1$ & $1,2$ & $2,1$ & $2,2$ \\
$s_1s_2$ & $-+$ & $--$ & $++$ & $+-$
\\\bottomrule
\end{tabular}
\caption{\label{Tab:vacuacorrespondenceI}Matching the four vacua in $Z_{S^2_{(R)}\cup S^2_{(L)}}^{(\protect\yngs(.4,1),\protect\yngs(.4,1))}$ with the four channels of the Liouville correlator.}
\end{table}
To present the precise identification, let us introduce the notation
\begin{equation}
Z_{\text{v}\otimes\text{v}}(x;j,k) = \sum_{p_j,q_k\geq 0} \Bigl[\Bigl( b^{-1}(i m_j^{(R)} + p_j) - b(i \anti m_k^{(L)} -q_k) \Bigr) x^{p_j} Z_{\text{vortex}|(j)}^{\mathbb R^2, (R)}(p_j) \ x^{q_k} Z_{\text{vortex}|(k)}^{\mathbb R^2, (L)}(q_k)\Bigr]\,,
\end{equation}
where we note that $Z_{\text{v}\otimes\text{v}}(x;j,j)$ has vanishing zeroth order term in $x$:
\begin{equation}
Z_{\text{v}\otimes\text{v}}(x;j,j)
= x\left( b Z_{\text{vortex}|(j)}^{\mathbb R^2, (L)}(1) + b^{-1}Z_{\text{vortex}|(j)}^{\mathbb R^2, (R)}(1)\right) + \mathcal O(x^2)\,.
\end{equation}
The gauge theory result \eqref{hbexpressionS2US2LiouvilleI} can then be reorganized in the following form
\begin{equation}\label{hbexpressionS2US2LiouvilletomatchI}
  \begin{aligned}
    Z_{S^2_{(R)}\cup S^2_{(L)}}^{(\yngs(.4,1),\yngs(.4,1))}
    & = \sum_{\substack{j,k=1 \\j\neq k}}^2 \Biggl[Z_{\text{1-loop}}^{(R)}(j) Z_{\text{1-loop}}^{(L)}(k)\Bigl( i b^{-1} m_j^{(R)}  - i b \anti m_k^{(L)}  \Bigr)^{2} \biggl|\frac{z^{i(m_j^{(R)} - \anti m_k^{(L)})}}{ i b^{-1} m_j^{(R)}  - i b \anti m_k^{(L)}}  Z_{\text{v}\otimes\text{v}}(z;j,k)\biggr|^2 \Biggr]
    \\
    & \quad +\sum_{\substack{j,k=1 \\j= k}}^2 \Biggl[Z_{\text{1-loop}}^{(R)}(j) Z_{\text{1-loop}}^{(L)}(k) \Bigl( b Z_{\text{vortex}|(j)}^{\mathbb R^2, (L)}(1) + b^{-1}Z_{\text{vortex}|(j)}^{\mathbb R^2, (R)}(1)\Bigr)^2
    \\[-3ex]
    & \qquad \qquad \qquad \qquad \qquad \qquad \qquad \qquad
    \times \biggl|\frac{z^{i(m_j^{(R)} -\anti m_k^{(L)})+1} z^{-1} Z_{\text{v}\otimes\text{v}}(z;j,k)}{b Z_{\text{vortex}|(j)}^{\mathbb R^2, (L)}(1) + b^{-1} Z_{\text{vortex}|(j)}^{\mathbb R^2, (R)}(1)}\biggr|^2 \Biggr]\,,
  \end{aligned}
\end{equation}
where $|\ldots|^2$ just means sending $z\rightarrow \bar z$. Each of the four summands of \eqref{hbexpressionS2US2LiouvilletomatchI} have the structure $[\ldots]\times |\ldots|^2$. These expressions, using \autoref{Tab:vacuacorrespondenceI}, can be matched to the four channels of the Liouville four-point function \eqref{Liouvillecorrelatorconfblock} as
\begin{equation}
  Z_{S^4_b}^{\text{free HM}} \times \big[ \ldots \big] = A\  \hat C \hat C\,, \qquad \text{and} \qquad \big|\ldots\big|^2 = |z|^{2\beta}\  |1-z|^{2\gamma}\  |G(z)|^2\,,
\end{equation}
where we used the parameters in \eqref{defAI}--\eqref{defbetaI}.  In the ancillary Mathematica file,\footnote{See the ancillary files available on arXiv for a Mathematica file checking that the conformal blocks are indeed equal.} we use contiguous relations on hypergeometric functions to prove the equality for conformal blocks.  It would be interesting to obtain a more straightforward proof.

\subsection{\label{appendix:gauge2}Gauge Theory Computation II}
Let us now compute the $S^4_b$ partition function of the theory described by the right quiver gauge theory in \autoref{fig:Liouville}. We denote parameters of the left theory with primes and the right theory without primes.  Omitting the 4d hypermultiplets, the partition function is
\begin{equation}\label{S2US2LiouvilleII}
\begin{aligned}
\widetilde Z_{S^2_{(R)}\cup S^2_{(L)}}^{(\yngs(.4,1),\yngs(.4,1))} & = \sum_{B \in \mathbb Z} \sum_{B' \in \mathbb Z} \int_{\text{JK}} \frac{\mathrm{d}\sigma}{2\pi} \frac{\mathrm{d}\sigma'}{2\pi} e^{-4\pi i \xi ( \sigma + \sigma') + i \vartheta (B + B')} \\
& \quad \times \prod_{j=1}^2 \frac{\Gamma\left(-i(\sigma-m_j) - B/2 \right)}{\Gamma\left(1+i(\sigma-m_j) - B/2 \right)} \ \prod_{s =1}^2 \frac{\Gamma\left(-i(-\sigma+\anti m_s) + B/2 \right)}{\Gamma\left(1+i(-\sigma+\anti m_s) + B/2 \right)}\\
& \quad \times \left[ \prod_{\pm} \left(\Delta^{\pm} + \frac{b+b^{-1}}{2}\right)\left(\Delta^{\pm} - \frac{b+b^{-1}}{2}\right) \right]^{-1} \\
& \quad \times \prod_{j=1}^2 \frac{\Gamma\left(-i(\sigma'-m_j') - B'/2 \right)}{\Gamma\left(1+i(\sigma'-m_j') - B'/2 \right)} \ \prod_{s =1}^2 \frac{\Gamma\left(-i(-\sigma'+\anti m_s') + B'/2 \right)}{\Gamma\left(1+i(-\sigma'+\anti m_s') + B'/2 \right)} \,,
\end{aligned}
\end{equation}
with $\Delta^\pm =b^{-1}(i\sigma \pm \frac{B}{2})- b (i\sigma' \pm \frac{B'}{2})$. Recall from \autoref{section:intersecting-partition-function-first-principle} the mass relations (with $c=0$)
\begin{equation}\label{paramidentificationS2US2II}
  b m_j' - b^{-1}m_j = -\frac{i}{2}(b-b^{-1})\,, \qquad
  b\anti m_s' - b^{-1}\anti m_s = \frac{i}{2}(b-b^{-1})\,.
\end{equation}
In \eqref{S2US2LiouvilleII} we have used $\xi_\text{FI}=\xi_\text{FI}'=\xi$ and similarly for $\vartheta$. We also define $z=e^{-2\pi\xi+i\vartheta}$.  For positive FI parameter, $\xi>0$, the Jeffrey--Kirwan-like residue prescription selects poles obtained by assigning to $\sigma'$ a pole position of the fundamental one-loop determinants in the third line of \eqref{S2US2LiouvilleII}. Taking into account cancellations with zeros, we thus have
\begin{equation}\label{oldpolesphere2II}
i \sigma' \pm \frac{B'}{2}= i m_k' + q_k^{\pm} \,, \ \text{with}\ q_k^{\pm}\geq 0 \ \text{for}\ k=1,2\,.
\end{equation}
For $\sigma$ there are various options
\begin{equation}
\begin{aligned}
i \sigma = &i m_j + l_j - \frac{B}{2} \,, \ \text{with}\ l_j\geq 0 \ \text{for}\ j=1,2\,,\\
i \sigma = &i m_k \pm \frac{B}{2}+ b^2 q_k^{\mp}-1\,, \\
i \sigma = &i m_k \pm \frac{B}{2}+ b^2 (q_k^{\mp}+1)\,.
\end{aligned}
\end{equation}
Here we used the relations among the fundamental mass parameters on the two spheres. Note that in the pole positions on the last two lines, the index $k$ takes the same value as in \eqref{oldpolesphere2II}. Also note that some of these poles collide, and some cancel against the zeros located at $i\sigma = i m_p + \frac{B}{2} - \lambda_p -1$ with $\lambda_p\geq 0$ for $p=1,2$. Among these poles, four particular classes of simple poles can be identified as follows, where $j,k=1,2$:
\begin{align}
\text{I}&:\begin{cases}
i \sigma' \pm \frac{B'}{2}= i m_k' + q_k^{\pm} \,, \ &\text{with}\ q_k^{\pm}\geq 0\,,  \\
i \sigma \pm \frac{B}{2}= i m_j + p_j^{\pm} \,, \ &\text{with}\ p_j^{\pm}\geq 0\,,
\end{cases}\nonumber \\
\text{II}&:\begin{cases}
i \sigma' \pm \frac{B'}{2}= i m_k' + q_k^{\pm} \,, \ &\text{with}\ q_k^+=0, q_k^{-}\geq 0\,,\\
i \sigma+ \frac{B}{2} = i m_k -1 \,, \ &\text{with}\ B<0\,,
\end{cases}\nonumber\\
\text{III}&:\begin{cases}
i \sigma' \pm \frac{B'}{2}= i m_k' + q_k^{\pm} \,, \ &\text{with}\ q_k^-=0, q_k^{+}\geq 0\,,\\
i \sigma- \frac{B}{2} = i m_k -1 \,, \ &\text{with}\ B>0\,,
\end{cases}\nonumber\\
\text{IV}&:\begin{cases}
i \sigma' = i m_k' \,, \ &\text{with}\ q_k^-=0, q_k^{+}= 0 \Longrightarrow B'=0\,,\\
i \sigma = i m_k -1 \,, \ &\text{with}\ B= 0\,.
\end{cases}
\label{JKlikepoles}
\end{align}
The sum of the residues of these poles will reproduce the Liouville correlator \eqref{Liouvillecorrelator}, while one can verify that all other series of poles cancel among themselves.  These poles can also be characterized as those for which $i\sigma\pm B/2\in\{im_1,im_2\}+\mathbb{Z}$ and $i\sigma'\pm B'/2\in\{im_1',im_2'\}+\mathbb{Z}$.

Computing the residues of the four classes of poles is straightforward.  One finds
\begin{multline}
\widetilde Z_{S^2_{(R)}\cup S^2_{(L)}}^{(\yngs(.4,1),\yngs(.4,1))}=  \sum_{j,k=1}^{2} \Biggl[ e^{-4\pi i \xi (m_j+m_k')}
\widetilde Z_{\text{1-loop}|(j)}(m, \anti m) \widetilde Z_{\text{1-loop}|(k)}(m', \anti m')
\\
\times \Biggl|\delta_{jk} z^{-1} \frac{\widetilde Z_{\text{vortex}|(j)}^{\mathbb R^2}(-1)}{(b+b^{-1}) b^{-1}}
+ \sum_{p,q\geq 0} \frac{z^p\widetilde Z_{\text{vortex}|(j)}^{\mathbb R^2}(p;m, \anti m) \ z^q\widetilde Z_{\text{vortex}|(k)}^{\mathbb R^2}(q; m', \anti m')}{\prod_{\pm}\left( b^{-1}(i m_j + p) - b(i m_k' + q) \pm \frac{b+b^{-1}}{2} \right)}\Biggr|^2 \Biggr]
\label{hbexpressionS2US2LiouvilleII}
\end{multline}
where $|\cdots|^2$ just involves $z\to\bar{z}$, in terms of
\begin{gather}
\widetilde Z_{\text{1-loop}|(j)}(m, \anti m) = \gamma\left(-i m_j + i m_{k\neq j}\right) \prod_{s=1}^{2} \gamma\left(i m_j - i \anti m_{s}\right)\,,\\
\widetilde Z_{\text{vortex}|(j)}^{\mathbb R^2}(\mathfrak m; m, \anti m) = \frac{\prod_{s=1}^2 \left( i m_j - i \anti m_{s}  \right)_\mathfrak{m}}{\mathfrak{m}! \left(1+im_j-im_{k\neq j}\right)_\mathfrak{m}}\,,\\
\widetilde Z_{\text{vortex}|(j)}^{\mathbb R^2}(-1) =\frac{(-im_j + i m_{k\neq j})}{\prod_{s=1}^2(im_j-i\anti m_{s}-1)}
\end{gather}
for $j=1,2$ and $\mathfrak m \in \mathbb Z_{\geq 0}$.

Next, we match this expression to the Liouville correlator, including the contribution of the four free four-dimensional hypermultiplets,
\begin{equation}\label{PFoffreeHMII}
\widetilde Z_{S^4_b}^{\text{free HM}} = \prod_{j,s=1}^2 \frac{1}{\Upsilon_b(\frac{b}{2}+\frac{1}{2b} -i M_{js})}\,,
\end{equation}
with masses fixed by \eqref{massrelaonesymm}--\eqref{massrelatwosymm}
\begin{equation}
\begin{aligned}
&\Bigl[\frac{M_{js}}{\sqrt{\ell\tilde \ell}}+\frac{i}{2\ell}+ \frac{i}{2\tilde \ell}  \Bigr]+\frac{-m_j+\anti m_s}{\ell}=\frac{i}{\ell}\\
&\Bigl[\frac{M_{js}}{\sqrt{\ell\tilde \ell}}+\frac{i}{2\ell}+ \frac{i}{2\tilde \ell}  \Bigr]+\frac{-m'_j+\anti m'_s}{\tilde \ell}=\frac{i}{\tilde \ell}\,.
\end{aligned}
\end{equation}

\subsection{\label{appendix:matching2}Matching Liouville to Gauge Theory II}
The precise match between the Liouville correlation function \eqref{Liouvillecorrelator} and the gauge theory quantities \eqref{hbexpressionS2US2LiouvilleII}--\eqref{PFoffreeHMII} is given as follows
\begin{equation}\label{matchII}
\widetilde Z_{S^4_b}^{\text{free HM}}\  \widetilde Z_{S^2_{(R)}\cup S^2_{(L)}}^{(\yngs(.4,1),\yngs(.4,1))} = \widetilde A \ |z|^{2\widetilde\beta}\  |1-z|^{2\widetilde\gamma}\  \langle \hat V_{-\frac{Q}{2}}(z,\bar z)\  \hat V_{\alpha_1}(0)\  \hat V_{\alpha_2}(1)\  \hat V_{\alpha_3}(\infty) \rangle\,,
\end{equation}
with the parameters $\alpha_i$ identified with two-dimensional masses as
\begin{align}\label{LiouvilleparamidentificationsII}
\alpha_1 - \frac{Q}{2} & = \frac{i}{2b}(m_1-m_2) = \frac{ib}{2}(m'_1-m'_2) \,, \nonumber\displaybreak[1]\\
\alpha_2 - \frac{Q}{2} & = \frac{1}{b} + \frac{i}{2b}(\anti m_1 + \anti m_2 - m_1 - m_2) = b + \frac{ib}{2}(\anti m'_1 + \anti m'_2 - m'_1 - m'_2) \,,\nonumber\displaybreak[1]\\
\alpha_3 - \frac{Q}{2} & = \frac{i}{2b}(\anti m_2 - \anti m_1) = \frac{ib}{2} (\anti{m}'_2 - \anti{m}'_1)\,,
\end{align}
where the two expressions of each momentum are related by \eqref{paramidentificationS2US2II}, and
\begin{align}
\widetilde A & = \frac{1}{Q^2}\  b^{- 4(b - b^{-1})(\alpha_2-Q/2) } \,,\label{defAII}\\
\widetilde \gamma & = Q\alpha_2-Q^2 \,, \label{defgammaII}\\
\widetilde \beta & = -\frac{Q}{2}\left(Q + \frac{1 - im_1 - im_2}{b}\right)\,. \label{defbetaII}
\end{align}

First of all, \eqref{hbexpressionS2US2LiouvilleII} contains a sum over four terms specified by the values of $j,k$. These choices of vacua $j=1,2$ and $k=1,2$ correspond to the four channels of \eqref{Liouvillecorrelatorconfblock} as in \autoref{Tab:vacuacorrespondence}.
\begin{table}[t]\centering
\begin{tabular}{ccccc}\toprule
$j,k$ & $1,1$ & $1,2$ & $2,1$ & $2,2$ \\
$s_1s_2$ & $--$ & $-+$ & $+-$ & $++$ \\ \bottomrule
\end{tabular}
\caption{\label{Tab:vacuacorrespondence}Matching the four vacua in $\widetilde Z_{S^2_{(R)}\cup S^2_{(L)}}^{(\protect\yngs(.4,1),\protect\yngs(.4,1))} $ with the four channels of the Liouville correlator.}
\end{table}
To simplify the details of the identification \eqref{matchII}, let us introduce a concise notation for the double sum over positive integers appearing in \eqref{hbexpressionS2US2LiouvilleII}
\begin{equation}
\widetilde Z_{\text{v}\otimes\text{v}}(x;j,k)=\sum_{p,q\geq 0} \frac{x^p \widetilde Z_{\text{vortex}|(j)}^{\mathbb R^2}(p; m, \anti m) \ x^q \widetilde Z_{\text{vortex}|(k)}^{\mathbb R^2}(q; m', \anti m')}{\prod_{\pm}\left( b^{-1}(i m_j + p) - b(i m_k' + q) \pm \frac{b+b^{-1}}{2} \right)}\,.
\end{equation}
We then rewrite \eqref{hbexpressionS2US2LiouvilleII} as
\begin{equation}\label{hbexpressionS2US2LiouvilletomatchII}
\begin{aligned}
\widetilde Z_{S^2_{(R)}\cup S^2_{(L)}}^{(\yngs(.4,1),\yngs(.4,1))} & = \sum_{\substack{j,k=1 \\j\neq k}}^2 \Biggl[\frac{\widetilde Z_{\text{1-loop}|(j)}(m, \anti m) \widetilde Z_{\text{1-loop}|(k)}(m', \anti m')}{\prod_{\pm}( i b^{-1} m_j  - i b m_k' \pm \frac{b+b^{-1}}{2} )^{2}} \\
& \qquad \qquad \times \biggl|z^{i(m_j + m_k')}\prod_{\pm}\Bigl( i b^{-1} m_j - i b m_k' \pm \frac{b+b^{-1}}{2} \Bigr)\widetilde Z_{\text{v}\otimes\text{v}}(z;j,k)\biggr|^2\Biggr]\\
& \quad +\sum_{\substack{j,k=1 \\j= k}}^2 \Biggl[\widetilde Z_{\text{1-loop}|(j)}(m, \anti m) \widetilde Z_{\text{1-loop}|(k)}(m', \anti m') \biggl(\frac{\widetilde Z_{\text{vortex}|(j)}^{\mathbb R^2}(-1)}{(b+b^{-1})b^{-1}}\biggr)^2 \\
& \qquad \qquad \times \biggl|z^{i(m_j + m_k^{\prime})-1}\biggl(1  + \frac{(b+b^{-1}) b^{-1}}{\widetilde Z_{\text{vortex}|(j)}^{\mathbb R^2, (R)}(-1)} z \widetilde Z_{\text{v}\otimes\text{v}}(z;j,k)\biggr)\biggr|^2\Biggr]\,,
\end{aligned}
\end{equation}
where $|\ldots|^2$ is again taken to mean to just replace $z\rightarrow \bar z$. The identification \eqref{matchII}, using \eqref{Liouvillecorrelatorconfblock}, is now straightforward. Each of the four terms obtained by summing over $j,k$, which as mentioned above are identified with the four channels of the Liouville correlator as in \autoref{Tab:vacuacorrespondence}, have the structure $[\ldots]\times |\ldots|^2$. These factors are identified concretely for each vacuum as:
\begin{equation}
\widetilde Z_{S^4_b}^{\text{free HM}} \times \big[ \ldots \big] = A\  \hat C \hat C\,, \qquad \text{and} \qquad \big|\ldots\big|^2 = |z|^{2\widetilde\beta}\  |1-z|^{2\widetilde\gamma}\  |G(z)|^2\,,
\end{equation}
where we used the parameters in \eqref{defAII}--\eqref{defbetaII} and the arguments of the brackets and moduli squared can be read off from \eqref{hbexpressionS2US2LiouvilletomatchII}.  This identification is a consequence of the identification in Appendix~\ref{appendix:matching1} and the equality of partition functions that we prove next.

\subsection{\label{appendix:dual-quivers}Seiberg Duality Between Quivers}

We prove here that the two quivers studied in previous sections have equal partition functions: we apply a 2d $\cN=(2,2)$ Seiberg-like duality to the left node of the first quiver and show how the 0d Fermi multiplet contribution transforms into a pair of 0d chiral multiplets.

Enrich the partition function~\eqref{S2US2LiouvilleI} by allowing independent left and right FI parameters, and write the 0d Fermi multiplet contribution as a differential operator:
\begin{equation}\label{S2US2LiouvilleIasdiffop}
  Z_{S^2_{(R)}\cup S^2_{(L)}}^{(\yngs(.4,1),\yngs(.4,1))} =
  \bigl|b^{-1} z^{(R)}\partial_{z^{(R)}} - b z^{(L)}\partial_{z^{(L)}}\bigr|^2
  Z^{(R)}_{\text{SQED}} Z^{(L)}_{\text{SQED}}
  \Bigr|_z
\end{equation}
where $|_z$ denotes taking $z^{(R)},z^{(L)}\to z$, and
\begin{multline}
  Z^{(L)}_{\text{SQED}} =
  \sum_{B^{(L)} \in \mathbb Z} \int \frac{\mathrm{d}\sigma^{(L)}}{2\pi} (z^{(L)})^{i\sigma^{(L)}+B^{(L)}/2} \bigl(\bar{z}^{(L)}\bigr)^{i\sigma^{(L)}-B^{(L)}/2} \\
  \times \prod_{j=1}^2 \frac{\Gamma\bigl(-i(\sigma^{(L)}-m_j^{(L)}) - \frac{B^{(L)}}{2} \bigr)}{\Gamma\bigl(1+i(\sigma^{(L)}-m_j^{(L)}) - \frac{B^{(L)}}{2} \bigr)}
  \prod_{s =1}^2 \frac{\Gamma\bigl(-i(-\sigma^{(L)}+\anti m_s^{(L)}) + \frac{B^{(L)}}{2} \bigr)}{\Gamma\bigl(1+i(-\sigma^{(L)}+\anti m_s^{(L)}) + \frac{B^{(L)}}{2} \bigr)}
\end{multline}
and similarly $Z^{(R)}_{\text{SQED}}$ in terms of $(m^{(R)},\anti{m}^{(R)},z^{(R)},\bar{z}^{(R)})$.

As shown in~\cite{Benini:2014mia,Gomis:2014eya}, the SQED partition function is invariant under Seiberg duality up to some factors,\footnote{Incidentally, this equality is a Weyl symmetry of the Liouville correlator corresponding to~$Z_{\text{SQED}}$.}
\begin{equation}
  Z_{\text{SQED}}^{(L)}
  =
  C \abs{z^{(L)}}^{2\delta_0} \abs{1-z^{(L)}}^{2\delta_1}
  Z_{\text{SQED}}(m',\anti{m}',z',\bar{z}') \,,
\end{equation}
with exponents $\delta_0=-1/2+i m^{(L)}_1+i m^{(L)}_2$ and $\delta_1=1+i\anti{m}^{(L)}_1+i\anti{m}^{(L)}_2-i m^{(L)}_1-i m^{(L)}_2$, coefficient $C = \prod_{j=1}^{2} \prod_{s=1}^{2} \gamma\bigl(i m^{(L)}_s-i\anti{m}^{(L)}_j\bigr)$, exponentiated FI parameter $z'=1/z^{(L)}$ and shifted twisted masses $m'_j = \anti{m}^{(L)}_j - \frac{i}{2}$ and $\anti{m}'_s = m^{(L)}_s + \frac{i}{2}$.
These parameters will turn out to be those of the left theory in the second quiver, as given in \eqref{Liouville-mrel1} and~\eqref{Liouville-mrel2} in the main text (the relation can also be seen by identifying Liouville and gauge theory data).
Parameters of the right theories are related as $z=z^{(R)}$ and $m_j = m^{(R)}_j - \frac{i}{2}$ and $\anti{m}_s = \anti{m}^{(R)}_s + \frac{i}{2}$.

Next we permute $\abs{z^{(L)}}^{2\delta_0} \abs{1-z^{(L)}}^{2\delta_1}$ and the differential operator of~\eqref{S2US2LiouvilleIasdiffop}:
\begin{multline}
  \bigl(b^{-1} z^{(R)}\partial_{z^{(R)}} - b z^{(L)}\partial_{z^{(L)}}\bigr)
  (z^{(L)})^{\delta_0}(1-z^{(L)})^{\delta_1}
  \\
  = (z')^{\frac{1}{2}-\delta_0-\delta_1}(z'-1)^{\delta_1-1}
  \Bigl( \bigl(b^{-1} z\partial_z + b z'\partial_{z'} -bi m^{(L)}_1-bi m^{(L)}_2 \bigr)(z')^{\frac{1}{2}}
  \\
  -\bigl(b^{-1} z\partial_z + b z'\partial_{z'}-bi\anti{m}^{(L)}_1-bi\anti{m}^{(L)}_2 \bigr)(z')^{-\frac{1}{2}} \Bigr) \,.
\end{multline}
When combined with its complex conjugate, this gives four terms
\begin{multline}
  Z_{S^2_{(R)}\cup S^2_{(L)}}^{(\yngs(.4,1),\yngs(.4,1))} =
  C \abs{z'}^{2(\frac{1}{2}-\delta_0-\delta_1)} \abs{z'-1}^{2(\delta_1-1)} \!\!
  \sum_{s_+,s_-=\pm 1}\Bigl[
  \\
  \bigl(b^{-1} z\partial_{z} + b z'\partial_{z'} -bi m^{(L),s_+} \bigr)
  \bigl(b^{-1} \bar{z}\partial_{\bar{z}} + b\bar{z}'\partial_{\bar{z}'} -bi m^{(L),s_-} \bigr)
  \\
  s_+ s_- (z')^{s_+/2} (\bar{z}')^{s_-/2}
  Z_{\text{SQED}}(m',\anti{m}',z',\bar{z}')
  Z_{\text{SQED}}^{(R)}\Bigr]_{z=z'}
\end{multline}
where $i m^{(L),-}=i\anti{m}^{(L)}_1+i\anti{m}^{(L)}_2$ and $i m^{(L),+}=i m^{(L)}_1+i m^{(L)}_2$.  The factors $(z')^{s_+/2}$ and $(\bar{z}')^{s_-/2}$ can be absorbed into the Coulomb branch expression of~$Z_{\text{SQED}}'$ by shifting $\sigma'\to\sigma'+i(s_++s_-)/4$ and $B'\to B'+(s_--s_+)/2$.  Pulling the differential into the integral as well we get
\begin{multline}\label{zboxbox-partially-dualized}
  Z_{S^2_{(R)}\cup S^2_{(L)}}^{(\yngs(.4,1),\yngs(.4,1))} =
  C \abs{z'}^{2(\frac{1}{2}-\delta_0-\delta_1)} \abs{z'-1}^{2(\delta_1-1)}
  \sum_{B' \in \mathbb Z} \int \frac{\mathrm{d}\sigma'}{2\pi} \biggl[
  (z')^{i\sigma'^+}\bigl(\bar{z}'\bigr)^{i\sigma'^-} \!\!
  \sum_{s_+=\pm 1}\sum_{s_-=\pm 1}\biggl[\\
  \prod_{j=1}^2 \frac{\Gamma\bigl(\frac{s_+}{2}-i\sigma'^++im_j' \bigr)}{\Gamma\bigl(-\frac{s_-}{2}+1+i\sigma'^- -im_j' \bigr)}
  \prod_{s =1}^2 \frac{\Gamma\bigl(-\frac{s_+}{2}+i\sigma'^+-i\anti m_s' \bigr)}{\Gamma\bigl(\frac{s_-}{2}+1-i\sigma'^-+i\anti m_s' \bigr)}
  \\
  \times
  b s_+ \bigl( i\sigma'^+ -i m^{(L),s_+} + b^{-2} z\partial_{z} \bigr)
  b s_- \bigl( i\sigma'^- -i m^{(L),s_-} + b^{-2} \bar{z}\partial_{\bar{z}} \bigr)
  Z_{\text{SQED}}^{(R)}\biggr]\biggr]_{z=z'}
\end{multline}
where we used the shorthand $i\sigma'^{\pm}=i\sigma'\pm B'/2$.
Using $\Gamma(x+1/2)=(x-1/2)\Gamma(x-1/2)$ and $i m'_j - \frac{1}{2} =  i\anti{m}^{(L)}_j$ and $i \anti{m}'_s + \frac{1}{2} = i m^{(L)}_s$, the $s_+$-dependent factors can be massaged into
\begin{multline}
  \sum_{s_+=\pm} \Bigl[ bs_+ \bigl( i\sigma'^+ -i m^{(L),s_+} + b^{-2} z\partial_{z} \bigr)
  \prod_{j=1}^2 \Gamma\bigl(im_j'-i\sigma'^+ +\tfrac{s_+}{2} \bigr)
  \prod_{s=1}^2\Gamma\bigl(i\sigma'^+ -i\anti m_s'-\tfrac{s_+}{2}\bigr) \Bigr]
  \\
  =
  b \bigl(i\sigma'^+ -b^{-2}z\partial_{z}\bigr)^{-1}
  \biggl[
  \prod_{j=1}^2 \Gamma\bigl(im_j'-i\sigma'^+ -\tfrac{1}{2} \bigr)
  \prod_{s=1}^2\Gamma\bigl(i\sigma'^+ -i\anti m_s'+\tfrac{1}{2}\bigr)
  \prod_{j=1}^{2} \bigl( b^{-2} z\partial_{z} - i m'_j + \tfrac{1}{2} \bigr)
  \\
  -
  \prod_{j=1}^2 \Gamma\bigl(im_j'-i\sigma'^+ +\tfrac{1}{2} \bigr) \prod_{s=1}^2\Gamma\bigl(i\sigma'^+ -i\anti m_s'-\tfrac{1}{2}\bigr)
  \prod_{s=1}^{2} \bigl(b^{-2} z\partial_{z} - i\anti{m}'_s-\tfrac{1}{2}\bigr) \biggr] \,.
\end{multline}
The Gamma functions which appear are the same as in~\eqref{zboxbox-partially-dualized}.  The factors linear in $z\partial_z$, when acting on the Coulomb branch representation of~$Z_{\text{SQED}}^{(R)}$, become
\begin{align}
  & b^{-2} z\partial_{z} - i m'_j + \tfrac{1}{2} \to b^{-2} (i\sigma^+-im^{(R)}_j) \,, \\
  & b^{-2} z\partial_{z} - i\anti{m}'_s - \tfrac{1}{2} \to b^{-2} (i\sigma^+-i\anti{m}^{(R)}_s) \,,
\end{align}
where $i\sigma^{\pm}=i\sigma\pm B/2$.  These factors simply shift arguments of Gamma functions.  An analogous expression holds for $s_-$-dependent factors and involves $\bar{z}\partial_{\bar{z}}$.

Altogether we get
\begin{multline}\label{sqedchirallast}
  Z_{S^2_{(R)}\cup S^2_{(L)}}^{(\yngs(.4,1),\yngs(.4,1))} =
  C b^{-4} \abs{z}^{2(\frac{1}{2}-\delta_0-\delta_1)} \abs{z-1}^{2(\delta_1-1)}
  \sum_{B,B' \in \mathbb Z} \int \frac{\mathrm{d}\sigma\mathrm{d}\sigma'}{(2\pi)^2} \Biggl[
  \frac{  z^{i\sigma^+ + i\sigma'^+}\bar{z}^{i\sigma^- + i\sigma'^-} }{\prod_{\pm} (bi\sigma'^{\pm} - i\sigma^{\pm}/b\bigr)}
  \\
  \times
  \sum_{s_+,s_-=\pm 1} \biggl[ s_+ s_-
  \prod_{j=1}^2 \frac{\Gamma\bigl(im_j -i\sigma^+ +\tfrac{s_+}{2} \bigr)}{\Gamma\bigl(1-im_j +i\sigma^- -\tfrac{s_-}{2} \bigr)}
  \frac{\Gamma\bigl(im_j'-i\sigma'^+ -\tfrac{s_+}{2} \bigr)}{\Gamma\bigl(1-im_j'+i\sigma'^- +\tfrac{s_-}{2}\bigr)}
  \\
  \times \prod_{s=1}^2 \frac{\Gamma\bigl(i\sigma^+-i\anti{m}_s -\tfrac{s_+}{2} \bigr)}{\Gamma\bigl(1-i\sigma^-+i\anti{m}_s +\tfrac{s_-}{2} \bigr)}
  \frac{\Gamma\bigl(i\sigma'^+ -i\anti{m}'_s+\tfrac{s_+}{2}\bigr)}{\Gamma\bigl(1-i\sigma'^- +i\anti m_s' -\tfrac{s_-}{2}\bigr)}
  \biggr] \Biggr] \,.
\end{multline}
For all four choices of $(s_+,s_-)$ the shifts by $\pm\frac{s_+}{2}$ and $\pm\frac{s_-}{2}$ in Gamma function arguments can be canceled by shifting $i\sigma^{\pm}\to i\sigma^{\pm}+s_{\pm}/2$ and $i\sigma'^{\pm}\to i\sigma'^{\pm}-s_{\pm}/2$.  The fluxes $B=i\sigma^+-i\sigma^-$ and $B'=i\sigma'^+-i\sigma'^-$ remain integers.  The exponents $\sigma^{\pm}+\sigma'^{\pm}$ of $z$ and $\bar{z}$ stay constant, but $bi\sigma'^{\pm}-i\sigma^{\pm}/b$ are shifted by $-s_{\pm}(b+b^{-1})/2$.  After these manipulations, $s_{\pm}$~only appear in
\begin{equation}\label{and-a-chiral-appears}
  \sum_{s_+=\pm 1} \frac{s_+}{(bi\sigma'^+ - b^{-1} i\sigma^+ - s_+(b+b^{-1})/2\bigr)}
  = \frac{b+b^{-1}}{\prod_{\pm} (bi\sigma'^+ - b^{-1} i\sigma^+ \pm (b+b^{-1})/2\bigr)}
\end{equation}
and a similar factor with $(s_+,\sigma^+,\sigma'^+)\to(s_-,\sigma^-,\sigma'^-)$.  Lo and behold, we have obtained the contribution of a pair of 0d chiral multiplets!\footnote{It was crucial that $z=z'$: otherwise, shifting $\sigma^{\pm}$ and $\sigma'^{\pm}$ would introduce a factor $(z/z')^{s_+/2}(\bar{z}/\bar{z}')^{s_-/2}$ which leads to corrections in~\eqref{and-a-chiral-appears}.}  All in all,
\begin{equation}\label{Zboxbox-duality1}
  Z_{S^2_{(R)}\cup S^2_{(L)}}^{(\yngs(.4,1),\yngs(.4,1))} =
  b^{-4} (b+b^{-1})^2 \biggl( \prod_{j=1}^{2} \prod_{s=1}^{2} \gamma\bigl(i m^{(L)}_s-i\anti{m}^{(L)}_j\bigr) \biggr)
  \abs{z}^{2(\frac{1}{2}-\delta_0-\delta_1)} \abs{z-1}^{2(\delta_1-1)}
  \widetilde{Z}_{S^2_{(R)}\cup S^2_{(L)}}^{(\yngs(.4,1),\yngs(.4,1))} \,.
\end{equation}

The 4d hypermultiplets to which the two 2d/0d quivers couple have slightly shifted masses \eqref{PFoffreeHMI} and~\eqref{PFoffreeHMII}.
Using $\Upsilon(x+1/b)/\Upsilon(x)=\gamma(x/b)b^{-1+2x/b}$,
\begin{equation}
  \frac{Z_{S^4_b}^{\text{free HM}}}{\widetilde{Z}_{S^4_b}^{\text{free HM}}}
  = \frac{b^{4 -4(\sum_{s=1}^{2}im^{(L)}_s) +4(\sum_{j=1}^{2}i\anti{m}^{(L)}_j)}}{\prod_{j,s=1}^2 \gamma(im^{(L)}_s-i\anti{m}^{(L)}_j)} \,.
\end{equation}
These factors cancel most factors in~\eqref{Zboxbox-duality1} and give
\begin{equation}
  Z_{S^4_b}^{\text{free HM}} Z_{S^2_{(R)}\cup S^2_{(L)}}^{(\yngs(.4,1),\yngs(.4,1))}
  =  \frac { (b+b^{-1})^2 \abs{z-1}^{2(i\anti{m}^{(L)}_1+i\anti{m}^{(L)}_2-i m^{(L)}_1-i m^{(L)}_2)} } { b^{4(\sum_{s=1}^{2}im^{(L)}_s) - 4(\sum_{j=1}^{2}i\anti{m}^{(L)}_j)} \abs{z}^{2(i\anti{m}^{(L)}_1+i\anti{m}^{(L)}_2)} }
  \widetilde{Z}_{S^4_b}^{\text{free HM}} \widetilde{Z}_{S^2_{(R)}\cup S^2_{(L)}}^{(\yngs(.4,1),\yngs(.4,1))} \,.
\end{equation}
The prefactors are consistent with the matchings of Appendix~\ref{appendix:matching1} and Appendix~\ref{appendix:matching2}.

Throughout this section the contour integrals surround poles such that $i\sigma\pm B/2$ (and its analogues for other gauge groups) is a twisted mass (times $i$) plus an integer, or half-integer in~\eqref{sqedchirallast}.  This reproduces the set of poles~\eqref{JKlikepoles} selected by the Jeffrey--Kirwan-like residue prescription for the quiver with chiral multiplets.

\section{\label{appendix:fermi-quiver-check}Quiver with 0d Fermi Multiplets}

In the main text we propose the equality~\eqref{general-matching-sphere} between a 4d/2d/0d partition function and a Toda CFT correlator.  We focus here on the case $\nu=\nu'=1$, namely each 2d theory has a single gauge group factor and the Toda CFT side involves a pair of antisymmetric degenerate operators (with coefficients $b$ and $1/b$).  The relation reads
\begin{equation}\label{asas-matching}
  Z\left[\mathcal{T}_{\text{Fermi},\nu=1}\right] = A_1(x,x';\bar{x},\bar{x}')
  \Bigl\langle\hat{V}_{\alpha_{\infty}}(\infty) \hat{V}_{\lambda\omega_1}(1) \hat{V}_{\alpha_{0}}(0)
  \hat{V}_{-b\omega_n}(x,\bar{x}) \hat{V}_{-b^{-1}\omega_{\nf-n'}}(x',\bar{x}')\Bigr\rangle \,.
\end{equation}
Complexified FI parameters of the two nodes are related to positions of punctures by
\begin{equation}
  x=\hat{z}^{(R)}=(-1)^{\nf+n-1}e^{-2\pi\xi^{(R)}+i\vartheta^{(R)}} \quad \text{and} \quad
  x'^{-1}=\hat{z}^{(L)}=(-1)^{\nf+n'-1}e^{-2\pi\xi^{(L)}+i\vartheta^{(L)}} \,.
\end{equation}
Twisted masses obey $b^{-1}m^{(R)}_j=b\anti{m}^{(L)}_j$ and $b^{-1}\anti{m}^{(R)}_s=bm^{(L)}_s$ and these $SU(\nf)\times SU(\nf)\times U(1)$ mass parameters are encoded in momenta as~\eqref{general-matching-alpha}
\begin{align}
  & \alpha_0 - Q = \sum_{j=1}^{\nf} b^{-1} i m^{(R)}_j h_j \,, \\
  & \alpha_\infty - Q = - \sum_{s=1}^{\nf} b^{-1} i\anti{m}^{(R)}_s h_s \,, \\
  & \lambda = n_{\nu} b + (\nf-n'_{\nu'})b^{-1} + \sum_{s=1}^{\nf} b^{-1} i\anti{m}^{(R)}_s - \sum_{j=1}^{\nf} b^{-1} i m^{(R)}_j \,.
\end{align}
The prefactor is
\begin{equation}
  A_1 = A \abs{x}^{2\beta_0} \abs{x'}^{2\beta'_0} \abs{1-x}^{2\beta_1} \abs{1-x'}^{2\beta'_1} \abs{x-x'}^{2\gamma'},
\end{equation}
with
\begin{align}
  A & = \frac{b^{n(2\sum i m^{(R)}-2\sum i\anti{m}^{(R)}-\nf+(\nf-n)b^2)}}{b^{n'(2\sum i m^{(L)}-2\sum i\anti{m}^{(L)}-\nf+(\nf-n')b^{-2})}}\,, \\
  \beta_1 & = -n b \bigl(b+b^{-1}-\lambda/\nf\bigr)\,, \\
  \beta'_1 & = -n' b^{-1}\lambda/\nf\,, \\
  \beta_0 & = -\vev{Q,b\omega_n} + \frac{n}{\nf} \sum_{j=1}^{\nf} i m^{(R)}_j\,, \\
  \beta'_0 & = -\vev{Q,b^{-1}\omega_{n'}} -\frac{n'}{\nf} \sum_{j=1}^{\nf} i\anti{m}^{(L)}_j\,, \\
  \gamma' & = n'n/\nf\,.
\end{align}
We normalize generic, semi-degenerate, and degenerate vertex operators as follows:
\begin{align}
  \hat{V}_{\alpha} & = \frac{\hat{\mu}^{\vev{\alpha-Q,\rho}}}{\prod_{s<t}^{\nf} \Upsilon(\vev{Q-\alpha,h_s-h_t})} V_{\alpha}\,, \\
  \hat{V}_{\lambda h_1} & = \frac{\hat{\mu}^{\vev{\lambda h_1,\rho}}}{\bigl(\Upsilon(b)\bigr)^{\nf-1} \Upsilon(\lambda)} V_{\lambda h_1} \,, \\
  \hat{V}_{-b\omega-\omega'/b} & = \hat{\mu}^{\vev{-b\omega-\omega'/b, \rho}} V_{-b\omega-\omega'/b}
\end{align}
where $\hat{\mu} = \bigl[\pi\mu\gamma(b^2)b^{2-2b^2}\bigr]^{1/b}$ is such that $(\hat{\mu},b)\to(\hat{\mu},1/b)$ is a symmetry of Toda CFT\@.  This normalization makes vertex operators invariant under Weyl symmetries (permutations of the $\vev{\alpha-Q,h_j}$).\footnote{For degenerate vertex operators the normalization differs from~\cite{Gomis:2014eya} by a power of $b^{2(b+\frac{1}{b})}$ to preserve $b\to 1/b$ invariance.  In the Liouville case generic and semi-degenerate vertex operators are the same and we chose a more symmetrical normalization \eqref{Liouville-normalization}; the differences cancel out in correlators we consider.}

The Toda correlator is not known explicitly.  In Appendix~\ref{appendix:asas-sch} we match the leading terms as $x\to 0$: on the gauge theory side this limits selects a solution of the Higgs branch equations for the $U(n)$ gauge group while on the Toda CFT side it selects one primary operator in the fusion of $\hat{V}_{\alpha_0}$ with $\hat{V}_{-b\omega_n}$.  The limits $x\to\infty$, $x'\to 0$ or $x'\to\infty$ are similar.  In Appendix~\ref{appendix:asas-braid} we show that braiding matrices match.\footnote{While these tests are very constraining they do not prove the equality~\eqref{asas-matching} as the two sides could differ by the ratio of two polynomials in $\hat{z}^{(R)}$ and $\hat{z}^{(L)}$ with equal leading term, constant term and value at~$1$.}  In Appendix~\ref{appendix:asas-tch} we match leading exponents in the limits $x\to 1$ (similarly, $x'\to 1$) and $x'\to x$.

\subsection{\label{appendix:asas-sch}Reduction to Four-Point Function}

In this section we explain how to expand the partition function~\eqref{asas-matching} in powers of $x$ and $\bar{x}$, assuming that $\abs{x}<1,\abs{x'}$.  Other orderings of $x$, $1$ and $x'$ are related by exchanging $\hat{z}^{(R,L)}$ or mapping them to their inverse by charge conjugation.

In full, the partition function is
\begin{equation}\label{asas-Z}
  \begin{multlined}
    Z =
    Z_{\text{4d}}
    \sum_{B^{(R)}} \int \frac{\mathrm{d}\sigma^{(R)}}{(2\pi)^n}
    \sum_{B^{(L)}} \int \frac{\mathrm{d}\sigma^{(L)}}{(2\pi)^{n'}}
    Z^{(R)}_{\text{2d}} Z^{(L)}_{\text{2d}} \prod_{a=1}^{n} \prod_{c=1}^{n'} \Biggl[ -\prod_{\pm} \biggl[ \\
    b^{-1}\biggl(i\sigma^{(R)}_a\pm\frac{B^{(R)}_a}{2}\biggr) - b\biggl(i\sigma^{(L)}_c\pm\frac{B^{(L)}_c}{2}\biggr)\biggr] \Biggr] \,,
  \end{multlined}
\end{equation}
where $Z_{\text{4d}}=\prod_{j=1}^{\nf}\prod_{s=1}^{\nf}\Upsilon\bigl((1+i\anti{m}^{(R)}_s-i m^{(R)}_j)/b\bigr)^{-1}$ is the 4d contribution, the last factor is the 0d contribution, and $Z^{(R/L)}_{\text{2d}}$ are given by
\begin{equation}
  Z_{\text{2d}} =
  \begin{aligned}[t]
    & \frac{1}{n!} \hat{z}^{\Tr(i\sigma+B/2)}
    \overline{\hat{z}}^{\Tr(i\sigma-B/2)}
    \prod_{1\leq a<c\leq n} \Bigl( \sigma_{ac}^+\sigma_{ac}^- \Bigr) \\
    & \times \prod_{a=1}^{n} \prod_{j=1}^{\nf} \frac{\Gamma(i m_j-i\sigma_a-B_a/2)}{\Gamma(1-i m_j+i\sigma_a-B_a/2)} \prod_{a=1}^{n} \prod_{s=1}^{\nf} \frac{\Gamma(-i\anti{m}_s+i\sigma_a-B_a/2)}{\Gamma(1+i\anti{m}_s-i\sigma_a-B_a/2)} \,.
  \end{aligned}
\end{equation}
with $i\sigma_{ac}^{\pm}=i\sigma_a\pm\frac{B_a}{2}-i\sigma_c\mp\frac{B_c}{2}$.
Since $\abs{\hat{z}^{(R)}}<1$ we close the $\mathrm{d}\sigma^{(R)}$ contours towards $-i\infty$ and sum residues at poles labeled by a choice of $n$~flavors $J\subseteq\{1,\ldots,\nf\}$ and of $2n$~vorticities $p^{\pm}_j\geq 0$ for $j\in J$, at
\begin{equation}
  \biggl(i\sigma^{(R)}_a\pm\frac{B^{(R)}_a}{2}\biggr)_{1\leq a\leq n}
  = \Bigl( i m^{(R)}_j + p^{\pm}_j \Bigr)_{j\in J}
\end{equation}
up to permutations which cancel a $1/n!$ factor.  The Toda CFT correlator is not known so we focus on leading terms only, namely all $p^{\pm}_j=0$.  The residue for a given~$J$ is then
\begin{equation}\label{asas-schres}
  \begin{aligned}
    & Z_{\text{4d}} \prod_{j\in J} \frac{\abs{\hat{z}^{(R)}}^{2i m^{(R)}_j} \prod_{k\not\in J}^{\nf} \gamma(i m^{(R)}_k-i m^{(R)}_j)}{\prod_{s=1}^{\nf}\gamma(1+i\anti{m}^{(R)}_s-i m^{(R)}_j)}
    \sum_{B^{(L)}} \int \frac{\mathrm{d}\sigma^{(L)}}{(2\pi)^{n'}} \Biggl[
    \\
    & \qquad\quad
    Z_{\text{2d}}^{(L)}
    \prod_{j\in J} \prod_{c=1}^{n'} \prod_{\pm} \mp \biggl[b^{-1}i m^{(R)}_j - b\biggl(i\sigma^{(L)}_c\pm\frac{B^{(L)}_c}{2}\biggr)\biggr] \Biggr]\,.
  \end{aligned}
\end{equation}
The 0d contribution combines with the one-loop determinant of antifundamental chiral multiplets of the left theory (using $i m^{(R)}_j/b=bi\anti{m}^{(L)}_j$)
\begin{equation}
  \begin{aligned}
    & b^{2nn'} \prod_{j\in J} \prod_{c=1}^{n'} \prod_{\pm} \biggl[\mp i\anti{m}^{(L)}_j \pm i\sigma^{(L)}_c - \frac{B^{(L)}_c}{2}\biggr]
    \prod_{c=1}^{n'} \prod_{j=1}^{\nf} \frac{\Gamma(-i\anti{m}^{(L)}_j+i\sigma^{(L)}_c-B^{(L)}_c/2)}{\Gamma(1+i\anti{m}^{(L)}_j-i\sigma^{(L)}_c-B^{(L)}_c/2)}
    \\
    & = b^{2nn'} \prod_{c=1}^{n'} \prod_{j=1}^{\nf} \frac{\Gamma(-i\anti{m}^{(L)}_j+\delta_{j\in J}+i\sigma^{(L)}_c-B^{(L)}_c/2)}{\Gamma(1+i\anti{m}^{(L)}_j-\delta_{j\in J}-i\sigma^{(L)}_c-B^{(L)}_c/2)} \,.
  \end{aligned}
\end{equation}
Namely we get the $S^2$~partition function $Z_{S^2}^{(L)}\sum_{B^{(L)}}\int\mathrm{d}\sigma^{(L)}\,Z_{\text{2d}}^{(L)}/(2\pi)^{n'}$ with shifted masses $i\anti{m}^{(L)}_j\to i\anti{m}^{(L)}_j-\delta_{j\in J}$ or equivalently $i m^{(R)}_j\to i m^{(R)}_j-\delta_{j\in J}b^2$.

The denominator~$\gamma$ functions combine with the 4d contribution thanks to $\Upsilon(x)\gamma(bx)=\Upsilon(x+b)b^{-1+2bx}$, and the numerator~$\gamma$ functions coincide with a Toda CFT three-point function of a degenerate operator.  Altogether, the residue~\eqref{asas-schres} is
\begin{equation}
  \abs{\hat{z}^{(R)}}^{2\sum_{j\in J}i m^{(R)}_j}
  \Bigl(b^{\cdots} \widehat{C}_{-b\omega_n,\alpha_0}^{\alpha_0-bh_J}\Bigr) \bigl[Z_{S^2}^{(L)}Z_{\text{4d}}\bigr]_{i m^{(R)}_j\to i m^{(R)}_j-\delta_{j\in J}b^2}
\end{equation}
where $h_J=\sum_{j\in J}h_j$.  The factor in square brackets is the partition function of the 4d/2d system obtained by only keeping the left 2d theory, known to match a Toda CFT four-point function: this is the special case $n=0$ in our matching~\eqref{asas-matching}.  Including the mass shifts, the residue is
\begin{equation}
  b^{\cdots} \abs{x'}^{2(\beta'_0+\gamma')} \abs{1-x'}^{2\beta'_1} \Bigl\langle \hat{V}_{\alpha_\infty}(\infty) \hat{V}_{\lambda\omega_1}(1) \hat{V}_{\alpha_0-bh_J}(0)
  \hat{V}_{-b^{-1}\omega_{\nf-n'}}(x',\bar{x}')\Bigr\rangle \,.
\end{equation}
We note in particular that only the momentum~$\alpha_0$ is shifted: the momentum $\lambda\omega_1$ is unchanged because mass shifts cancel the change in~$nb$.  The exponent of $\abs{x'}^2$ is also shifted by an amount which turns out to coincide with~$\gamma'$, the exponent of $\abs{x'-x}^2$ in the full matching, as expected in the $x\to 0$ limit.

The structure we find is consistent with the Toda CFT $x\to 0$ OPE namely a sum over weights~$h_J$ of the $n$-th antisymmetric representation,
\begin{equation}
  \hat{V}_{-b\omega_n}(x,\bar{x}) \hat{V}_{\alpha_0}(0)
  \sim \sum_{J}
  \abs{x}^{2[\dimToda(\alpha_0-bh_J)-\dimToda(\alpha_0)-\dimToda(-b\omega_n)]} \widehat{C}_{-b\omega_n,\alpha_0}^{\alpha_0-bh_J} \hat{V}_{\alpha_0-bh_J}(0) + \cdots
\end{equation}
Powers of $\abs{x}^2=\abs{\hat{z}}^2$ work out, namely $\dimToda(\alpha_0-bh_J)-\dimToda(\alpha_0)-\dimToda(-b\omega_n)+\beta_0 = \sum_{j\in J} i m^{(R)}_j$, and constant factors too.  More precisely, we compared the contribution of primary operators to the zero-vorticity terms in the gauge theory expansion.  The gauge theory results provide a prediction for conformal blocks of this Toda CFT five-point function.

\subsection{\label{appendix:asas-braid}Partition Function and Braiding Matrices}

In this section we explain how to expand the partition function~\eqref{asas-matching} for $\abs{\hat{z}^{(R)}}\lessgtr 1$ and $\abs{\hat{z}^{(L)}}\lessgtr 1$ and how these expansions are related by analytic continuation.

The $S^2$~partition function $Z_{S^2}=\sum_B\int\mathrm{d}\sigma\,Z_{\text{2d}}/(2\pi)^n$ of $U(n)$ SQCD can be written as a differential operator acting on that of $n$~copies of SQED (defined as SQCD with $n=1$).  This involves additional K\"ahler parameters $\hat{z}_a$ all set equal to $\hat{z}$ eventually:\footnote{We omit the labels $(R)$ and~$(L)$ when expressions apply equally to both 2d theories: masses $(m,\anti{m})$ stand for $(m^{(R)},\anti{m}^{(R)})$ or $(m^{(L)},\anti{m}^{(L)})$.}
\begin{equation}
  Z_{S^2}
  =
  \begin{aligned}[t]
    \frac{1}{n!}
    \prod_{a<c}^{n} -\abs{\zdz_a-\zdz_c}^2
    \prod_{a=1}^{n} Z_{S^2}^{\text{SQED}}\bigl(\hat{z}_a,\overline{\hat{z}}_a,m,\anti{m}\bigr)
    \Biggr|{}_{\substack{\hat{z}_a=\hat{z}\\\overline{\hat{z}}_a=\overline{\hat{z}}}}
  \end{aligned}
\end{equation}
where $\zdz_a=z_a\partial/\partial z_a$ and so on, and $\abs{\zdz_a-\zdz_c}^2=(\zdz_a-\zdz_c)(\bar{\zdz}_a-\bar{\zdz}_c)$.  We introduce in this way $\hat{z}^{(R)}_a$ for $1\leq a\leq n$ and $\hat{z}^{(L)}_c$ for $1\leq c\leq n'$.  The 0d contribution can then be written as a differential operator $-\abs{b^{-1}\zdz^{(R)}_a - b\zdz^{(L)}_c}^2$ for each $1\leq a\leq n$ and $1\leq c\leq n'$ acting on the product of $n+n'$ SQED partition functions.  All in all,
\begin{equation}\label{asas-Zdiffop}
  \begin{aligned}
    Z =
    &
    \frac{Z_{S^4_b}^{\text{hyper}}}{n!n'!}
    \prod_{a=1}^{n} \prod_{c=1}^{n'} - \abs{b^{-1}\zdz^{(R)}_a - b\zdz^{(L)}_c}^2
    \prod_{a<c}^{n} - \abs{\zdz^{(R)}_a-\zdz^{(R)}_c}^2
    \prod_{a<c}^{n'} - \abs{\zdz^{(L)}_a-\zdz^{(L)}_c}^2
    \\[-2pt]
    & \left.
      \prod_{a=1}^{n} Z_{S^2}^{\text{SQED}}\bigl(\hat{z}^{(R)}_a,\overline{\hat{z}}^{(R)}_a,m^{(R)},\anti{m}^{(R)}\bigr)
      \prod_{c=1}^{n'} Z_{S^2}^{\text{SQED}}\bigl(\hat{z}^{(L)}_c,\overline{\hat{z}}^{(L)}_c,m^{(L)},\anti{m}^{(L)}\bigr)
    \right|{\substack{\hat{z}^{(R)}_a=\hat{z}^{(R)}\\\overline{\hat{z}}^{(R)}_a=\overline{\hat{z}}^{(R)}\\\hat{z}^{(L)}_c=\hat{z}^{(L)}\\\overline{\hat{z}}^{(L)}_c=\overline{\hat{z}}^{(L)}}}
    \,.
  \end{aligned}
\end{equation}

The SQED partition function admits factorized expansions
\begin{equation}\label{SQED-factor}
  Z_{S^2}^{\text{SQED}}(\hat{z},\overline{\hat{z}})
  = \sum_{j=1}^{\nf} \Bigl[ c^\sch_j F^\sch_j\bigl(\hat{z}\bigr) F^\sch_j\bigl(\overline{\hat{z}}\bigr) \Bigl]
  = \sum_{s=1}^{\nf} \Bigl[c^\uch_s F^\uch_s\bigl(\hat{z}\bigr) F^\uch_s\bigl(\overline{\hat{z}}\bigr) \Bigl]
\end{equation}
in terms of holomorphic functions $F^\sch_j(\hat{z}) = (-\hat{z})^{i m_j}(1+\cdots)$ and $F^\uch_s(\hat{z}) = (-\hat{z})^{i\anti{m}_s}(1+\cdots)$, with (hypergeometric) series in powers of $\hat{z}$ and of $\hat{z}^{-1}$ respectively converging for $\abs{\hat{z}}\lessgtr 1$.  Both the series and the constants $c^\sch_j$ and~$c^\uch_s$ are known explicitly and coincide (up to powers of $\hat{z}$ and $1-\hat{z}$) with s-channel and u-channel conformal blocks and three-point functions of a Toda CFT four-point function with a fundamental degenerate insertion~$\hat{V}_{-b\omega_1}$.  We suppress the dependence on masses to keep notations short.  The choice of sign ensures that the functions have the same branch cut, namely the positive real axis.

In the sphere partition function~\eqref{asas-Zdiffop} we can expand both sets of $S^2$~partition functions using~\eqref{SQED-factor}.  Each set can be expanded in the s-channel or the u-channel according to whether $\abs{\hat{z}^{(R)}}\lessgtr 1$ or whether $\abs{\hat{z}^{(L)}}\lessgtr 1$.  We denote the four cases by (s,s)-channel for $\abs{\hat{z}^{(R)}},\abs{\hat{z}^{(L)}}<1$, (u,s)-channel for $\abs{\hat{z}^{(L)}}<1<\abs{\hat{z}^{(R)}}$, (s,u)-channel for $\abs{\hat{z}^{(R)}}<1<\abs{\hat{z}^{(L)}}$, and (u,u)-channel for $1<\abs{\hat{z}^{(R)}},\abs{\hat{z}^{(L)}}$.  In each case, antisymmetry in permuting the $\hat{z}^{(R)}_a$ or $\hat{z}^{(L)}_c$ (and not their complex conjugates) reduces the sum to a sum over choices of an $n$-element and an $n'$-element subsets of $\{1,\ldots,\nf\}$.  For example, the (s,s)-channel is as follows, omitting the 4d contribution and a sign $(-1)^{(n+n')(n+n'-1)/2}$ as they are constant:
\begin{gather}
  \label{asas-Zssch}
  Z \simeq
  \sum_{j_1<\cdots<j_n}^{\nf} \sum_{s_1<\cdots<s_{n'}}^{\nf}
  \Biggl[\biggl(\prod_{a=1}^{n} c^\sch_{j_a}\biggr)
  \biggl(\prod_{c=1}^{n'} c'^\sch_{s_c}\biggr)
  F^\ssch_{\{j\},\{s\}}(\hat{z}^{(R)},\hat{z}^{(L)})
  F^\ssch_{\{j\},\{s\}}\bigl(\overline{\hat{z}}^{(R)},\overline{\hat{z}}^{(L)}\bigr)\Biggr]
  \\
  \label{asas-Fssch}
  \begin{aligned}
    F^\ssch_{\{j\},\{s\}}(\hat{z}^{(R)},\hat{z}^{(L)})
    & =
    \prod_{a=1}^{n} \prod_{c=1}^{n'} (b^{-1}\zdz^{(R)}_a - b\zdz^{(L)}_c)
    \prod_{a<c}^{n} (\zdz^{(R)}_a-\zdz^{(R)}_c)
    \prod_{a<c}^{n'} (\zdz^{(L)}_a-\zdz^{(L)}_c)
    \\
    & \qquad\times
    \prod_{a=1}^{n} F^{\sch(R)}_{j_a}(\hat{z}^{(R)}_a)
    \prod_{c=1}^{n'} F^{\sch(L)}_{s_c}(\hat{z}^{(L)}_c)
    \Biggr|{}_{\substack{\hat{z}^{(R)}_a=\hat{z}^{(R)}\\\hat{z}^{(L)}_c=\hat{z}^{(L)}}} \,.
  \end{aligned}
\end{gather}
where $F^{\sch(R)}$ and~$F^{\sch(L)}$ differ in which twisted masses they involve.

The holomorphic blocks $F^\sch_j(\hat{z})$ and $F^\uch_j(\hat{z})$ of the SQED partition function are related by analytic continuation
\begin{equation}
  F^\sch_j(\hat{z}) \stackrel{\text{braid}}{=} \sum_{s=1}^{\nf} B_{js} F^\uch_s(\hat{z}) \,.
\end{equation}
The braiding matrix is explicitly $B_{js} = D_j \check{B}_{js} \widetilde{D}_s$ in terms of
\begin{equation}\label{BDB}
  \begin{aligned}
    \check{B}_{js} & = \frac{\pi}{\sin\pi(i m_j-i\anti{m}_s)} \\
    D_j & = \frac{\prod_{k\neq j}^{\nf} \Gamma(1-i m_k+i m_j)}{\prod_{t=1}^{\nf} \Gamma(-i\anti{m}_t+i m_j)} \\
    \widetilde{D}_s & = \frac{\prod_{t\neq s}^{\nf} \Gamma(-i\anti{m}_t+i\anti{m}_s)}{\prod_{k=1}^{\nf}\Gamma(1-i m_k+i\anti{m}_s)} \,.
  \end{aligned}
\end{equation}

We deduce braiding matrices relating the various expansions of the 4d/2d/0d partition function.  For definiteness we analytically continue from the (s,s) channel $\abs{\hat{z}^{(R)}},\abs{\hat{z}^{(L)}}<1$ to the (u,s) channel $\abs{\hat{z}^{(R)}}<1<\abs{\hat{z}^{(L)}}$.  For this, apply the SQED braiding $F^{\sch(R)}_{j_a}\bigl(\hat{z}^{(R)}\bigr)\stackrel{\text{braid}}{=}\sum_{t_a=1}^{\nf} B^{(R)}_{j_at_a}F^{\uch(R)}_{t_a}\bigl(\hat{z}^{(R)}\bigr)$ to~\eqref{asas-Fssch} and note that antisymmetry in the~$j_a$ forces all~$t_a$ to be distinct:
\begin{equation}
  F^{\ssch}_{\{j\},\{s\}}(\hat{z}^{(R)},\hat{z}^{(L)})
  \stackrel{\text{braid}}{=}
  \sum_{t_1\neq\cdots\neq t_n}^{\nf} \Biggl[ \biggl(\prod_{a=1}^{n} B^{(R)}_{j_a t_a}\biggr)
  (-1)^{\operatorname{sign}(t)} F^{\usch}_{\{t\},\{s\}}(\hat{z}^{(R)},\hat{z}^{(L)}) \Biggr]
\end{equation}
where the signature of~$t$ (as a permutation of $\{t\}$) is due to the antisymmetry of the differential operator $\prod_{a<c}^{n} (\zdz^{(R)}_a-\zdz^{(R)}_c)$ appearing in the construction of~$F^{\usch}$.  The braiding matrix is thus an antisymmetrized product of SQED braiding matrices.  However, to compare with the relevant Toda CFT braiding matrix we need to normalize the series $F^{\ssch}$ and~$F^{\usch}$ by
their leading coefficients $F^{\ssch}_{\text{lead}}$ and~$F^{\usch}_{\text{lead}}$: the braiding matrix is then
\begin{equation}\label{braid-1}
  B_{\{j\}\{s\},\{t\}\{s\}}
  = \sum_{\sigma\in S_n} \Bigl[ (-1)^{\operatorname{sign}(\sigma)} \prod_{a=1}^{n} B^{(R)}_{j_a t_{\sigma(a)}} \Bigr] \frac{F^{\usch}_{\text{lead}}}{F^{\ssch}_{\text{lead}}} \,.
\end{equation}

The leading term of~$F^{\ssch}$ is simply obtained by applying the differential operator to leading terms of each series $F^\sch$ and~$F'^\sch$: it is $\bigl(-\hat{z}^{(R)}\bigr)^{\sum_{a=1}^n i m^{(R)}_{j_a}} (-\hat{z}^{(L)})^{\sum_{c=1}^{n'}i m^{(L)}_{s_c}}$ times a leading coefficient
\begin{equation}\label{asas-Fssch-lead}
  F^{\ssch}_{\text{lead}} =
  \prod_{a=1}^{n} \prod_{c=1}^{n'} (b^{-1}i m^{(R)}_{j_a} - bi m^{(L)}_{s_c})
  \prod_{a<c}^{n} (i m^{(R)}_{j_a}-i m^{(R)}_{j_c})
  \prod_{a<c}^{n'} (i m^{(L)}_{s_a}-i m^{(L)}_{s_c})
  \,.
\end{equation}
However, the same procedure yields zero for~$F^{\usch}$ if any $t_a=s_c$ because $b^{-1}\zdz^{(R)}_a-b\zdz^{(L)}_c$ then annihilates $\bigl(-\hat{z}^{(R)}_a\bigr)^{i\anti{m}^{(R)}_{t_a}}\bigl(-\hat{z}^{(L)}_c\bigr)^{i m^{(L)}_{s_c}}$.  To get a non-zero result one must consider higher order terms $\bigl(-\hat{z}^{(R)}_a\bigr)^{i\anti{m}^{(R)}_{t_a}-k^{(R)}_a}\bigl(-\hat{z}^{(L)}_c\bigr)^{i m^{(L)}_{s_c}+k^{(L)}_c}$ with $k^{(R)}_a,k^{(L)}_c\geq 0$ not both zero.  Depending on whether $1<\abs{\hat{z}^{(R)}}<\abs{\hat{z}^{(L)}}^{-1}$ or $1<\abs{\hat{z}^{(L)}}^{-1}<\abs{\hat{z}^{(R)}}$ a different term dominates: $(k^{(R)}_a,k^{(L)}_c)=(1,0)$ or $(0,1)$ respectively.  Thus the holomorphic blocks in these two channels have different normalizations to ensure that their leading coefficient is~$1$.  We will be interested in the first of these channels; denoting $i\anti{m}^{(R)\{s\}}_t = i\anti{m}^{(R)}_t-\delta_{t\in\{s\}}$ the leading term of~$F^{\usch}$ is then $\bigl(-\hat{z}^{(R)}\bigr)^{\sum_{a=1}^n i\anti{m}^{(R)\{s\}}_{t_a}} \bigl(-\hat{z}^{(L)}\bigr)^{\sum_{c=1}^{n'}i m^{(L)}_{s_c}}$ times
\begin{multline}\label{asas-Fusch-lead}
  F^{\usch}_{\text{lead}} =
  \prod_{a=1}^{n} \prod_{c=1}^{n'} \bigl(b^{-1}i\anti{m}^{(R)\{s\}}_{t_a} - bi m^{(L)}_{s_c}\bigr)
  \prod_{a<c}^{n} \bigl(i\anti{m}^{(R)\{s\}}_{t_a}-i\anti{m}^{(R)\{s\}}_{t_c}\bigr)
  \prod_{a<c}^{n'} \bigl(i m^{(L)}_{s_a}-i m^{(L)}_{s_c}\bigr)
  \\
  \times \prod_{u\in\{s\}\cap\{t\}}
  \frac{- \prod_{j=1}^{\nf} (i m^{(R)}_j-i\anti{m}^{(R)}_u)}
  {\prod_{v=1}^{\nf} (1+i\anti{m}^{(R)}_v-i\anti{m}^{(R)}_u)}
\end{multline}
where the sign $(-1)^{\#(\{s\}\cap\{t\})}$ is due to the choice of branch cut.  We decompose $B=D\check{B}\widetilde{D}$ in~\eqref{braid-1} according to~\eqref{BDB} and
massage the normalization as
\begin{equation}
  \frac{F^{\usch}_{\text{lead}}}{F^{\ssch}_{\text{lead}}}
  \prod_{\mathclap{j\in\{j\}}} D^{(R)}_j\prod_{\mathclap{t\in\{t\}}} \widetilde{D}^{(R)}_t
  =
  \Biggl[\frac{\prod_{a<c}^{n} (i\anti{m}^{(R)}_{t_a}-i\anti{m}^{(R)}_{t_c})}
  {\prod_{a<c}^{n} (i m^{(R)}_{j_a}-i m^{(R)}_{j_c})}
  \prod_{\mathclap{j\in\{j\}}} D^{(R)}_j\prod_{\mathclap{t\in\{t\}}} \widetilde{D}^{(R)}_t\Biggr]_{i\anti{m}^{(R)}\to i\anti{m}^{(R)\{s\}}}
  \mspace{-25mu}
  (-1)^{\#(\{s\}\cap\{t\})}
  \,.
\end{equation}
Since $\check{B}$ is antiperiodic under integer shifts of $i\anti{m}^{(R)}$ we conclude that the braiding matrix of our 4d/2d/0d sphere partition function is
\begin{equation}
  B_{\{j\}\{s\},\{t\}\{s\}}
  =
  \Biggl[
  \frac{\prod_{a<c}^{n} (i\anti{m}^{(R)}_{t_a}-i\anti{m}^{(R)}_{t_c})}
  {\prod_{a<c}^{n} (i m^{(R)}_{j_a}-i m^{(R)}_{j_c})}
  \sum_{\sigma\in S_n} \Bigl[ (-1)^{\operatorname{sign}(\sigma)} \prod_{a=1}^{n} B^{(R)}_{j_a t_{\sigma(a)}} \Bigr]
  \Biggr]_{i\anti{m}^{(R)}\to i\anti{m}^{(R)\{s\}}} \,.
\end{equation}
Strikingly, the dependence on~$n'$ and on the choice of $n'$~antifundamental flavors~$\{s\}$ is restricted to a shift of mass parameters.  Therefore the braiding matrix is equal to that of a similar 4d/2d/0d setup with the left 2d theory removed and twisted masses shifted.  It was shown in~\cite[Appendix A.3]{Gomis:2014eya} that this gauge theory (SQCD) braiding matrix is equal to the Toda CFT braiding matrix we expect, with momenta $\alpha_0$, the degenerate $-b\omega_{n}$, the semi-degenerate $\lambda\omega_1$ including an $n'b^{-1}$ shift, and $\alpha_{\infty}+b^{-1}\sum_{c=1}^{n'}h_{s_c}$.

To recapitulate, the differential operator introducing 0d fields only affects braiding matrices through a change in normalization; braiding matrices thus essentially coincide with those of a pure 2d theory, known to match with Toda CFT braiding matrices; the normalization change is reproduced by a momentum shift $\alpha_{\infty}\to\alpha_{\infty}+b^{-1}h'$ on the Toda CFT side, itself due to the additional $-b^{-1}\omega_{\nf-n'}$ degenerate insertion.

\subsection{\label{appendix:asas-tch}Channels \texorpdfstring{$\hat{z}^{(R)}\to 1$}{z\003\002R \041\222 1} and \texorpdfstring{$\hat{z}^{(R)}\hat{z}^{(L)}\to 1$}{z\003\002R z\003\002L \041\222 1}}

So far we have focused on expansions corresponding to taking the OPE of degenerate and generic punctures.  We now consider the $x=\hat{z}^{(R)}\to 1$ limit, corresponding to the fusion rule
\begin{equation}\label{asas-tch-fusion}
  \hat{V}_{-b\omega_n} \hat{V}_{\lambda\omega_1} = \hat{V}_{(\lambda+b)\omega_1-b\omega_{n+1}} + \hat{V}_{\lambda\omega_1-b\omega_n}
\end{equation}
derived in~\cite{Gomis:2014eya}.  Since three-point functions of two generic vertex operators and $\hat{V}_{\lambda\omega_1-b\omega_n}$ or $\hat{V}_{(\lambda+b)\omega_1-b\omega_{n+1}}$ are unwieldy we only compare powers of $\abs{1-x}^2$.  On the Toda CFT side these are
\begin{gather}
  \dimToda((\lambda+b)\omega_1-b\omega_{n+1})-\dimToda(\lambda\omega_1)-\dimToda(-b\omega_n)+\beta_1
  = n(b^2+1) + \beta_1 = 0
  \\
  \dimToda(\lambda\omega_1-b\omega_n)-\dimToda(\lambda\omega_1)-\dimToda(-b\omega_n)+\beta_1
  = b\lambda + \beta_1
  = \zeta-n'-n
\end{gather}
where we introduced $\zeta=\nf + \sum_{s=1}^{\nf} i\anti{m}^{(R)}_s - \sum_{j=1}^{\nf} i m^{(R)}_j$ for convenience.

On the gauge theory side we expand each $Z_{S^2}^{\text{SQED}}\bigl(\hat{z}^{(R)}_a,\overline{\hat{z}}^{(R)}_a\bigr)$ in the representation~\eqref{asas-Zdiffop} near $\hat{z}^{(R)}=1$:
\begin{equation}\label{asas-SQEDtch}
  Z_{S^2}^{\text{SQED}}\bigl(\hat{z}^{(R)},\overline{\hat{z}}^{(R)}\bigr)
  = G\bigl(1-\hat{z}^{(R)},1-\overline{\hat{z}}^{(R)}\bigr)
  + \abs{1-\hat{z}^{(R)}}^{2(\zeta-1)} H\bigl(1-\hat{z}^{(R)},1-\overline{\hat{z}}^{(R)}\bigr)
\end{equation}
where $G$ and~$H$ are series in non-negative powers of $(1-\hat{z}^{(R)})$ and $\bigl(1-\overline{\hat{z}}^{(R)}\bigr)$ and it turns out that~$H$ factorizes into a holomorphic times an antiholomorphic series.  When combining such decompositions of $n$~SQED partition functions one would expect $2^n$~terms; however antisymmetry of the holomorphic differential operator $\prod_{a=1}^{n} \prod_{c=1}^{n'} (b^{-1}\zdz^{(R)}_a - b\zdz^{(L)}_c) \prod_{a<c}^{n} (\zdz^{(R)}_a-\zdz^{(R)}_c) \prod_{a<c}^{n'} (\zdz^{(L)}_a-\zdz^{(L)}_c)$ under permuting the $\hat{z}^{(R)}_a$ eliminates all terms involving more than one~$H$.

Acting with a derivative $\zdz^{(R)}=\hat{z}^{(R)}\partial/\partial\hat{z}^{(R)}$ and $\overline{\zdz}^{(R)}$ on~\eqref{asas-SQEDtch} turns the series in $(1-\hat{z}^{(R)})$ and $\bigl(1-\overline{\hat{z}}^{(R)}\bigr)$ into other such series and subtracts one from the exponent $\zeta-1$.  Since the holomorphic differential operator involves at most $n'+n-1$ derivatives~$\zdz^{(R)}_a$ for any given~$\hat{z}^{(R)}_a$, we obtain a decomposition
\begin{equation}
  Z =
  K\bigl(\hat{z}^{(L)},\overline{\hat{z}}^{(L)}; 1-\hat{z}^{(R)},1-\overline{\hat{z}}^{(R)}\bigr)
  + \abs{1-\hat{z}^{(R)}}^{2[\zeta-1-(n'+n-1)]} L\bigl(\hat{z}^{(L)},\overline{\hat{z}}^{(L)}; 1-\hat{z}^{(R)},1-\overline{\hat{z}}^{(R)}\bigr)
\end{equation}
for some series $K$ and~$L$ in non-negative powers of $(1-\hat{z}^{(R)})$ and $\bigl(1-\overline{\hat{z}}^{(R)}\bigr)$ whose coefficients are functions of $\hat{z}^{(L)}$ and $\overline{\hat{z}}^{(L)}$.  This precisely reproduces the Toda CFT exponents.

When $n=1$ we can analyze the leading term in the series~$L$ more precisely.  It must come from acting on $\abs{1-\hat{z}^{(R)}}^{2(\zeta-1)}H(0,0)$ with $n+n'-1=n'$ derivatives $\zdz^{(R)}$ and $n'$~derivatives $\overline{\zdz}^{(R)}$.  In particular for each factor $(b^{-1}\zdz^{(R)} - b\zdz^{(L)}_c)$ the derivative $b\zdz^{(L)}_c$ does not contribute to this leading term.  We obtain
\begin{equation}
  L\bigl(\hat{z}^{(L)},\overline{\hat{z}}^{(L)};0,0\bigr)
  =
  \biggl(H(0,0) \prod_{k=1}^{n'}\bigl(-b^{-2}(k-\zeta)^2\bigr)\biggr)
  Z_{S^4_b}^{\text{hyper}} Z_{S^2}^{\text{$U(n')$ SQCD}} \,.
\end{equation}
The first factor is ($A$~times) $\widehat{C}_{-b\omega_1,\lambda\omega_1}^{(\lambda-b)\omega_1}$ and the other two factors are the Toda CFT correlator $\vev{\hat{V}_{\alpha_\infty}(\infty) \hat{V}_{(\lambda-b)\omega_1}(1) \hat{V}_{\alpha_0}(0) \hat{V}_{-b^{-1}\omega_{\nf-n'}}(x',\bar{x}')}$ as expected for this term of the fusion~\eqref{asas-tch-fusion}.

Let us return to the case of general $n$ and~$n'$ and consider the limit $\hat{z}^{(R)}\hat{z}^{(L)}\to 1$ namely $x^{(L)}\to x^{(R)}$.  On the Toda CFT side the OPE involves a single conformal family
\begin{equation}
  \hat{V}_{-b^{-1}\omega_{\nf-n'}}(x') \hat{V}_{-b\omega_n}(x)
  \sim \abs{x'-x}^{-2\vev{\omega_{\nf-n'},\omega_n}} \Bigl(\hat{V}_{-b\omega_n-b^{-1}\omega_{\nf-n'}}(x) + \cdots\Bigr) \,.
\end{equation}
Taking the factor $\abs{x'-x}^{2\gamma'}$ into account we find the exponent
\begin{equation}
  \gamma'-\vev{\omega_{\nf-n'},\omega_n} = \max(0,n+n'-\nf) \,,
\end{equation}
a non-negative integer.  On the gauge theory side the limit $\hat{z}^{(L)}\to 1/\hat{z}^{(R)}$ is smooth, as can be seen for example from the (u,s)-channel (resp.\@ (s,u)-channel) expansion of the partition function in non-negative (resp.\@ non-positive) powers of $\hat{z}^{(L)}$ and $1/\hat{z}^{(R)}$ explained above~\eqref{asas-Zssch}.  This is consistent with the Toda CFT result, but does not explain the positive exponent when $n+n'>\nf$.  For this, recall first that the partition function is written in the (u,s)-channel as a sum, over subsets $\{t\}$ and $\{s\}$ of $\{1,\ldots,\nf\}$ with $n$ and~$n'$ elements, of series whose first non-zero term is at degree $d=\#(\{t\}\cap\{s\})$.  The holomorphic series are
\begin{equation}\label{asas-tch-lead}
  \bigl(c_{d,0} (\hat{z}^{(R)})^{-d} + \cdots + c_{0,d} (\hat{z}^{(L)})^d\bigr)
\end{equation}
plus terms of higher homogeneous degree in $1/\hat{z}^{(R)}$ and~$\hat{z}^{(L)}$.  Notice now that the Toda CFT exponent $\max(0,n+n'-\nf)$ is the minimal possible value of~$d$ over all subsets $\{t\}$ and~$\{s\}$.  It is plausible that the leading polynomial~\eqref{asas-tch-lead} and all higher order terms are divisible by $(\hat{z}^{(L)}-1/\hat{z}^{(R)})^{\max(0,n+n'-\nf)}$.  Presumably there exists a Seiberg dual of the $U(n)\times U(n')$ quiver, with $n\to\nf-n$ and $n'\to\nf-n'$, whose partition function differs from the original quiver's by a power of $\abs{\hat{z}^{(L)}-1/\hat{z}^{(R)}}^2$ in such a way as to make manifest the factor $(\hat{z}^{(L)}-1/\hat{z}^{(R)})^{n+n'-\nf}$ when $n+n'>\nf$.

\section{\label{appendix:prefactors}Prefactors}

This appendix lists prefactors $A_1$ and~$A_2$ appearing in the equalities that we propose in the main text, relating 4d/2d/0d partition functions and Toda CFT degenerate correlators.  These factors can be absorbed as ambiguities of the partition function,\footnote{Factors independent of mass parameters are a renormalization scheme ambiguity, powers of $\abs{x_{\kappa}}^2$ and $\abs{x'_{\kappa}}^2$ are absorbed in a shift of vector multiplet scalars, and the remaining factors can be canceled by turning on FI parameters for $U(1)$ flavor symmetries.} but can be useful for extracting Toda CFT results (such as new conformal blocks) from the partition functions obtained by localization.  For the matching~\eqref{general-matching-sphere} between a quiver with 0d~Fermi multiplets and a Toda CFT correlator with antisymmetric degenerate operators, the coefficient $A_1(x,x';\bar{x},\bar{x}')=Aa(x,x')a(\bar{x},\bar{x}')$ is given by (we recall $K_{\iota}=n_{\iota}-n_{\iota-1}$ and $K'_{\iota}=n'_{\iota}-n'_{\iota-1}$)
\begin{gather}\label{conj1-axx}
  \!\begin{aligned}
  A & = b^{n_{\nu}(2\sum i m^{(R)}-2\sum i\anti{m}^{(R)}
                         -\nf+(\nf-n_{\nu})b^2)
                 -2\vev{Q,b\sum_{\kappa}\omega_{K_\kappa}}}\\
  & \quad\times b^{n'_{\nu'}(2\sum i m^{(L)}-2\sum i\anti{m}^{(L)}
                -\nf+(\nf-n'_{\nu'})b^{-2})
      -2\vev{Q,b^{-1}\sum_{\kappa}\omega_{K'_{\kappa}}}}
  \end{aligned}
\\
  \begin{aligned}[t]
    a(x,x') & =
    \prod_{\iota=1}^{\nu} \Bigl[ (x_{\iota})^{\beta_{0\iota}} (1-x_{\iota})^{\beta_{1\iota}}\Bigr]
    \prod_{\iota=1}^{\nu'} \Bigl[ (x'_{\iota})^{\beta'_{0\iota}} (1-x'_{\iota})^{\beta'_{1\iota}} \Bigr]
    \\
    &\quad
    \times \prod_{\iota<\kappa}^{\nu} (x_{\iota}-x_{\kappa})^{\gamma_{\iota\kappa}}
    \prod_{\iota=1}^{\nu'}\prod_{\kappa=1}^{\nu} (x'_{\iota}-x_{\kappa})^{\gamma'_{\iota\kappa}}
    \prod_{\iota<\kappa}^{\nu'} (x'_{\iota}-x'_{\kappa})^{\gamma''_{\iota\kappa}}
  \end{aligned}
\end{gather}
where
\begin{align}
  \beta_{0\kappa} & = -\vev{Q,b\omega_{K_{\kappa}}} + \frac{K_{\kappa}}{\nf} \biggl(\sum_{j=1}^{\nf} i m^{(R)}_j\biggr) - \bigl(n_{\kappa-1}+K_{\kappa}(\nu-\kappa)\bigr)\frac{b^2}{2} \\
  \beta'_{0\kappa} & = -\vev{Q,b^{-1}\omega_{K'_{\kappa}}} -\frac{K'_{\kappa}}{\nf} \biggl(\sum_{j=1}^{\nf} i\anti{m}^{(L)}_j\biggr) - \bigl(n'_{\kappa-1}+K'_{\kappa}(\nu'-\kappa)\bigr)\frac{b^{-2}}{2} \\
  \beta_{1\kappa} & = -K_{\kappa} b \bigl(b+b^{-1}-\lambda/\nf\bigr) \\
  \beta'_{1\kappa} & = -K'_{\kappa} b^{-1}\lambda/\nf \\
  \gamma_{\iota\kappa} & = b^2 \bigl(\nf - K_{\max(\iota,\kappa)}\bigr) K_{\min(\iota,\kappa)}/\nf\\
  \gamma'_{\iota\kappa} & = K'_{\iota} K_{\kappa}/\nf \\
  \gamma''_{\iota\kappa} & = b^{-2} K'_{\min(\iota,\kappa)} \bigl(\nf - K'_{\max(\iota,\kappa)}\bigr)/\nf
  \,.
\end{align}

For the quiver with 0d chiral multiplets, we recall that the two FI parameters must be equal.  The prefactor is then $A_2(x,\bar{x})=\widetilde{A}\abs{x}^{2\widetilde{\beta}}\abs{1-x}^{2\widetilde{\gamma}}$ given by (neglecting powers of~$b$ in~$\widetilde{A}$)
\begin{align}
  \label{conj2-A2}
  \widetilde{A} & = b^{\cdots} \frac{\Upsilon'(0)}{\Upsilon'(-nb-n'b^{-1})} = b^{\cdots} \frac{\prod_{k=1}^{n} \gamma(-kb^2) \prod_{k'=1}^{n'} \gamma(-k'b^{-2})}{\prod_{k=1}^{n}\prod_{k'=1}^{n'} (-kb-k'b^{-1})}
  \displaybreak[1]\\
  \widetilde{\beta} & = (nb+n'b^{-1})\biggl(- \tfrac{1}{2} (\nf-2) (b+b^{-1}) - \tfrac{1}{2} (nb+ n'b^{-1}) + \frac{1}{\nf} \sum_{j=1}^{\nf} b^{-1}\bigl(i m_j-\tfrac{1}{2}\bigr)\biggr)
  \displaybreak[1]\\
  \widetilde{\gamma} & = -(nb+n'b^{-1})(b+b^{-1}-\lambda/\nf) \\\nonumber
  & = (nb+n'b^{-1}) \Biggl[
\frac{n - \nf}{\nf} b + \frac{n'-\nf}{\nf} b^{-1} + \frac{1}{\nf}\sum_{s=1}^{\nf} b^{-1} \bigl(i\anti{m}_s+\tfrac{1}{2}\bigr)
- \frac{1}{\nf}\sum_{j=1}^{\nf} b^{-1} \bigl(i m_j-\tfrac{1}{2}\bigr)
\Biggr]\,.
\end{align}

\clearpage

{
\linepenalty=1000
\bibliographystyle{utphys}
\bibliography{ref}
}

\end{document}